\documentclass[longauth]{aa} 

\usepackage{graphicx}
\usepackage{txfonts}
\usepackage{xcolor}
\usepackage{multirow}
\setcitestyle{notesep={ }}
\usepackage[]{hyperref}
\usepackage{caption}

\newcommand{\ttGProll}{GP(\texttt{roll})}
\newcommand{\ttt}{\texttt{t}}

\newcommand{\ttxy}{(\texttt{xy})}
\newcommand{\ttdx}{\texttt{dx}}
\newcommand{\ttdy}{\texttt{dy}}
\newcommand{\ttsme}{\texttt{smear}}
\newcommand{\ttsky}{\texttt{sky}}
\newcommand{\Rslope}{\mathrm{d}\log{R_{p,{\mathrm{valley}}}}/\mathrm{d}\log{P}}
\newcommand{\RMslope}{\mathrm{d}\log{R_{p,{\mathrm{valley}}}}/\mathrm{d}\log{M_{\star}}}
\newcommand{\Rhoslope}{\mathrm{d}\log{\hat{\rho}_{\mathrm{valley}}}/\mathrm{d}\log{P}}
\begin{document}

   \title{Characterising TOI-732\,b and c: New insights into the M-dwarf radius and density valley\thanks{This article uses data from CHEOPS programme CH\_PR100031}}



   \author{A. Bonfanti\inst{1} $^{\href{https://orcid.org/0000-0002-1916-5935}{\includegraphics[scale=0.5]{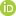}}}$\and
M. Brady\inst{2}$^{\href{https://orcid.org/0000-0003-2404-2427}{\includegraphics[scale=0.5]{figures/orcid.jpg}}}$\and
T. G. Wilson\inst{3} $^{\href{https://orcid.org/0000-0001-8749-1962}{\includegraphics[scale=0.5]{figures/orcid.jpg}}}$\and
J. Venturini\inst{4} $^{\href{https://orcid.org/0000-0001-9527-2903}{\includegraphics[scale=0.5]{figures/orcid.jpg}}}$\and
J. A. Egger\inst{5} $^{\href{https://orcid.org/0000-0003-1628-4231}{\includegraphics[scale=0.5]{figures/orcid.jpg}}}$\and
A. Brandeker\inst{6} $^{\href{https://orcid.org/0000-0002-7201-7536}{\includegraphics[scale=0.5]{figures/orcid.jpg}}}$\and
S. G. Sousa\inst{7} $^{\href{https://orcid.org/0000-0001-9047-2965}{\includegraphics[scale=0.5]{figures/orcid.jpg}}}$\and
M. Lendl\inst{4} $^{\href{https://orcid.org/0000-0001-9699-1459}{\includegraphics[scale=0.5]{figures/orcid.jpg}}}$\and
A. E. Simon\inst{5,8} $^{\href{https://orcid.org/0000-0001-9773-2600}{\includegraphics[scale=0.5]{figures/orcid.jpg}}}$\and
D. Queloz\inst{9,10} $^{\href{https://orcid.org/0000-0002-3012-0316}{\includegraphics[scale=0.5]{figures/orcid.jpg}}}$\and
G. Olofsson\inst{6} $^{\href{https://orcid.org/0000-0003-3747-7120}{\includegraphics[scale=0.5]{figures/orcid.jpg}}}$\and
V. Adibekyan\inst{7} $^{\href{https://orcid.org/0000-0002-0601-6199}{\includegraphics[scale=0.5]{figures/orcid.jpg}}}$\and
Y. Alibert\inst{8,5} $^{\href{https://orcid.org/0000-0002-4644-8818}{\includegraphics[scale=0.5]{figures/orcid.jpg}}}$\and
L. Fossati\inst{1} $^{\href{https://orcid.org/0000-0003-4426-9530}{\includegraphics[scale=0.5]{figures/orcid.jpg}}}$\and
M. J. Hooton\inst{10} $^{\href{https://orcid.org/0000-0003-0030-332X}{\includegraphics[scale=0.5]{figures/orcid.jpg}}}$\and
D. Kubyshkina\inst{1}\and
R. Luque\inst{2} $^{\href{https://orcid.org/0000-0002-4671-2957}{\includegraphics[scale=0.5]{figures/orcid.jpg}}}$\and
F. Murgas\inst{11,12} $^{\href{https://orcid.org/0000-0001-9087-1245}{\includegraphics[scale=0.5]{figures/orcid.jpg}}}$\and
A. J. Mustill\inst{13} $^{\href{https://orcid.org/0000-0002-2086-3642}{\includegraphics[scale=0.5]{figures/orcid.jpg}}}$\and
N. C. Santos\inst{7,14} $^{\href{https://orcid.org/0000-0003-4422-2919}{\includegraphics[scale=0.5]{figures/orcid.jpg}}}$\and
V. Van Grootel\inst{15} $^{\href{https://orcid.org/0000-0003-2144-4316}{\includegraphics[scale=0.5]{figures/orcid.jpg}}}$\and
R. Alonso\inst{16,17} $^{\href{https://orcid.org/0000-0001-8462-8126}{\includegraphics[scale=0.5]{figures/orcid.jpg}}}$\and
J. Asquier\inst{18}\and
T. Bandy\inst{5}\and
T. Bárczy\inst{19} $^{\href{https://orcid.org/0000-0002-7822-4413}{\includegraphics[scale=0.5]{figures/orcid.jpg}}}$\and
D. Barrado Navascues\inst{20} $^{\href{https://orcid.org/0000-0002-5971-9242}{\includegraphics[scale=0.5]{figures/orcid.jpg}}}$\and
S. C. C. Barros\inst{7,14} $^{\href{https://orcid.org/0000-0003-2434-3625}{\includegraphics[scale=0.5]{figures/orcid.jpg}}}$\and
W. Baumjohann\inst{1} $^{\href{https://orcid.org/0000-0001-6271-0110}{\includegraphics[scale=0.5]{figures/orcid.jpg}}}$\and
J. Bean\inst{2}\and
M. Beck\inst{4} $^{\href{https://orcid.org/0000-0003-3926-0275}{\includegraphics[scale=0.5]{figures/orcid.jpg}}}$\and
T. Beck\inst{5}\and
W. Benz\inst{5,8} $^{\href{https://orcid.org/0000-0001-7896-6479}{\includegraphics[scale=0.5]{figures/orcid.jpg}}}$\and
M. Bergomi\inst{21} $^{\href{https://orcid.org/0000-0001-7564-2233}{\includegraphics[scale=0.5]{figures/orcid.jpg}}}$\and
N. Billot\inst{4} $^{\href{https://orcid.org/0000-0003-3429-3836}{\includegraphics[scale=0.5]{figures/orcid.jpg}}}$\and
L. Borsato\inst{21} $^{\href{https://orcid.org/0000-0003-0066-9268}{\includegraphics[scale=0.5]{figures/orcid.jpg}}}$\and
C. Broeg\inst{5,8} $^{\href{https://orcid.org/0000-0001-5132-2614}{\includegraphics[scale=0.5]{figures/orcid.jpg}}}$\and
A. Collier Cameron\inst{22} $^{\href{https://orcid.org/0000-0002-8863-7828}{\includegraphics[scale=0.5]{figures/orcid.jpg}}}$\and
Sz. Csizmadia\inst{23} $^{\href{https://orcid.org/0000-0001-6803-9698}{\includegraphics[scale=0.5]{figures/orcid.jpg}}}$\and
P. E. Cubillos\inst{24,1}\and
M. B. Davies\inst{25} $^{\href{https://orcid.org/0000-0001-6080-1190}{\includegraphics[scale=0.5]{figures/orcid.jpg}}}$\and
M. Deleuil\inst{26} $^{\href{https://orcid.org/0000-0001-6036-0225}{\includegraphics[scale=0.5]{figures/orcid.jpg}}}$\and
A. Deline\inst{4}\and
L. Delrez\inst{27,15} $^{\href{https://orcid.org/0000-0001-6108-4808}{\includegraphics[scale=0.5]{figures/orcid.jpg}}}$\and
O. D. S. Demangeon\inst{7,14} $^{\href{https://orcid.org/0000-0001-7918-0355}{\includegraphics[scale=0.5]{figures/orcid.jpg}}}$\and
B.-O. Demory\inst{8,5} $^{\href{https://orcid.org/0000-0002-9355-5165}{\includegraphics[scale=0.5]{figures/orcid.jpg}}}$\and
D. Ehrenreich\inst{4,28} $^{\href{https://orcid.org/0000-0001-9704-5405}{\includegraphics[scale=0.5]{figures/orcid.jpg}}}$\and
A. Erikson\inst{23}\and
A. Fortier\inst{5,8} $^{\href{https://orcid.org/0000-0001-8450-3374}{\includegraphics[scale=0.5]{figures/orcid.jpg}}}$\and
M. Fridlund\inst{29,30} $^{\href{https://orcid.org/0000-0002-0855-8426}{\includegraphics[scale=0.5]{figures/orcid.jpg}}}$\and
D. Gandolfi\inst{31} $^{\href{https://orcid.org/0000-0001-8627-9628}{\includegraphics[scale=0.5]{figures/orcid.jpg}}}$\and
M. Gillon\inst{27} $^{\href{https://orcid.org/0000-0003-1462-7739}{\includegraphics[scale=0.5]{figures/orcid.jpg}}}$\and
M. Güdel\inst{32}\and
M. N. Günther\inst{18} $^{\href{https://orcid.org/0000-0002-3164-9086}{\includegraphics[scale=0.5]{figures/orcid.jpg}}}$\and
A. Heitzmann\inst{4}\and
Ch. Helling\inst{1,33}\and
S. Hoyer\inst{26} $^{\href{https://orcid.org/0000-0003-3477-2466}{\includegraphics[scale=0.5]{figures/orcid.jpg}}}$\and
K. G. Isaak\inst{18} $^{\href{https://orcid.org/0000-0001-8585-1717}{\includegraphics[scale=0.5]{figures/orcid.jpg}}}$\and
D. Kasper\inst{2} $^{\href{https://orcid.org/0000-0003-0534-6388}{\includegraphics[scale=0.5]{figures/orcid.jpg}}}$\and
L. L. Kiss\inst{34,35}\and
K. W. F. Lam\inst{23} $^{\href{https://orcid.org/0000-0002-9910-6088}{\includegraphics[scale=0.5]{figures/orcid.jpg}}}$\and
J. Laskar\inst{36} $^{\href{https://orcid.org/0000-0003-2634-789X}{\includegraphics[scale=0.5]{figures/orcid.jpg}}}$\and
A. Lecavelier des Etangs\inst{37} $^{\href{https://orcid.org/0000-0002-5637-5253}{\includegraphics[scale=0.5]{figures/orcid.jpg}}}$\and
D. Magrin\inst{21} $^{\href{https://orcid.org/0000-0003-0312-313X}{\includegraphics[scale=0.5]{figures/orcid.jpg}}}$\and
P. F. L. Maxted\inst{38} $^{\href{https://orcid.org/0000-0003-3794-1317}{\includegraphics[scale=0.5]{figures/orcid.jpg}}}$\and
C. Mordasini\inst{5,8}\and
V. Nascimbeni\inst{21} $^{\href{https://orcid.org/0000-0001-9770-1214}{\includegraphics[scale=0.5]{figures/orcid.jpg}}}$\and
R. Ottensamer\inst{32}\and
I. Pagano\inst{39} $^{\href{https://orcid.org/0000-0001-9573-4928}{\includegraphics[scale=0.5]{figures/orcid.jpg}}}$\and
E. Pallé\inst{16,17} $^{\href{https://orcid.org/0000-0003-0987-1593}{\includegraphics[scale=0.5]{figures/orcid.jpg}}}$\and
G. Peter\inst{40} $^{\href{https://orcid.org/0000-0001-6101-2513}{\includegraphics[scale=0.5]{figures/orcid.jpg}}}$\and
G. Piotto\inst{21,41} $^{\href{https://orcid.org/0000-0002-9937-6387}{\includegraphics[scale=0.5]{figures/orcid.jpg}}}$\and
D. Pollacco\inst{3}\and
R. Ragazzoni\inst{21,41} $^{\href{https://orcid.org/0000-0002-7697-5555}{\includegraphics[scale=0.5]{figures/orcid.jpg}}}$\and
N. Rando\inst{18}\and
H. Rauer\inst{23,42,43} $^{\href{https://orcid.org/0000-0002-6510-1828}{\includegraphics[scale=0.5]{figures/orcid.jpg}}}$\and
I. Ribas\inst{44,45} $^{\href{https://orcid.org/0000-0002-6689-0312}{\includegraphics[scale=0.5]{figures/orcid.jpg}}}$\and
G. Scandariato\inst{39} $^{\href{https://orcid.org/0000-0003-2029-0626}{\includegraphics[scale=0.5]{figures/orcid.jpg}}}$\and
D. Ségransan\inst{4} $^{\href{https://orcid.org/0000-0003-2355-8034}{\includegraphics[scale=0.5]{figures/orcid.jpg}}}$\and
A. Seifahrt\inst{2}\and
A. M. S. Smith\inst{23} $^{\href{https://orcid.org/0000-0002-2386-4341}{\includegraphics[scale=0.5]{figures/orcid.jpg}}}$\and
M. Stalport\inst{15,27}\and
G. Stefánsson\inst{46}\thanks{NASA Sagan Fellow} $^{\href{https://orcid.org/0000-0001-7409-5688}{\includegraphics[scale=0.5]{figures/orcid.jpg}}}$\and
M. Steinberger\inst{1}\and
J. Stürmer\inst{47} $^{\href{https://orcid.org/0000-0002-4410-4712}{\includegraphics[scale=0.5]{figures/orcid.jpg}}}$\and
Gy. M. Szabó\inst{48,49} $^{\href{https://orcid.org/0000-0002-0606-7930}{\includegraphics[scale=0.5]{figures/orcid.jpg}}}$\and
N. Thomas\inst{5}\and
S. Udry\inst{4} $^{\href{https://orcid.org/0000-0001-7576-6236}{\includegraphics[scale=0.5]{figures/orcid.jpg}}}$\and
E. Villaver\inst{16,17}\and
N. A. Walton\inst{50} $^{\href{https://orcid.org/0000-0003-3983-8778}{\includegraphics[scale=0.5]{figures/orcid.jpg}}}$\and
K. Westerdorff\inst{40}\and
T. Zingales\inst{41,21} $^{\href{https://orcid.org/0000-0001-6880-5356}{\includegraphics[scale=0.5]{figures/orcid.jpg}}}$
}

\institute{\label{inst:1} Space Research Institute, Austrian Academy of Sciences, Schmiedlstrasse 6, A-8042 Graz, Austria \\
     \email{andrea.bonfanti@oeaw.ac.at} \and
\label{inst:2} Department of Astronomy \& Astrophysics, University of Chicago, Chicago, IL 60637, USA \and
\label{inst:3} Department of Physics, University of Warwick, Gibbet Hill Road, Coventry CV4 7AL, United Kingdom \and
\label{inst:4} Observatoire astronomique de l'Université de Genève, Chemin Pegasi 51, 1290 Versoix, Switzerland \and
\label{inst:5} Weltraumforschung und Planetologie, Physikalisches Institut, University of Bern, Gesellschaftsstrasse 6, 3012 Bern, Switzerland \and
\label{inst:6} Department of Astronomy, Stockholm University, AlbaNova University Center, 10691 Stockholm, Sweden \and
\label{inst:7} Instituto de Astrofisica e Ciencias do Espaco, Universidade do Porto, CAUP, Rua das Estrelas, 4150-762 Porto, Portugal \and
\label{inst:8} Center for Space and Habitability, University of Bern, Gesellschaftsstrasse 6, 3012 Bern, Switzerland \and
\label{inst:9} ETH Zurich, Department of Physics, Wolfgang-Pauli-Strasse 2, CH-8093 Zurich, Switzerland \and
\label{inst:10} Cavendish Laboratory, JJ Thomson Avenue, Cambridge CB3 0HE, UK \and
\label{inst:11} Instituto de Astrof\'isica de Canarias (IAC), E-38205 La Laguna, Tenerife, Spain \and
\label{inst:12} Departamento de Astrof\'isica, Universidad de La Laguna (ULL), E-38206 La Laguna, Tenerife, Spain \and
\label{inst:13} Lund Observatory, Dept. of Astronomy and Theoretical Physics, Lund University, Box 43, 22100 Lund, Sweden \and
\label{inst:14} Departamento de Fisica e Astronomia, Faculdade de Ciencias, Universidade do Porto, Rua do Campo Alegre, 4169-007 Porto, Portugal \and
\label{inst:15} Space sciences, Technologies and Astrophysics Research (STAR) Institute, Université de Liège, Allée du 6 Août 19C, 4000 Liège, Belgium \and
\label{inst:16} Instituto de Astrofisica de Canarias, Via Lactea s/n, 38200 La Laguna, Tenerife, Spain \and
\label{inst:17} Departamento de Astrofisica, Universidad de La Laguna, Astrofísico Francisco Sanchez s/n, 38206 La Laguna, Tenerife, Spain \and
\label{inst:18} European Space Agency (ESA), European Space Research and Technology Centre (ESTEC), Keplerlaan 1, 2201 AZ Noordwijk, The Netherlands \and
\label{inst:19} Admatis, 5. Kandó Kálmán Street, 3534 Miskolc, Hungary \and
\label{inst:20} Depto. de Astrofisica, Centro de Astrobiologia (CSIC-INTA), ESAC campus, 28692 Villanueva de la Cañada (Madrid), Spain \and
\label{inst:21} INAF, Osservatorio Astronomico di Padova, Vicolo dell'Osservatorio 5, 35122 Padova, Italy \and
\label{inst:22} Centre for Exoplanet Science, SUPA School of Physics and Astronomy, University of St Andrews, North Haugh, St Andrews KY16 9SS, UK \and
\label{inst:23} Institute of Planetary Research, German Aerospace Center (DLR), Rutherfordstrasse 2, 12489 Berlin, Germany \and
\label{inst:24} INAF, Osservatorio Astrofisico di Torino, Via Osservatorio, 20, I-10025 Pino Torinese To, Italy \and
\label{inst:25} Centre for Mathematical Sciences, Lund University, Box 118, 221 00 Lund, Sweden \and
\label{inst:26} Aix Marseille Univ, CNRS, CNES, LAM, 38 rue Frédéric Joliot-Curie, 13388 Marseille, France \and
\label{inst:27} Astrobiology Research Unit, Université de Liège, Allée du 6 Août 19C, B-4000 Liège, Belgium \and
\label{inst:28} Centre Vie dans l’Univers, Faculté des sciences, Université de Genève, Quai Ernest-Ansermet 30, 1211 Genève 4, Switzerland \and
\label{inst:29} Leiden Observatory, University of Leiden, PO Box 9513, 2300 RA Leiden, The Netherlands \and
\label{inst:30} Department of Space, Earth and Environment, Chalmers University of Technology, Onsala Space Observatory, 439 92 Onsala, Sweden \and
\label{inst:31} Dipartimento di Fisica, Università degli Studi di Torino, via Pietro Giuria 1, I-10125, Torino, Italy \and
\label{inst:32} Department of Astrophysics, University of Vienna, Türkenschanzstrasse 17, 1180 Vienna, Austria \and
\label{inst:33} Institute for Theoretical Physics and Computational Physics, Graz University of Technology, Petersgasse 16, 8010 Graz, Austria \and
\label{inst:34} Konkoly Observatory, Research Centre for Astronomy and Earth Sciences, 1121 Budapest, Konkoly Thege Miklós út 15-17, Hungary \and
\label{inst:35} ELTE E\"otv\"os Lor\'and University, Institute of Physics, P\'azm\'any P\'eter s\'et\'any 1/A, 1117 Budapest, Hungary \and
\label{inst:36} IMCCE, UMR8028 CNRS, Observatoire de Paris, PSL Univ., Sorbonne Univ., 77 av. Denfert-Rochereau, 75014 Paris, France \and
\label{inst:37} Institut d'astrophysique de Paris, UMR7095 CNRS, Université Pierre \& Marie Curie, 98bis blvd. Arago, 75014 Paris, France \and
\label{inst:38} Astrophysics Group, Lennard Jones Building, Keele University, Staffordshire, ST5 5BG, United Kingdom \and
\label{inst:39} INAF, Osservatorio Astrofisico di Catania, Via S. Sofia 78, 95123 Catania, Italy \and
\label{inst:40} Institute of Optical Sensor Systems, German Aerospace Center (DLR), Rutherfordstrasse 2, 12489 Berlin, Germany \and
\label{inst:41} Dipartimento di Fisica e Astronomia "Galileo Galilei", Università degli Studi di Padova, Vicolo dell'Osservatorio 3, 35122 Padova, Italy \and
\label{inst:42} Zentrum für Astronomie und Astrophysik, Technische Universität Berlin, Hardenbergstr. 36, D-10623 Berlin, Germany \and
\label{inst:43} Institut fuer Geologische Wissenschaften, Freie Universitaet Berlin, Maltheserstrasse 74-100,12249 Berlin, Germany \and
\label{inst:44} Institut de Ciencies de l'Espai (ICE, CSIC), Campus UAB, Can Magrans s/n, 08193 Bellaterra, Spain \and
\label{inst:45} Institut d’Estudis Espacials de Catalunya (IEEC), Gran Capità 2-4, 08034 Barcelona, Spain \and
\label{inst:46} Department of Astrophysical Sciences, Princeton University, 4 Ivy Lane, Princeton, NJ 08540, USA \and
\label{inst:47} Landessternwarte, Zentrum für Astronomie der Universität Heidelberg, Königstuhl 12, D-69117 Heidelberg, Germany \and
\label{inst:48} ELTE E\"otv\"os Lor\'and University, Gothard Astrophysical Observatory, 9700 Szombathely, Szent Imre h. u. 112, Hungary \and
\label{inst:49} HUN-REN-ELTE Exoplanet Research Group, Szent Imre h. u. 112., Szombathely, H-9700, Hungary \and
\label{inst:50} Institute of Astronomy, University of Cambridge, Madingley Road, Cambridge, CB3 0HA, United Kingdom     
}

   \date{}

 
  \abstract
   {TOI-732 is an M dwarf hosting two transiting planets that are located on the two opposite sides of the radius valley. Inferring a reliable demographics for this type of systems is key to understanding their formation and evolution mechanisms.}
   {By doubling the number of available space-based observations and increasing the number of radial velocity (RV) measurements, we aim at refining the parameters of TOI-732\,b and c. We also use the results to study the slope of the radius valley and the density valley for a well-characterised sample of M-dwarf exoplanets.}
   {We performed a global Markov chain Monte Carlo analysis by jointly modelling ground-based light curves and CHEOPS and TESS observations, along with RV time series both taken from the literature and obtained with the MAROON-X spectrograph. The slopes of the M-dwarf valleys were quantified via a support vector machine (SVM) procedure.}
   {TOI-732\,b is an ultrashort-period planet ($P=0.76837931_{-0.00000042}^{+0.00000039}$ d) with a radius $R_b=1.325_{-0.058}^{+0.057}\,R_{\oplus}$, a mass $M_b=2.46\pm0.19\,M_{\oplus}$, and thus a mean density $\rho_b=5.8_{-0.8}^{+1.0}$\,g\,cm$^{-3}$, while the outer planet at $P=12.252284\pm0.000013$ d has $R_c=2.39_{-0.11}^{+0.10}\,R_{\oplus}$, $M_c=8.04_{-0.48}^{+0.50}\,M_{\oplus}$, and thus $\rho_c=3.24_{-0.43}^{+0.55}$\,g\,cm$^{-3}$. Even with respect to the most recently reported values, this work yields uncertainties on the transit depths and on the RV semi-amplitudes that are smaller up to a factor of $\sim$\,1.6 and $\sim$\,2.4 for TOI-732\,b and c, respectively. Our calculations for the interior structure and the location of the planets in the mass-radius diagram lead us to classify TOI-732\,b as a super-Earth and TOI-732\,c as a mini-Neptune. Following the SVM approach, we quantified $\Rslope=-0.065_{-0.013}^{+0.024}$, which is flatter than for Sun-like stars. In line with former analyses, we note that the radius valley for M-dwarf planets is more densely populated, and we further quantify the slope of the density valley as $\Rhoslope=-0.02_{-0.04}^{+0.12}$.}
   {Compared to FGK stars, the weaker dependence of the position of the radius valley on the orbital period might indicate that the formation shapes the radius valley around M dwarfs more strongly than the evolution mechanisms.}


   \maketitle
%
\section{Introduction}
M dwarfs are the most common stars in the Universe \citep[e.g.][]{bastian2010}. Because of their low mass and small radius, they are quite attractive in the domain of exoplanetology: It is easier to detect low-mass planets with the transit method \citep[e.g.][]{winn2010} and the radial velocity (RV) technique \citep[e.g.][]{hatzes2016}. In addition, the habitable zone \citep[HZ;][]{kasting1993,kopparapu2013} around M dwarfs is located closer to the host star than in stars of earlier spectral type, and therefore, it is more likely that planets in the habitable zone (HZ) of M dwarfs are found because both the transit method and the RV technique preferentially detect close-in planets. 

A few mechanisms may act against the possibility that M dwarfs harbour life, such as their strong magnetic activity \citep[e.g.][]{saar1985,reiners2009,shulyak2019}, accompanied by flares and high-energy emission that induce atmospheric escape \citep[e.g.][]{luger2015,tilley2019}, or the likely scenario of tidally locked close-in exoplanets that lead to extreme surface temperature gradients \citep[e.g.][]{barnes2017}. However, different works highlighted ways in which planets might be able to become and remain habitable despite the unfavourable stellar environment \citep{kay2016,sergeev2020,childs2022,ojha2022,lobo2023}. 

The first exoplanet discovered around an M dwarf is GJ\,876\,b. This Jovian-mass planet was independently detected via the RV technique by \citet{delfosse1998} and \citet{marcy1998}. Subsequent studies have also revealed smaller exoplanets \citep[e.g. GJ\,436\,b;][]{butler2004,gillon2007} or multiplanet systems containing Neptune-sized planets and super-Earths, such as GJ\,581 \citep{bonfils2005,udry2007}. Since then, the search for exoplanets orbiting M dwarfs has been rather prolific; the community has never lost interest in M-dwarf exoplanets, as proven by several recent discoveries, such as TOI-244\,b \citep{castroGonzalez2023}, TOI-715\,b \citep{dransfield2023}, K2-416\,b and K2-417\,b \citep{incha2023}, TOI-3785\,b \citep{powers2023}, TOI-3984\,A\,b and TOI-5293\,A\,b \citep{canas2023}, TOI-1680\,b \citep{ghachoui2023}, and TOI-2084\,b and TOI-4184\,b \citep{barkaoui2023}.

In this work, we characterise TOI-732. This M dwarf is orbited by an ultrashort-period planet (TOI-732\,b, $P$\,$\sim$\,0.77 d) and an outer planet (TOI-732\,c, $P$\,$\sim$\,12.25 d). The system has been studied by \citet[][hereafter C20]{cloutier2020}, \citet[][hereafter N20]{nowak2020}, and \citet[][hereafter L22]{luque2022}. We add a new TESS \citep[Transiting Exoplanet Survey Satellite;][]{ricker2015} sector, 25 novel space-based light curves (LCs) observed by CHEOPS \citep[Characterising Exoplanet Satellite;][]{benz2021}, and 39 RV data points taken with the high-precision echelle spectrograph MAROON-X \citep{Seifahrt18,Seifahrt22} to the already published data. Even with respect to the most recent analysis by L22, we almost doubled the number of space-based transit events, and we significantly increased the number of RV data points, which allowed us to significantly reduce the uncertainties in the planetary parameters.

From the point of view of planet formation and evolution, TOI-732 is an interesting system because the planets are located on the two opposite sides of the radius valley \citep{fulton2017}. The paucity of exoplanets with orbital periods $P<100$\,d in the $R_p$\,1.5--2.0\,$R_{\oplus}$ radius range determines a bimodal $R_p$ distribution that peaks at $\sim$\,1.3 and $\sim$2.4\,$R_{\oplus}$ \citep[e.g.][]{fulton2017,fulton2018,vanEylen2018}. One interpretation of this distribution suggests that sub-Neptunes likely form with rocky cores with sizes 1.5\,$R_{\oplus}$ or smaller, surrounded by an envelope with a low mean molecular weight that is subject to photo-evaporation \citep[e.g.][]{lammer2003,lopez2013,owen2013,chen2016}. However, atmospheric erosion may also be driven by cooling of rocky cores \citep[core-powered mass loss; e.g.][]{ginzburg2018,gupta2019} and by energy release following cohesive giant impacts during proto-planet formation \citep[impact erosion;e.g. ][]{kegerreis2020}. Finally, the smallest sub-Neptunes might also be the result of late formation within gas-depleted discs \citep[gas-poor formation; e.g.][]{lee2014,lee2021,lee2022}.

These mechanisms that account for the radius valley assume that super-Earths and sub-Neptunes assembled from the same underlying population of dry rocky cores, which might or might not retain a H-He atmosphere. However, global planet formation models show that migration is a key mechanism delivering water-rich sub-Neptunes at short orbital periods \citep[e.g.][]{Alibert2013, venturini2020, Emsenhuber21}, especially for low-mass planets around M dwarfs \citep{Alibert17, Miguel2020,burn2021}. In particular, \citet{venturini2020} showed that the radius valley emerges from a combination of formation and evolution processes that separate small rocky from larger water-rich- planets that formed beyond the ice line.
Observational support for this scenario was recently found by L22, who studied a sub-sample of M-dwarf exoplanets and reported a clear density gap that separated super-Earths (identified as rocky planets) from mini-Neptunes (identified as water-ice-rich worlds and not as rocky cores surrounded by H-He). They also concluded that the radius dispersion, especially among puffy exoplanets, may be the consequence of the different accretion histories of H-He envelopes and not of the atmospheric mass loss.

Obtaining observational data is key to investigating the relative importance of the different formation and evolution scenarios. So far, most of the studies have investigated the nature of the radius valley by focusing on FGK stars \citep[e.g.][]{vanEylen2018,macdonald2019,martinez2019,ho2023}, while only a few works specifically drew attention to low-mass stars \citep{cloutier2020valley,vanEylen2021,luque2022}. The discoveries of M-dwarf systems in which planets straddle the radius gap have steadily increased. They comprise, for example, TOI-776 \citep{luque2021,fridlund2023}, TOI-1634 \citep{cloutier2021,hirano2021}, TOI-270 \citep{vanEylen2021}, TOI-1468 \citep{chaturvedi2022}, K2-3 \citep{diamondLowe2022}, TOI-2096 \citep{pozuelos2023}, and LHS\,1903 \citep{wilson2023}. More generally, the parameters of planets orbiting low-mass stars are progressively known with increasingly better precision. This work therefore also aims at describing the characteristics of the radius valley better for planets orbiting M dwarfs.

This paper is organised as follows. Sect.~\ref{sec:star} presents the stellar properties, Sects.~\ref{sec:photometry} and \ref{sec:RV} describe the photometric and RV data, that were used to characterise the system as explained in Sect.~\ref{sec:results}. Sect.~\ref{sec:Rvalley} investigates the radius and density valleys of M-dwarf exoplanets from a quantitative perspective by using the most precise collection of planetary parameters available so far. Finally, Sect.~\ref{sec:conclusions} gathers our conclusions.

\section{Host star properties}\label{sec:star}
TOI-732 is an M4\,V star \citep{scholz2005} located $\sim$\,22 pc away from us \citep{GaiaDR3}, with magnitudes $V=13.14\pm0.04$ \citep{zacharias2012} and $K=8.204\pm0.021$ \citep{cutri2003}. It is part of a visual binary system, and its companion is known as LP\,729-55 and is located at an angular separation $\theta=15.81\pm0.15''$, which implies a projected orbital distance of $348\pm3$ AU (N20). LP\,729-55 is fainter by $\sim$\,2 $K$-band mag than TOI-732, and its spectral type has been estimated by N20 as M5.0\,V.

To estimate the stellar effective temperature $T_{\mathrm{eff}}$ and metallicity [Fe/H] of TOI-732, we used the ODUSSEAS\footnote{\url{https://github.com/AlexandrosAntoniadis/ODUSSEAS}} code \citep[][]{Antoniadis-2020}, and we input spectroscopic data taken from the ESO archive. Although we were able to combine data from ESPRESSO \citep{pepe2021} and HARPS \citep{mayor2003}, we used the latter because it provided a higher signal-to-noise ratio (S/N) of the combined spectrum. ODUSSEAS uses the ridge regression \citep{hoerlKennard1970} implemented via the machine-learning \textsc{Python} package \textit{scikit-learn} \citep{sklearn2011}, which is trained to measure the pseudo-equivalent widths of more than 4000 stellar absorption lines. Using a library of HARPS spectra for several M stars with well-defined reference parameters from interferometric calibrations (Antoniadis-Karnavas et al. in prep.), ODUSSEAS derived $T_{\mathrm{eff}}=3358\pm92$\,K and [Fe/H]\,=\,$0.06\pm0.11$\,dex. The trigonometric surface gravity was estimated using $T_{\mathrm{eff}}$ and [Fe/H] in combination with the Gaia parallax \citep{GaiaDR3} and photometry, following the same procedure as described in \citet[][]{Sousa-2021}, which yields $\log{g}=4.85\pm0.11$.

Because of the heavy line blending, the determination of the individual elemental abundances of M dwarfs from visible spectra is challenging \citep[e.g.][]{Maldonado-20}. In this work, we estimated the abundance of Mg and Si following the procedure presented in \citet{Demangeon-21}. In brief, we used the systemic radial velocity ($RV_{\mathrm{sys}}$), parallax ($\pi$), right ascension ($\alpha$), declination ($\delta$), and proper motions ($\mu_\alpha$ and $\mu_\delta$) from \textit{Gaia} DR3 \citep{GaiaDR3} to derive the Galactic space velocity UVW of TOI-732 via the \texttt{GalVel\_Pop.py} routine\footnote{\url{https://github.com/vadibekyan/GalVel_Pop/blob/main/GalVel_Pop.py}}. We obtained $U = 4.0 \pm 0.1$ km\,s$^{-1}$, $V = -10.3 \pm 0.3$ km\,s$^{-1}$, and $W = -27.5 \pm 0.2$ km\,s$^{-1}$ with respect to the local standard of rest (LSR), adopting the solar peculiar motion from \citet{Schonrich-10}. Based on these velocities, adopting the characteristic parameters of Galactic stellar populations from \citet{Reddy-06}, and following \citet{Adibekyan-12b}, we estimated that the star belongs to the Galactic thin disc with a 97\% probability. Then, from the APOGEE DR17 \citep{Abdurrouf-22}, we selected cool stars with metallicities similar to that of TOI-732 that belong to the chemically defined Galactic thin disc. We obtained a sample of several thousand stars, for which we calculated the mean abundance of Mg and Si, and their standard deviation (star-to-star scatter). After taking the stellar metallicity into account, we obtained [Mg/H] = 0.04 $\pm$ 0.20 dex and [Si/H] = 0.02 $\pm$ 0.21 dex. 

We computed the infrared flux method (IRFM) \citep{Blackwell1977} radius of TOI-732 using a modified Markov chain Monte Carlo (MCMC) approach \citep{Schanche2020}. We constructed spectral energy distributions (SEDs) by constraining stellar atmospheric models from three catalogues \citep{Kurucz1993,Castelli2003,Allard2014} with the results of our spectral analysis. From these, we calculated the stellar bolometric flux via comparison of synthetic and observed broadband photometry in the following bandpasses: {\it Gaia} $G$, $G_\mathrm{BP}$, and $G_\mathrm{RP}$, 2MASS $J$, $H$, and $K$, and \textit{WISE} $W1$ and $W2$ \citep{GaiaDR3,Skrutskie2006,Wright2010}. The bolometric flux was first converted into effective temperature and angular diameter and then into stellar radius using the offset-corrected \textit{Gaia} parallax \citep{lindegren2021}. The stellar atmospheric modelling uncertainties were accounted for by using a Bayesian modelling that averaged the radius posterior distributions. The complex spectral features of M-dwarfs can cause degeneracies in the strengths of molecular lines and thus in the bolometric flux computation within the MCMC when using different atmospheric models. This propagates to large errors on M-dwarf IRFM radii compared to using empirical relations (see C20 and N20). The consistency between our estimate and the outcomes in both C20 and N20 is well below 1\,$\sigma$ and therefore, we attributed the typical uncertainty to $R_{\star}$ as derived from empirical relations. We obtained $R_{\star}=0.380\pm0.012\,R_{\odot}$.

We used $T_{\mathrm{eff}}$, [Fe/H], and $R_{\star}$ along with their error bars to then derive the stellar mass $M_{\star}$ from two different evolutionary models. In detail, we applied the isochrone placement algorithm \citep{bonfanti2015,bonfanti2016}, which is designed to interpolate the set of input parameters within pre-computed grids of PARSEC\footnote{\textsl{PA}dova and T\textsl{R}ieste \textsl{S}tellar \textsl{E}volutionary \textsl{C}ode: \url{http://stev.oapd.inaf.it/cgi-bin/cmd}} v1.2S \citep{marigo2017} isochrones and tracks, and we obtained a first estimate for the mass. A second estimate was instead obtained via the Code Liègeois d'Évolution Stellaire \citep[CLES;][]{scuflaire2008}, which builds the best-fit evolutionary tracks on the fly following the Levenberg-Marquadt minimisation scheme \citep{salmon2021}.
As outlined in \citet{bonfanti2021}, the consistency of the two results is checked through a $\chi^2$-based criterion, after which the mass distributions inferred from the two different evolutionary models are merged together. We finally obtained $M_{\star}=0.381_{-0.034}^{+0.024}\,M_{\odot}$.

Both C20 and N20 have derived the stellar mass and obtained $M_{\star,C20}=0.401\pm0.012\,M_{\odot}$ and $M_{\star,N20}=0.379\pm0.016\,M_{\odot}$. These estimates are consistent with ours, but are more precise by a factor of $\sim$\,2, but the uncertainties appear to be underestimated. In detail, C20 used the mass-luminosity relation from \citet{benedict2016}. Even considering the K-band luminosity, which yields the most satisfactory fit, the average root mean square (rms) of the residuals is 0.014\,$M_{\odot}$, which is larger than the reported estimate. Furthermore, the mass residuals in the neighbourhood of the TOI-732 absolute stellar magnitude (i.e. $M_K=6.494\pm0.021$\,mag) as displayed in \citet[Fig.~23, right panel]{benedict2016} are higher than the average value by about a factor of two. Instead, N20 used the mass-radius relation from \citet{schweitzer2019}, whose rms inherent to the fit is 0.02\,$M_{\odot}$. When the fit-related source of errors is accounted for, the uncertainties on $M_{\star}$ from both C20 and N20 become similar to ours. Therefore, our mass uncertainty is probably genuine and robust, also considering that it comes from evolutionary models employing different physical ingredients and was inferred using different derivation algorithms \citep[see][for further details]{bonfanti2021}.

As is well known, M dwarfs evolve very slowly. Any age inference via isochrone fitting is therefore inconclusive. However, due to stellar interactions that manifest themselves as kinematic disturbances over the lifetimes of stars, we can estimate the stellar age based on kinematics alone\citep{Wielen1977,Nordstrom2004,Casagrande2011,Maciel2011}. We used the method of \citet{AlmeidaFernandes2018}, which allows for age estimates based on kinematic-age probability distributions that were formalised and bench-marked using a sample of 9000 stars in the Geneva-Copenhagen Survey whose isochronal ages are known. For this study, we computed the age of TOI-732 using the Galactic $U$, $V$, and $W$ velocities and {\it Gaia} DR3 Galactic reference coordinates\citep{GaiaDR3}, and we obtained an age of $3.10^{+6.20}_{-0.98}$\,Gyr. All the relevant stellar parameters are reported in Table~\ref{tab:star}.

We further investigated the evolutionary stage of TOI-732 by computing the equivalent width (EW) of the H$\alpha$ emission component, which has been related to the age of M dwarfs by \citet{kiman2021}.
To this end, we used the HARPS and ESPRESSO combined spectra. Following the procedure described in \citet[][]{Schmidt-2015} and \citet[][]{West-2011}, we calculated a H$\alpha$ EW of 0.64 \AA{} from the HARPS spectra and 0.52 \AA{} from the ESPRESSO spectra.

\citet{kiman2021} defined a boundary that separates active from inactive M dwarfs, which latter have an H$\alpha$ EW below a colour-dependent threshold value. Given the $G-G_{\mathrm{RP}}=1.197$ colour \citep{GaiaDR3} of TOI-732, the corresponding activity boundary is H$\alpha$-EW$_{\mathrm{bound}}=0.85$ \AA. Because both our H$\alpha$-EW estimates derived from HARPS and ESPRESSO are below the threshold value, TOI-732 can be categorised as inactive. \citet{kiman2021} pointed out that inactive stars can be found at different evolutionary stages, and their age therefore cannot be well constrained in this way. However, they noted an increasing number of stars with low H$\alpha$ EW as age increases. In particular, mid-M-type stars show a strong H$\alpha$ decline after 1 Gyr, and TOI-732 is therefore likely to be older than one billion years, which is consistent with our kinematic age estimate. 

A further indication for the evolutionary stage of TOI-732 may come from M-dwarf gyrochronology. \citet{pass2022} found that M dwarfs usually start spinning down at about 2-3 Gyr. Given the H$\alpha$-based inactivity of TOI-732, it is likely that the star is older than the turning-point age of 2-3 Gyr, which again agrees with the stellar evolutionary stage we inferred from kinematics.

\begin{table}
\caption{Stellar properties}
\renewcommand{\arraystretch}{1.2}
\label{tab:star}
\centering
\begin{tabular}{l c r}
\hline
\hline
\noalign{\smallskip}
\multicolumn{1}{l}{\multirow{5}*{Star names}} & \multicolumn{2}{c}{TOI-732} \\
\multicolumn{1}{l}{}               & \multicolumn{2}{c}{TIC 36724087} \\
\multicolumn{1}{l}{}               & \multicolumn{2}{c}{LTT 3780} \\
\multicolumn{1}{l}{}               & \multicolumn{2}{c}{LP 729-54} \\
\multicolumn{1}{l}{}               & \multicolumn{2}{c}{Gaia DR2 3767281845873242112} \\
\noalign{\smallskip}
\hline
\noalign{\smallskip}
Parameter & Value & Source \\
\noalign{\smallskip}
\hline
\noalign{\smallskip}
   $\alpha$\; [$^\circ$]            & 154.64485         & Gaia DR3  \\
   $\delta$\; [$^\circ$]            & $-11.71784$       & Gaia DR3  \\
   $\mu_\alpha$\; [mas\,yr$^{-1}$]   & $-341.537\pm0.032$& Gaia DR3  \\
   $\mu_\delta$\; [mas\,yr$^{-1}$]   & $-247.747\pm0.032$& Gaia DR3  \\
   $RV_{\mathrm{sys}}$ [km\,s$^{-1}$]& $+0.27\pm0.34$    & Gaia DR3  \\
   $\pi$\; [mas]                    & $45.382\pm0.030$  & Gaia DR3\tablefootmark{(a)}    \\
   $T_{\mathrm{eff}}$\; [K]         & $3358\pm92$ & Spectroscopy \\
   $\log{g}$                        & $4.85\pm0.11$ & Trigonometric      \\\relax
   [Fe/H]                           & $0.06\pm0.11$ & Spectroscopy     \\\relax
   [Mg/H]                           & $0.04\pm0.20$  & Thin disc    \\\relax
   [Si/H]                           & $0.02\pm0.21$  & Thin disc    \\
   $R_{\star}$\; [$R_{\odot}$]      & $0.380\pm0.012$   & IRFM  \\
   $M_{\star}$\; [$M_{\odot}$]     & $0.381_{-0.034}^{+0.024}$   & Isochrones    \\
   $t_{\star}$\; [Gyr]             & $3.10^{+6.20}_{-0.98}$       & Kinematics \\
   $L_{\star}$\; [$L_{\odot}$]     & $0.0165\pm0.0021$   & $R_{\star}$ \& $T_{\mathrm{eff}}$ \\
   $\rho_{\star}$\; [$\rho_{\odot}$] & $6.94\pm0.84$ & $R_{\star}$ \& $M_{\star}$ \\
\noalign{\smallskip}
\hline
\end{tabular}
\tablefoot{\tablefoottext{a}{Correction from \citet{lindegren2021} applied.}
}
\end{table}

\section{Photometric data}\label{sec:photometry}

\begin{table}
\caption{Photometric observations from ground-based facilities.}
\label{tab:groundLog}
\centering
\begin{tabular}{c c c c c}
\hline\hline
Telescope & Planet & Start Date & Duration & Filter \\
          &        & [UTC]      & [h]      &         \\
\hline
   CTIO & b & 2019-06-09 & 2.4 & z' \\
   CTIO & b & 2019-06-16 & 2.7 & z' \\
   SAAO & b & 2019-06-17 & 3.1 & g' z' \\
   SSO  & c & 2020-01-04 & 3.8 & B \\
   Trappist-N & c & 2019-11-12 & 4.4 & z' \\
   OSN  & c & 2019-11-12 & 3.5 & V R \\
   OAA  & b & 2020-01-31 & 6.5 & I \\
   MEarth & c & 2020-02-10 & 3.8 & RG715 \\
   MuSCAT2 & b & 2019-12-29 & 2.7 & g' r' i' z' \\
   MuSCAT2 & b & 2020-01-25 & 2.7 & g' r' i' z' \\
   MuSCAT2 & b & 2020-01-28 & 3.1 & g' r' i' z' \\
   MuSCAT2 & b & 2020-01-31 & 1.9 & g' r' i' z' \\
   MuSCAT2 & c & 2019-12-11 & 3.4 & g' r' i' z' \\
   MuSCAT2 & c & 2020-01-29 & 6.0 & g' r' i' z' \\
\hline
\end{tabular}
\end{table}

Both C20 and N20 have performed a photometric analysis of the system based on one TESS sector and several observations taken with ground-based facilities (see Table~\ref{tab:groundLog}). L22 used exclusively space-based observations instead, but added a second TESS sector for the photometric characterisation. In addition to using all the ground- and space-based data that were published in the literature, we added a significant number of space-based data as we benefited from a further TESS sector and collected 25 CHEOPS visits, 17 of which contain transit events, while the remaining 8 are short observations that were not time constrained (fillers) with the aim of monitoring stellar activity (see Table~\ref{tab:cheopsLog}). Therefore, our photometric analysis is based on a total of $\sim$\,140 transit events (spread over 132 different LCs), which enabled us to considerably improve the photometric properties of the system. All details of the available LCs are given below.

\begin{table*}
\caption{Details of the CHEOPS visits.}             
\label{tab:cheopsLog}      
\centering          
\begin{tabular}{l c c c c}
\hline\hline
Counter & File key & Planet & Start date [UTC] & Duration [h] \\
\hline                    
CH 1 & CH\_PR100031\_TG047301\_V0200 & b c & 2022-01-08T14:21:12.4 & 6.79 \\
CH 2 & CH\_PR100031\_TG047201\_V0200 & b & 2022-01-18T12:45:11.5 & 7.10 \\
CH 3 & CH\_PR100031\_TG047601\_V0200 & b c & 2022-01-20T20:47:11.6 & 6.92 \\
CH 4 & CH\_PR100031\_TG047202\_V0200 & b & 2022-01-24T15:56:12.3 & 6.50 \\
CH 5 & CH\_PR100031\_TG047203\_V0200 & b & 2022-01-25T10:22:12.3 & 6.50 \\
CH 6 & CH\_PR100031\_TG048301\_V0200 & 0 & 2022-01-26T15:15:11.5 & 2.57 \\
CH 7 & CH\_PR100031\_TG047204\_V0200 & b & 2022-01-26T23:29:11.5 & 5.85 \\
CH 8 & CH\_PR100031\_TG047205\_V0200 & b & 2022-01-30T01:40:11.5 & 5.89 \\
CH 9 & CH\_PR100031\_TG048501\_V0200 & b c & 2022-02-01T23:28:12.4 & 12.71 \\
CH 10 & CH\_PR100031\_TG048302\_V0200 & 0 & 2022-02-02T14:28:12.6 & 3.20 \\
CH 11 & CH\_PR100031\_TG048401\_V0200 & b & 2022-02-03T12:49:11.6 & 10.05 \\
CH 12 & CH\_PR100031\_TG047206\_V0200 & b & 2022-02-04T10:14:11.6 & 6.05 \\
CH 13 & CH\_PR100031\_TG048303\_V0200 & 0 & 2022-02-26T11:40:12.5 & 3.20 \\
CH 14 & CH\_PR100031\_TG049701\_V0200 & b & 2022-03-04T18:42:11.4 & 9.64 \\
CH 15 & CH\_PR100031\_TG048304\_V0200 & 0 & 2022-03-05T04:32:12.5 & 3.20 \\
CH 16 & CH\_PR100031\_TG049501\_V0200 & b c & 2022-03-10T20:47:12.6 & 11.66 \\
CH 17 & CH\_PR100031\_TG049702\_V0200 & b & 2022-03-12T09:46:12.6 & 9.79 \\
CH 18 & CH\_PR100031\_TG049703\_V0200 & b & 2022-03-16T07:34:12.6 & 11.26 \\
CH 19 & CH\_PR100031\_TG049704\_V0200 & b & 2022-03-19T09:33:12.5 & 10.14 \\
CH 20 & CH\_PR100031\_TG048305\_V0200 & 0 & 2022-03-22T04:34:12.5 & 3.20 \\
CH 21 & CH\_PR100031\_TG049705\_V0200 & b & 2022-03-22T11:22:12.5 & 10.17 \\
CH 22 & CH\_PR100031\_TG048306\_V0200 & 0 & 2022-03-26T22:23:12.5 & 3.20 \\
CH 23 & CH\_PR100031\_TG048307\_V0200 & b & 2022-03-29T11:42:12.4 & 3.20 \\
CH 24 & CH\_PR100031\_TG048308\_V0200 & 0 & 2022-04-01T05:19:13.0 & 3.20 \\
CH 25 & CH\_PR100031\_TG048309\_V0200 & 0 & 2022-04-10T15:48:12.5 & 3.08 \\
\hline                  
\end{tabular}
\tablefoot{Within the visits targeting TOI-732\,c, also transits of the ultra-short period planet TOI-732\,b are present. The ``0'' flag in the third column indicates a filler visit.}
\end{table*}

\subsection{TESS observations}
The Transiting Exoplanet Survey Satellite \citep[TESS,][]{ricker2015} observed the system in cycle 1 (Sector 8; from 28 February to 26 March 2019) in cycle 3 (Sector 35; from 9 February to 7 March 2021), and in cycle 5 (Sector 62; from 12 February to 10 March 2023). The data were downloaded from the Mikulski Archive for Space Telescopes (MAST)\footnote{\url{https://mast.stsci.edu/portal/Mashup/Clients/Mast/Portal.html}}, and we used the pre-search data conditioned simple aperture photometry (PDCSAP) LCs, as processed by the Science Processing Operation Center \citep[SPOC;][]{jenkins2016}.

After rejecting data with a poor-quality flag and performing a five median-absolute-deviation (MAD) clipping on the flux values to discard the outliers, we extracted the temporal windows that were centred on each transit event containing $\sim$\,4 hours of out-of-transit data both before and after the transit for detrending purposes. Following this procedure, we obtained 81 TESS LCs, 5 of which contain the transits of both planets because their transits are very close in time. Each LC lists the epoch of observation (\texttt{t}), the normalised PDCSAP flux with its uncertainty, and other parameters that are available from the TESS data products, such as \textsc{mom\_centr1}, \textsc{mom\_centr2} (hereafter denoted with \texttt{x} and \texttt{y}, respectively), and \textsc{pos\_corr1} and \textsc{pos\_corr2} (hereafter denoted with \texttt{dx} and \texttt{dy}, respectively)\footnote{See \url{https://tasoc.dk/docs/EXP-TESS-ARC-ICD-TM-0014-Rev-F.pdf} for further details about the TESS Science Data Products.}.

\subsection{CHEOPS observations}
The Characterising Exoplanet Satellite \citep[CHEOPS,][]{benz2021} collected 25 LCs of TOI-732 from 8 January to 10 April 2022. Because nearby bright background stars strongly contaminate the aperture photometry, we opted to extract point-spread function (PSF) photometry using the \texttt{PIPE} package\footnote{\url{https://github.com/alphapsa/PIPE}} \citep{morris2021, brandeker2022}. The raw CHEOPS LCs are shown in Appendix~\ref{sec:rawCHEOPS}.

In addition to the parameters given by \texttt{PIPE} (i.e. the stellar flux and the \texttt{x}- and \texttt{y}-location of the target PSF centroid on the detector), we added a few more vectors to the information comprising the CHEOPS LCs that were to be used for the following data detrending. In detail, these vectors are produced by the default data reduction pipeline \citep[DRP,][]{hoyer2020} v.13.1, and they are the spacecraft roll angle (\texttt{roll}), the flux due to contaminating background stars (\texttt{conta}), the smearing effect that is seen as trails on the CCD (\texttt{smear}), and the background flux (\texttt{bg}) due to zodiacal light, for example.

Among these data products, the stellar flux measured by CHEOPS usually exhibits a highly variable pattern against the roll angle \citep[see e.g.][for a broader discussion about this topic]{bonfanti2021}. As our global LC+RV modelling (see Sec.~\ref{sec:globalModel}) only accounts for polynomials when the time series is decorrelated, it would be hard to model the flux versus \texttt{roll} pattern. Therefore, after masking out the in-transit data points, we preliminary detrended the \texttt{PIPE} flux against roll angle via Gaussian processes \citep[GPs;][]{rasmussen2005} using a Matérn 3/2 kernel \citep{foreman2017}. We duly increased the error bars of the flux by adding the standard deviation of the GP model in quadrature.

\subsection{Ground-based observations}

Several LCs taken with ground-based facilities from 2019 up to 2020 are available on the \textsc{Exofop} webpage\footnote{\url{https://exofop.ipac.caltech.edu/tess/target.php?id=36724087}}. In particular, we downloaded the data obtained with the following one-meter-class telescopes that are part of the Las Cumbres Observatory Global Telescope (LCOGT) network \citep{brown2013}, which are located at the Cerro Tololo Inter-American Observatory (CTIO), the South Africa Astronomical Observatory (SAAO), and the Siding Spring Observatory (SSO).
In addition, we downloaded LCs acquired with: (i) the 60\,cm Trappist-North telescope \citep{jehin2011,barkaoui2019} at Oukaimeden Observatory in Morocco; (ii) the 150\,cm (T150) telescope at the Observatorio de Sierra Nevada (OSN)\footnote{\url{https://www.osn.iaa.csic.es/en/}} in Granada, Spain; (iii) the 40\,cm telescope at the Observatori Astronòmic Albanyà (OAA)\footnote{\url{https://www.observatorialbanya.com/en/albanya-astronomical-observatory}} in Catalonia, Spain; (iv) the 40\,cm telescope array at the Fred Lawrence Whipple Observatory (FLWO) in Arizona \citep[MEarth project;][]{charbonneau2008}.

We also retrieved four transit LCs of TOI-732\,b and two transit LCs of TOI-732\,c each observed in four different filters with the MuSCAT2 multi-colour imager \citep{narita2019} installed on the 1.5 m Telescopio Carlos Sánchez (TCS) at the Teide Observatory in Tenerife, Spain. MuSCAT2 is equipped with four CCDs, each of which has $1024 \times 1024$ pixels with a field of view of $7.4 \times 7.4$ square arcmin. The instrument is capable of obtaining simultaneous images in the $g'$, $r'$, $i'$, and $z_s$ bandpasses. The basic data reduction (i.e. dark and flat-field calibrations) was performed by the MuSCAT2 pipeline \citep{Parviainen2019}. This pipeline is also capable of fitting a transit model including instrumental systematics and a photometric aperture optimised to reduce the light-curve scatter. For all the transits of planets b and c observed by MuSCAT2, we found an optimal aperture for the target star of 13.92 arcsec.

Finally, all LCs that were taken with a time cadence shorter than one minute were downsampled by binning the data to a one-minute cadence. Further details about the telescope properties and the observational setups can be found in C20 and N20.

\section{Radial velocity data}\label{sec:RV}
Both C20 and N20 combined photometric and RV data. In particular, C20 analysed the RV time series obtained with HARPS and HARPS-N \citep{cosentino2012}, while N20 separately analysed the RV time series obtained with CARMENES \citep{quirrenback2014,quirrenback2018}, IRD \citep{kotani2018}, and iSHELL \citep{rayner2016,rayner2022}. L22 performed an RV analysis using the data points coming from these five spectrographs together, while in our case, we further added MAROON-X observations as detailed below for a total of 165 data points. A summary of the RV data employed in our global analysis is given in Table~\ref{tab:RVlog}.

\begin{table}
\caption{RV data employed in the combined analysis.}
\label{tab:RVlog}
\centering
\begin{tabular}{c c c c}
\hline\hline
Instrument & Start date & Time span & Data points \\
           & [UTC]      & [d]       & [\#]         \\
\hline
   HARPS    & 2019-06-21 & 247 & 33 \\
   IRD      & 2019-12-10 & 1   & 4 \\
   HARPS-N  & 2019-12-14 & 92  & 30 \\
   CARMENES & 2019-12-27 & 54  & 52 \\
   iSHELL   & 2020-01-25 & 37  & 8 \\
   MAROON-X & 2021-02-22 & 102 & 38 \\
\hline
\end{tabular}
\end{table}

\subsection{Literature radial velocity data}\label{sec:RVliterature}
The RV measurements available in the literature were retrieved directly from C20 and N20, who also provided a detailed description of the RV data reduction. We briefly recall here that C20 obtained 33 spectra with the HARPS echelle spectrograph mounted at the ESO 3.6\,m telescope at the La Silla Observatory in Chile and 30 spectra with the HARPS-N echelle spectrograph mounted at Telescopio Nazionale Galileo \citep[TNG;][]{cosentino2000,oliva2006} in the Canary Islands, Spain. The corresponding RV measurements along with their error bars were extracted using the \texttt{TERRA} reduction pipeline \citep{angladaEscude2012}.

Instead, N20 obtained 52 spectra with the CARMENES spectrograph mounted on the 3.5\,m Calar Alto Observatory in Almería, Spain. They obtained the RV measurements using the \texttt{serval} code \citep{zechmeister2018} and applied the necessary corrections following \citet{trifonov2018} and \citet{kaminski2018}. N20 also took five spectra with the IRD instrument mounted on the Subaru 8.2\,m telescope in Mauna Kea, Hawaii. After discarding one low-quality observation, they reduced the spectra using \texttt{iraf} \citep{tody1986,tody1993} and extracted the RV measurements through the Subaru/IRD dedicated pipeline \citep{hirano2020}. Using the iSHELL spectrometer mounted on the NASA Infrared Facility (IRTF) in Mauna Kea, Hawaii, N20 further collected eight RV measurements by applying the spectral reduction method presented in \citet{cale2019}.

Each of the five instruments is characterised by its own offset and is affected by a different jitter term. We therefore organised these RV time series as five independent data sets. These five RV time series contain the vectors of epochs, RV measurements, and RV error bars as found in the literature.

\subsection{MAROON-X}

We observed TOI-732 with MAROON-X, which is a high-precision echelle spectrograph installed on the 8.1 m telescope Gemini-North \citep{Seifahrt18, Seifahrt22}, 19 times between February and June 2021. The MAROON-X data were reduced with a \texttt{python3} pipeline based on the pipeline originally used for the CRIRES instrument \citep{Bean10}, and the RVs were calculated with a version of \texttt{serval} \citep{Zechmeister20} modified to work on MAROON-X data.  \texttt{serval} calculates RVs by least-squares fitting each individual spectrum to a template created by co-adding all spectra together. The \texttt{serval} routine also extracts the chromatic index (\texttt{crx}), the differential line width (\texttt{dlw}), and the H$\alpha$ index, which may be useful for data detrending. The wavelength calibration is accomplished by simultaneously observing the science target with an etalon spectra, and the etalons themselves are calibrated using a ThAr lamp.  

MAROON-X has two separate CCDs, each with slightly different wavelength coverages, which are exposed simultaneously.  The blue channel (500\,---\,670\,nm) and the red channel (650\,---\,920\,nm) were treated as two separate instruments for the purposes of this analysis because they have a different wavelength coverage and thus capture different stellar signals. We achieved a median S/N of 200 in the red channel and 77 in the blue channel, which corresponded to median RV uncertainties of 0.5 m\,s$^{-1}$ in the red channel and 1 m\,s$^{-1}$ in the blue channel. The higher signal in the red channel is expected for the late stellar spectral type.

MAROON-X is a visitor instrument on Gemini-North, and it is thus connected and disconnected multiple times over the course of a semester. It organises its data into discretised runs. In particular, the TOI-732 data were collected over the course of three runs in 2021 (one in February, one in April, and one in May). Combined with the roughly 2.5 cm\,s$^{-1}$\,d$^{-1}$ RV drift of the etalon calibrations, this results in small offsets between the RVs of MAROON-X data taken in separate runs.  
We therefore treated each run of the MAROON-X data as an independent RV time series and further distinguished the data taken via the red and blue channel. That is, we fit six independent MAROON-X RV time series. Accounting for the five RV time series described in Sec.~\ref{sec:RVliterature}, we analysed a total of 11 RV time series.

\section{Methods and results}\label{sec:results}
\subsection{Global light-curve and radial-velocity modelling}\label{sec:globalModel}
We jointly analysed the 132 LCs and 11 RV time series using the MCMCI code \citep{bonfanti2020}, where we switched off the interaction with stellar evolutionary models to avoid a dramatic increase in computational time due to the large data sets. In short, the code fit the LCs against the photometric model of \citet{mandel2002} and the RV data against a Keplerian model using an MCMC approach.

On the stellar side, we adopted $T_{\mathrm{eff}}$, [Fe/H], $M_{\star}$, and $R_{\star}$ as jump parameters that were subject to Gaussian priors based on the values reported in Table~\ref{tab:star}. The reason for this choice is twofold. On the one hand, both $M_{\star}$ and $R_{\star}$ induce a prior on the mean stellar density $\rho_{\star}$, which better constrains the transit fitting. On the other hand, stellar parameters are the starting point for interpolating within \textsc{Atlas9}\footnote{\url{http://kurucz.harvard.edu/grids.html}} grids of quadratic limb-darkening (LD) coefficients $(u_1, u_2)$, which were set up for each of the 11 photometric filters using the code by \citet{espinoza2015}. Gaussian priors were then imposed on the 11 interpolated pairs $(u_1, u_2)$ as summarised in Table~\ref{tab:LD}, but the actual LD-related jump parameters were derived from a linear combination of $(u_1, u_2)$ following \citet{holman2006} to reduce their mutual correlation. 

For each planet, the jump parameters were the transit depth d$F\equiv\left(\frac{R_p}{R_{\star}}\right)^2$, the impact parameter $b$, the mid-transit time $T_0$, the orbital period $P$, and the RV semi-amplitude $K$. We assumed a circular orbit for TOI-732\,b as its ultrashort orbital period implies a tide-induced circularisation timescale of $\sim$\,15 Myr \citep{matsumura2008}. We instead fit for the eccentricity of TOI-732\,c using the parametrisation $(\sqrt{e}\cos{\omega}, \sqrt{e}\sin{\omega})$, where $e$ is the eccentricity and $\omega$ is the argument of periastron. In the case of TESS observations, N20 noted that a close-in star, namely TIC\,36724077, was located within the aperture mask. We therefore fitted for a dilution factor following their estimate. All planetary jump parameters were subject to uniform unbounded priors (except for the physical limits). For details about the adopted parametrisations, we refer to \citet[\S~2.1.2]{bonfanti2020} and references therein.

The MCMCI tool is able to detrend data against time and the ancillary vectors of the LC and RV time series along the MCMC process via polynomial interpolation. To find the best polynomial order for each detrending parameter of each LC and RV time series, we launched several preliminary MCMC runs and changed the polynomial order of one detrending parameter at a time. We finally selected the best detrending polynomial baseline (see Table~\ref{tab:LCdetrending}) according to the Bayesian information criterion \citep[BIC;][]{schwarz1978}.

We then launched a first MCMC run of 200\,000 steps (burn-in: 40\,000 steps) to evaluate the impact of the white and red noise as detailed in \citet{pont2006} and \citet{bonfanti2020} to properly rescale the photometric errors and provide reliable uncertainties on the fitted parameters. After this, we performed the final MCMCI analysis made of two independent runs (each comprising 200\,000 steps with a burn-in of 40\,000 steps) to check the posterior distribution convergence through the Gelman-Rubin (GR) test \citep{gelman1992}.

The chains converged according to the GR statistic ($\hat{R}\lesssim 1.006$ for all the jump parameters), and we obtained planetary radii of $R_b=1.325_{-0.058}^{+0.057}\,R_{\oplus}$ and $R_c=2.39_{-0.11}^{+0.10}\,R_{\oplus}$, masses of $M_b=2.46\pm0.19\,M_{\oplus}$ and $M_c=8.04_{-0.48}^{+0.50}\,M_{\oplus}$, and thus densities of $\rho_b=5.8_{-0.8}^{+1.0}$ g\,cm$^{-3}$ and $\rho_c=3.24_{-0.43}^{+0.55}$ g\,cm$^{-3}$. All relevant system parameters as derived from our MCMC global analysis are listed in Tables~\ref{tab:planetsMCMC}, \ref{tab:LD}, and \ref{tab:RVjitter}. The phase-folded and detrended LCs of both TOI-732 b and c, as observed by both TESS and CHEOPS, are shown in Fig.~\ref{fig:TESS-CHEOPS-LCs}, while the LCs taken by ground-based facilities are shown in Appendix~\ref{sec:groundLC}. Finally, the phase-folded and detrended RV time-series of both TOI-732 b and c are displayed in Fig.~\ref{fig:RVphaseFolded}.

\begin{table}
\renewcommand{\arraystretch}{1.3}
\caption{Parameters of the TOI-732 system.}
\label{tab:planetsMCMC}
\centering
\begin{tabular}{l c c}
\hline\hline
\noalign{\smallskip}
Parameter & TOI-732\,b & TOI-732\,c \\
\noalign{\smallskip}
\hline
\noalign{\smallskip}
$P$\; [d] & $0.76837931_{-0.00000042}^{+0.00000039}$ & $12.252284\pm0.000013$ \\
$T_0$\tablefootmark{(a)} [BJD] & $9606.58098_{-0.00040}^{+0.00032}$ & $9600.54227_{-0.00065}^{+0.00066}$ \\
$b$  & $0.462_{-0.094}^{+0.063}$ & $0.794_{-0.027}^{+0.023}$ \\
d$F$\; [ppm] & $1032_{-45}^{+44}$ & $3355_{-130}^{+140}$ \\
$\frac{R_p}{R_{\star}}$ & $0.03212_{-0.00072}^{+0.00068}$ & $0.0579_{-0.0011}^{+0.0012}$ \\
$W$ [min] & $47.90\pm0.73$ & $92.5_{-1.6}^{+1.7}$ \\
$i$\; [$^{\circ}$] & $86.10_{-0.68}^{+0.92}$ & $88.958_{-0.068}^{+0.074}$ \\
$a$\; [AU] & $0.01195_{-0.00029}^{+0.00028}$ & $0.0757\pm0.0018$ \\
$\frac{a}{R_{\star}}$ & $6.79_{-0.25}^{+0.29}$ & $43.0_{-1.6}^{+1.8}$ \\
$K$\; [m\,s$^{-1}$] & $3.24\pm0.20$ & $4.22\pm0.16$ \\
$e$ & 0 (fixed) & $0.024_{-0.017}^{+0.032}$ \\
$\omega$\; [$^{\circ}$] & 90 (fixed) & $-66_{-50}^{+110}$ \\
$T_{\mathrm{eq}}$\tablefootmark{(b)}\,[K] & $903\pm26$ & $359\pm10$ \\
$S$\; [$S_{\oplus}$] & $111_{-12}^{+13}$ & $2.76_{-0.31}^{+0.33}$ \\
$R_p$\; [$R_{\oplus}$] & $1.325_{-0.058}^{+0.057}$ & $2.39_{-0.11}^{+0.10}$ \\
$M_p$\; [$M_{\oplus}$] & $2.46\pm0.19$ & $8.04_{-0.48}^{+0.50}$ \\
$\rho_p$\; [g\,cm$^{-3}$] & $5.8_{-0.8}^{+1.0}$ & $3.24_{-0.43}^{+0.55}$ \\
\noalign{\smallskip}
\hline
\end{tabular}
\tablefoot{Uncertainties are defined as the 68.3\% credible intervals of the posterior distributions. All fitted parameters, that is $P$, $T_0$, $b$, d$F$, $K$, $e$, and $\omega$, were subject to uniform unbounded priors (except for physical limits) following the parameterisations detailed in \citet{bonfanti2020}. \\
\tablefoottext{a}{Shifted by $-$\,2\,450\,0000.} \tablefoottext{b}{Assuming zero albedo.}
}
\end{table}

The bulk densities obtained for both planets are at the $\sim$\,15\% precision level, and only $\sim$\,20\% of all known planets orbiting M dwarfs have been characterised to a similar or better precision\footnote{Source: Nasa Exoplanet Archive, \url{https://exoplanetarchive.ipac.caltech.edu/}}. This is a consequence of the precision we reached on both the transit depths of TOI-732\,b and TOI-732\,c (4.4\% and 4.0\%, respectively) and the radial velocity semi-amplitudes (6.2\% and 3.8\%, respectively), which marks a significant improvement over what was reported so far in the literature, as summarised in Table~\ref{tab:uncertaintiesCmp}.

\begin{table}
\renewcommand{\arraystretch}{1.2}
\caption{Comparison between literature uncertainties and those derived in this work on the orbital periods $P$, the transit depths d$F$, and the RV semi-amplitudes $K$ of the planets.}
\label{tab:uncertaintiesCmp}
\centering
\begin{tabular}{llcccc}
\hline\hline
Planet & Uncertainty & C20 & N20 & L22 & This work \\
\hline
\noalign{\smallskip}
\multirow{3}*{TOI-732\,b} & 
   $\Delta P$ [s] & 4.7 & 0.12 & 0.045 & 0.035 \\
& $\frac{\Delta\mathrm{d}F}{\mathrm{d}F}$ [\%] & 9.3   & 6.5 & 6.9 & 4.3 \\
& $\frac{\Delta K}{K}$ [\%] & 18 & 10 & 8.5 & 6.2 \\
\noalign{\smallskip}
\hline
\noalign{\smallskip}
\multirow{3}*{TOI-732\,c} & 
   $\Delta P$ [s] & 251 & 5.9 & 3.0 & 1.1 \\
&  $\frac{\Delta\mathrm{d}F}{\mathrm{d}F}$ [\%] & 12    & 5.5 & 9.6 & 4.0 \\
&  $\frac{\Delta K}{K}$ [\%] & 17 & 10 & 8.0 & 3.8 \\
\noalign{\smallskip}
\hline
\end{tabular}
\end{table}

Based on the large amount of available data and the broad temporal baseline spanning four years, we were able to reduce the uncertainties on the orbital periods of both planets by more than two orders of magnitude with respect to what was reported by C20. Even comparing our results with those of L22, who derived the most precise ephemerides so far, we improved the uncertainty on the planetary orbital periods by a factor of $\sim$\,1.3 and $\sim$\,2.7 for planets b and c, respectively (see Table~\ref{tab:uncertaintiesCmp}). By propagating our ephemerides, we computed that the 1$\sigma$ uncertainties on the transit timings of the two planets are comparable to the respective transit durations after $\sim$\,170 years from now.

\begin{figure*}
\centering
\includegraphics[width=0.495\textwidth]{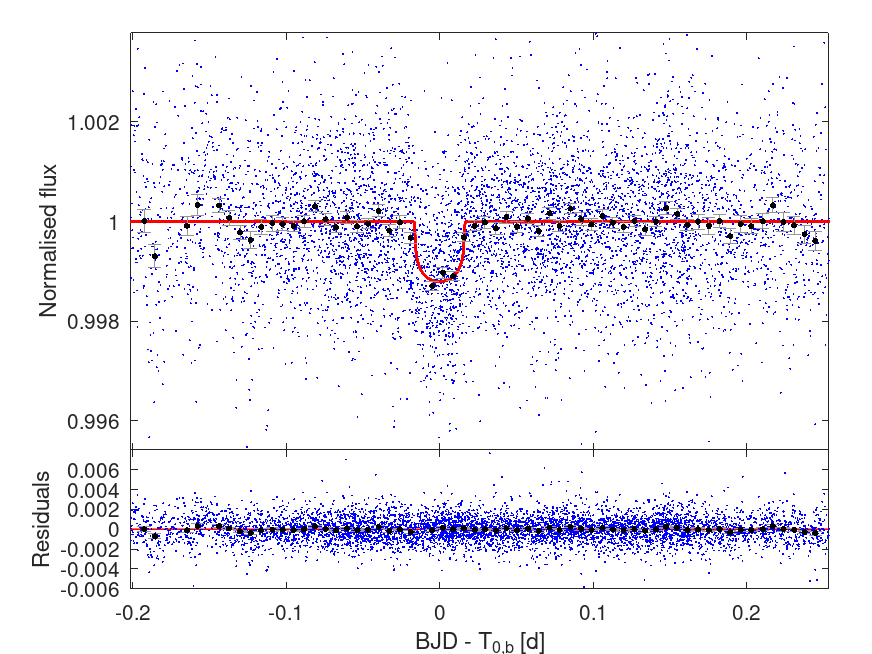}
\includegraphics[width=0.495\textwidth]{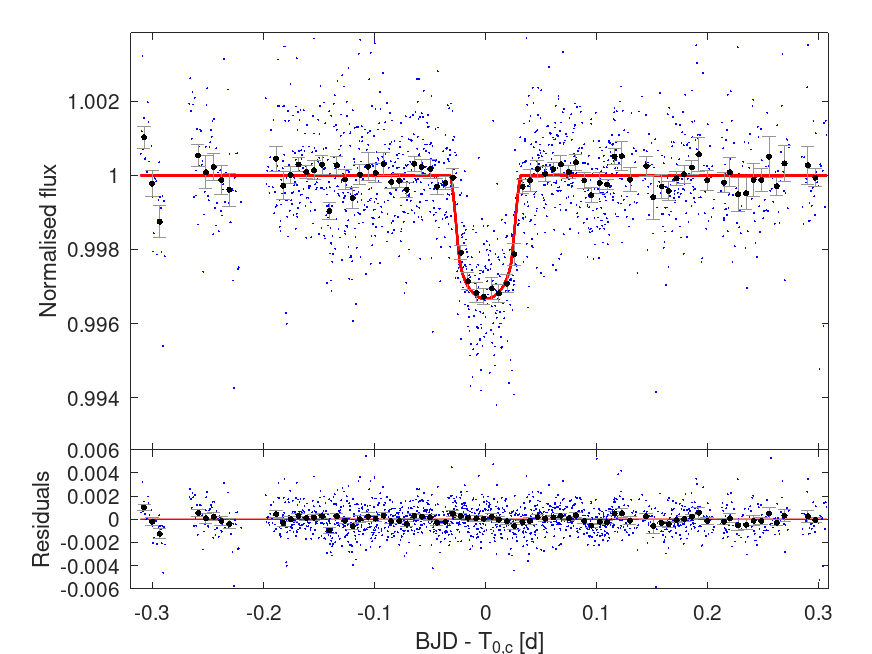} \\
\includegraphics[width=0.495\textwidth]{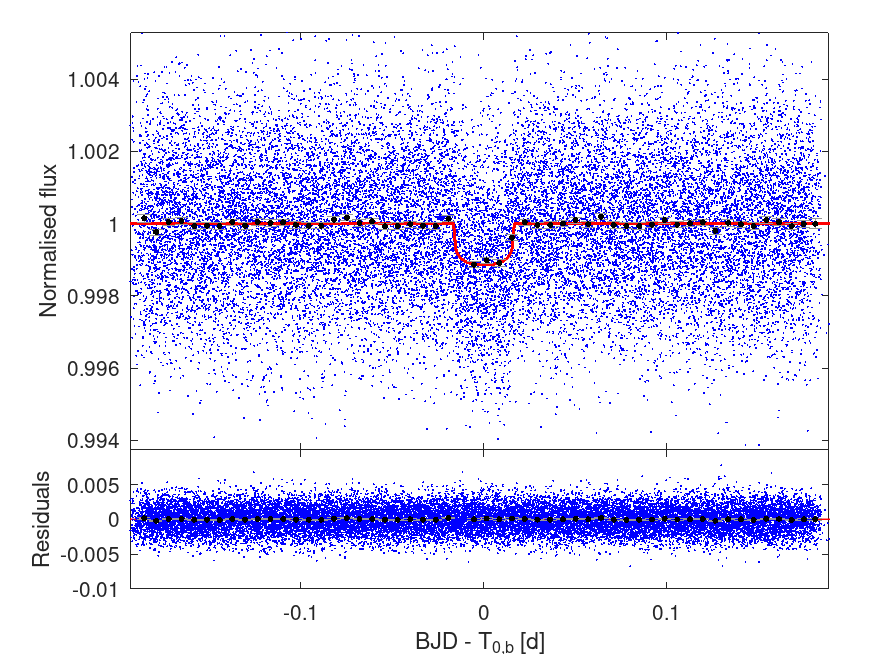}
\includegraphics[width=0.495\textwidth]{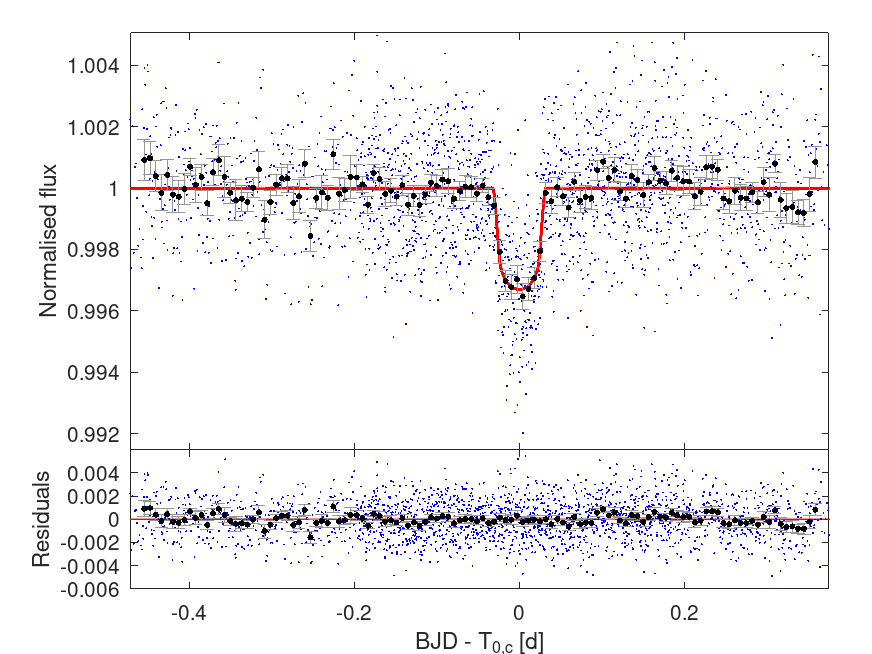}
\caption{Phase-folded and detrended LCs showing the transit of TOI-732\,b (first column) and TOI-732\,c (second column) as observed by CHEOPS (first row) and TESS (second row). The original data points are shown in blue, the binned data points are shown in black (binning of 10 min), and the transit model is displayed in red.}
\label{fig:TESS-CHEOPS-LCs}
\end{figure*}

\begin{figure}
\centering
\includegraphics[width=\columnwidth]{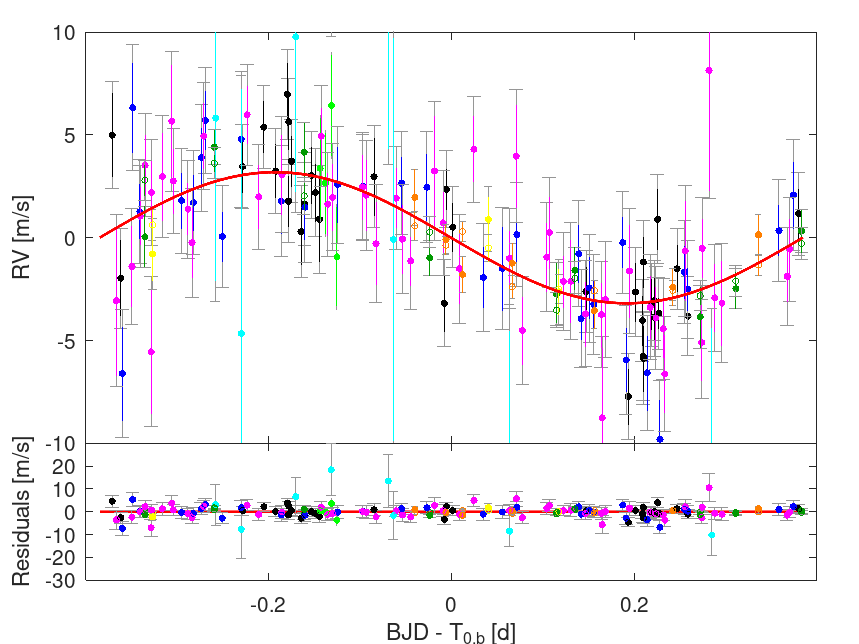} \\
\includegraphics[width=\columnwidth]{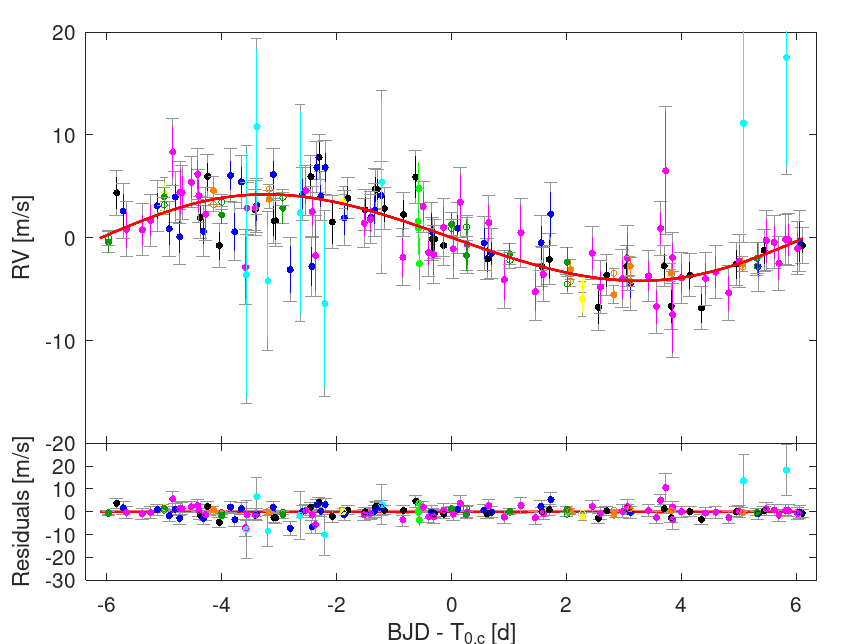}
\caption{Phase-folded and detrended RV time series of TOI-732\,b (\textit{Top}) and TOI-732\,c (\textit{Bottom}), obtained after subtracting the signal of the other planet. The corresponding Keplerian model is superimposed in red. The different colours mark different instruments, namely HARPS (black), IRD (light green), HARPS-N (blue), CARMENES (magenta), iSHELL (cyan), MAROON-X (yellow, orange, and deep green for observations taken in February, April, and May 2021, respectively). As MAROON-X has two different channels, full and empty symbols represent data acquired using the blue and red channel, respectively (see text for further details).}
\label{fig:RVphaseFolded}
\end{figure}

\subsection{Internal structure of the planets}\label{sec:internalStructure}

We modelled the internal structure of both TOI-732\,b\,and\,c using a neural-network-based Bayesian inference scheme following the method that was described in detail in \citet{Leleu2021} and is based on \citet{Dorn2017}. As input parameters, we used transit depths, periods, and the mass relative to that of the star for both planets, as well as some of the stellar parameters, namely mass, radius, age, effective temperature, [Si/H], [Mg/H], and [Fe/H]. We modelled both planets simultaneously, assuming that they consist of four fully distinct layers that we modelled according to the equations of state of \citet{Hakim2018} (an inner iron core with up to 19\% sulphur), \citet{Sotin2007} (a silicate mantle consisting of Si, Mg, and Fe) and \citet{Haldemann2020} (a condensed water layer), with a H-He envelope modelled following \citep{LopezFortney2014} on top. Furthermore, we assumed that the Si, Mg, and Fe ratios of both planets match those of the star \citep{Thiabaud2015}, even if we note that despite an expected trend between stellar and planetary composition, the correlation might not necessarily be strict \citep{adibekyan2021}.

As the problem of determining the internal structure of a planet is highly degenerate, the results of our analysis depend on our choice of prior. For the mass fractions of the inner iron core (i.e. the mantle layer and the water layer), all calculated with respect to the inner part of the planet without the H-He layer, we sampled from a prior that is uniform on the simplex on which they all add up to 1. Furthermore, we implemented an upper limit for the water-mass fraction of 0.5, in accordance with \citet{Thiabaud2014} and \citet{Marboeuf2014}. We also used a prior that is log-uniform for the mass of the H-He envelope.

\begin{figure}
	\centering
	\includegraphics[width=\columnwidth]{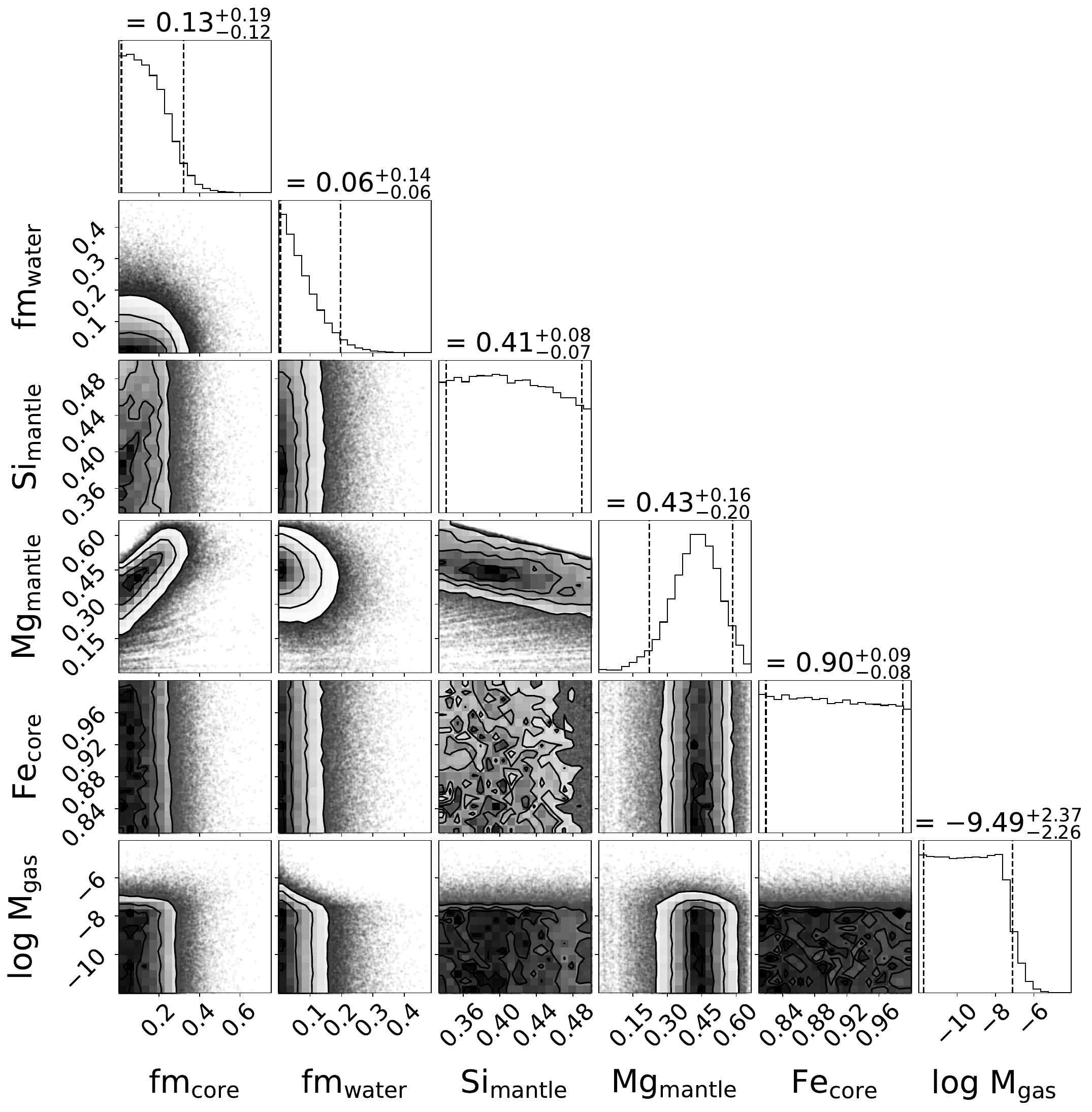}
	\caption{Corner plot showing the posteriors of the main parameters of our internal structure analysis for TOI-732\,b. The titles of each column correspond to the median of the distribution, with the 5$^\textrm{th}$ and 95$^\textrm{th}$ percentiles as the uncertainty values. From left to right, the depicted internal structure parameters are the mass fractions of the inner iron core and of the water layer (both calculated with respect to the condensed part of the planet without the H-He layer), the molar fractions of Si and Mg in the mantle, the molar fraction of Fe in the inner core, and the total mass of H-He in Earth masses on a logarithmic scale. The mass fractions of the inner core and the water layer add up to one, with the mass fraction of the mantle layer (not shown) by construction.}
	\label{internalstructure_b}
\end{figure}

\begin{figure}
	\centering
	\includegraphics[width=\columnwidth]{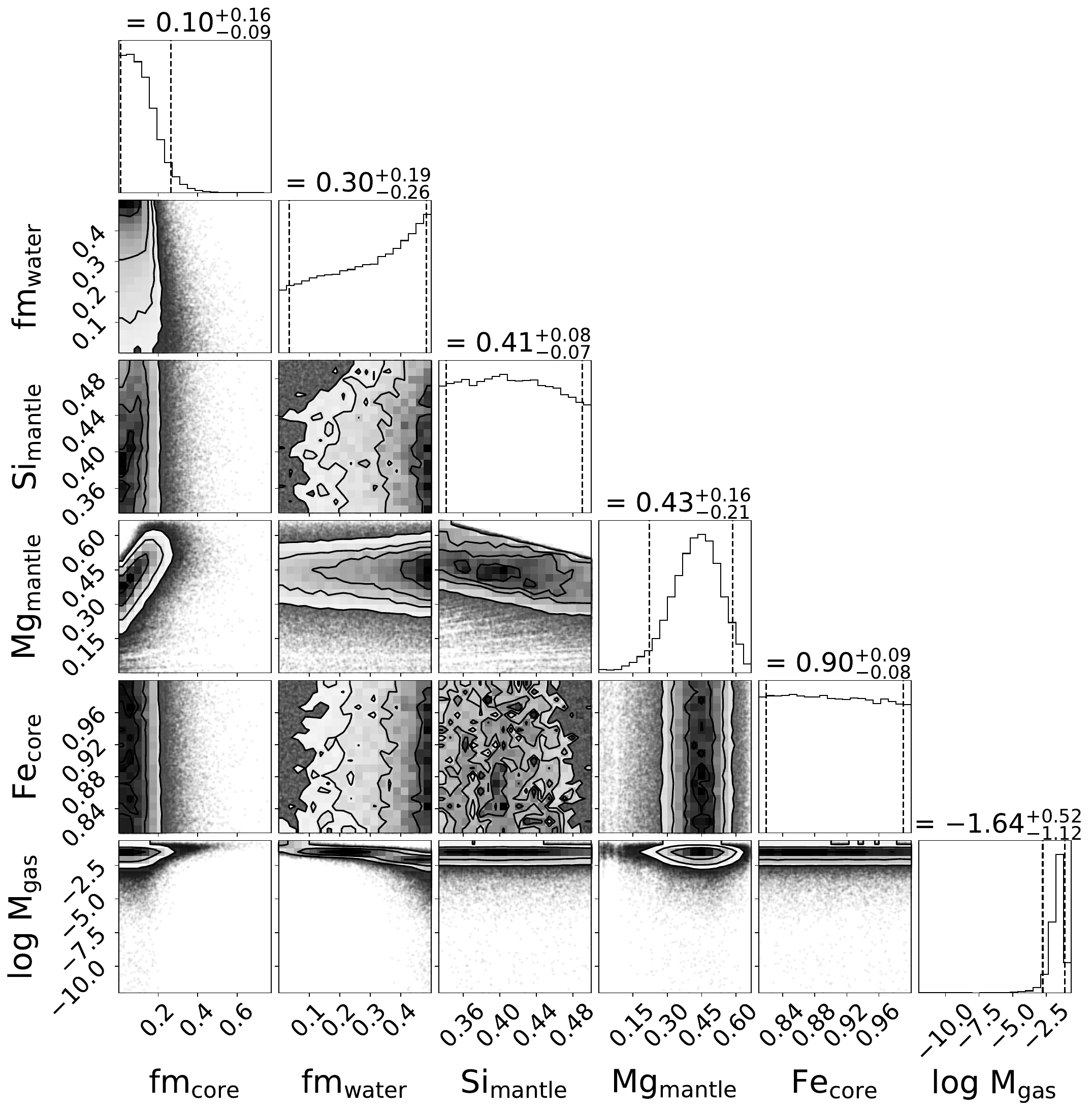}
	\caption{Same as Figure \ref{internalstructure_b}, but for TOI-732\,c.}
	\label{internalstructure_c}
\end{figure}

The results of our analysis are summarised in Figures \ref{internalstructure_b} and \ref{internalstructure_c}. The derived posteriors of the internal structure parameters show us that TOI-732\,b is unlikely to host a H-He layer given its density. Meanwhile, the presence of a water layer is possible, but not necessary, as the derived mass and radius values also agree with a purely rocky structure. For TOI-732\,c, the posterior distribution of the gas mass is instead quite well constrained, with a median of $M_{\mathrm{gas,c}}=0.02^{+0.05}_{-0.02}$ $M_{\oplus}$, which corresponds to a thickness of $R_{\mathrm{gas,c}}=0.40_{-0.27}^{+0.24}\,R_{\oplus}$ (errors are the 5th and 95th percentile of the distribution). However, the presence of a water layer is completely unconstrained.

\begin{figure}
    \includegraphics[width=1.07\columnwidth]{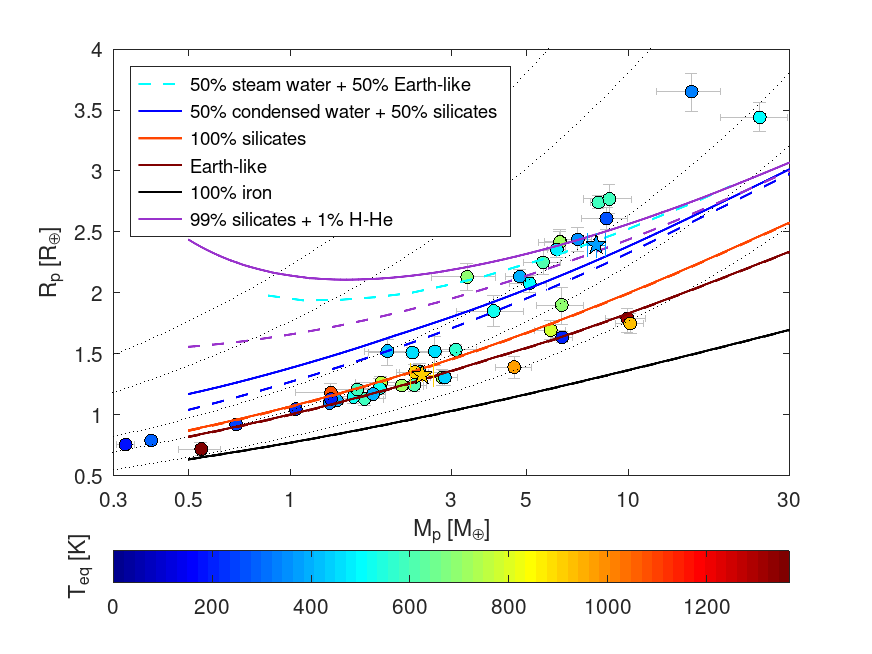}
    \caption{Mass-radius diagram of M-dwarf planets with $R_p<4\,R_{\oplus}$ and $M_p<30\,M_{\oplus}$ whose $R_p$ and $M_p$ precision is better than 8\% and 25\%, respectively. All planets are colour-coded with respect to their equilibrium temperature ($T_{\mathrm{eq}}$) and in particular, TOI-732\,b and TOI-732\,c are marked by a star. Following the colour scheme given in the legend, two different sets of planet composition models generated with BICEPS \citep{haldemann2023} are displayed. The solid and dashed lines are obtained for $T_{\mathrm{eq}}=T_{\mathrm{eq,b}}=900$ K and $T_{\mathrm{eq}}=T_{\mathrm{eq,c}}=360$ K, respectively. The 50\% steam + 50\% Earth-like line corresponds to the model of \citet{aguichine2021} for $T_{\rm eq} = 400$~K. An Earth-like composition implies a mixture of 32.5\% iron and 67.5\% silicates. The dotted black lines correspond to the loci of constant density, that is, 0.5, 1, 3, 5, and 10 g\,cm$^{-3}$ (from top to bottom).}
    \label{fig:MR}
\end{figure}

Figure~\ref{fig:MR} locates TOI-732\,b and TOI-732\,c on the mass-radius (MR) diagram along with M-dwarf planets with $R_p<4\,R_{\oplus}$ and $M_p<30\,M_{\oplus}$ whose precision on the radius and mass are better than 8\% and 25\%, respectively. When TOI-732\,b and c are included, this exoplanet sample (hereafter denoted as \texttt{Msample}\footnote{Planetary data of M dwarfs (that is stars with $T_{\mathrm{eff}}<4000$ K) were properly filtered and downloaded from the \href{https://exoplanetarchive.ipac.caltech.edu/}{NASA Exoplanet Archive} as of 27 July 2023}) is made of 45 well-characterised planets (a mean planetary bulk density above the 3\,$\sigma$ level). The main parameters of the \texttt{Msample} are listed in Tab.~\ref{tab:Msample}. 
Along with the planets belonging to the \texttt{Msample}, Fig.~\ref{fig:MR} also displays two sets of theoretical models for a planet composition that correspond to $T_{\mathrm{eq}}=T_{\mathrm{eq,b}}=900$ K (solid lines) and $T_{\mathrm{eq}}=T_{\mathrm{eq,c}}=360$ K (dashed lines) using the BICEPS model \citep{haldemann2023}. In addition, we further collected the MR model as computed by \citet{aguichine2021} for steam worlds made of 50\% water + 50\% rocks with $T_{\mathrm{eq}}=400\,\mathrm{K}\approx T_{\mathrm{eq,c}}$ (the dashed cyan line). Theoretical models of rocky and/or iron worlds do not depend upon $T_{\mathrm{eq}}$, but differences become noticeable when water and/or H-He envelopes are added to the planet structure. The MR diagram confirms that TOI-732\,b is likely rocky with a possible iron core, while TOI-732\,c is likely rich in volatiles. As shown above, inferring the internal planet structure from observables is a degenerate problem and other mixtures of silicates, gas, and water (which is indeed unconstrained according to our modelling of TOI-732\,c) may produce a ($M_p$, $R_p$) pair consistent with the observations. For example, we note that the MR location of TOI-732\,c is compatible with either a rocky planet surrounded by a H-He envelope (1\% by mass) or a steam world consisting of water and rocks in the same proportion by mass.

\section{Radius valley of M-dwarf planets}\label{sec:Rvalley}

According to their radii, TOI-732\,b and TOI-732\,c are located on the two opposite sides of the radius valley. Although some degeneracy is expected when modelling the internal structures of planets, we concluded in Sect.~\ref{sec:internalStructure} that TOI-732\,b is not likely to hold any gaseous envelope, while TOI-732\,c cannot be just purely rocky. When we also consider the mean planetary densities, $\rho_b>\rho_c$, which can lead to a classification of the inner planet as a super-Earth and of the outer one as a mini-Neptune, the TOI-732 system has a quite common architecture \citep[e.g.][]{ciardi2013,weiss2018,mishra2023}.

\subsection{Radius valley dependence on orbital period}
Taking a step further, we studied the radius valley $R_{p,\mathrm{valley}}$ for M dwarfs as a function of planet orbital period $P$ by using our \texttt{Msample}. Several theoretical studies \citep[e.g.][]{owen2017,lopez2018,gupta2019,wyatt2020,lee2021,rogers2021,affolter2023} have quantified different $\Rslope$ slopes characterising the radius valley depending on the specific formation and evolution mechanisms causing it (e.g. impact erosion, photo-evaporation, core-powered mass loss, or late planet formation in either gas-poor or even gas-empty discs). It is worth emphasising that planets formed in a gas-poor environment may also be subject to thermally driven mechanisms (i.e. photo-evaporation and core-powered mass loss). Hereafter, the discussion of thermally driven mechanisms is intended to involve planets that have not formed in a gas-poor environment, unless stated otherwise.

As summarised in Tab.~\ref{tab:slopeTheory}, a negative slope is theoretically expected for both impact erosion and thermally driven mass-loss mechanisms, with the slope becoming milder when passing from the former to the latter. 
Furthermore, in the case of late-time planet formation within a gas-poor environment, the slope is even shallower (but still negative) when photo-evaporation is considered to be at play afterwards. As emphasised by \citet{lee2021}, a positive $\Rslope$ is sometimes incorrectly associated to late-time planet formation, according to the work by \citet{lopez2018}. However, \citet{lopez2018} computed the expected scaling between $R_p$ and $P$ assuming a gas-empty scenario, and the positive slope line they derived therefore just corresponds to the maximum radius that can be reached by a purely rocky planet. Therefore, this locus of points does not trace the radius valley dividing rocky planets from sub-Neptune simply because no sub-Neptunes may form in a gas-empty environment. Nonetheless, we kept the $\Rslope_{L18}=+0.11$ in Table~\ref{tab:slopeTheory} because it sets the upper limit of the radius valley slope for a sample of heterogeneous exoplanets in the $R_p$-$P$ plane. On the one hand, the purely rocky exoplanets that are born in a gas-empty disc would be distributed following a positive trend, whose upper limit is given by $\Rslope_{L18}$. On the other hand, from a disc with gas, both super-Earth and sub-Neptunes would be generated, and they would appear on the two opposite sides of a descending radius valley. The full picture that we would see a posteriori in the $R_p$-$P$ plane would be the overlap of these two groups of exoplanets, which would show a radius valley with an intermediate slope, possibly even positive, depending on the weights of the formation mechanisms at play.

\begin{table}
\caption{Radius valley slopes $m\equiv\Rslope$ as predicted from theory for different scenarios.}
\label{tab:slopeTheory}
\centering
\begin{tabular}{l c l}
\hline\hline
Model & $m$ & Reference \\
\hline
 Impact erosion    & $-0.33$                              & \citet{wyatt2020} \\
 Photo-evaporation & [$-0.25,\,-0.16$] & \citet{owen2017}    \\
 Thermally-driven  & \multirow{2}*{$-0.10$}               & \multirow{2}*{\citet{affolter2023}}     \\
 \quad mass loss   &                                      &                     \\
 Photo-evaporation in & \multirow{2}*{[$-0.15,\,-0.08$]} & \multirow{2}*{\citet{lee2021}} \\
 \quad gas-poor discs & & \\
 Gas-empty formation & $+0.11$                             & \citet{lopez2018}    \\
\hline
\end{tabular}
\tablefoot{The photo-evaporation model has been developed using the energy-limited formula \citep[e.g.][]{watson1981,erkaev2007} and accounting for different efficiency values of stellar high-energy photons in the atmospheric mass removal. Instead, the thermally-driven mechanisms have been modelled via hydrodynamic simulations that couple photo-evaporation and core-powered mass loss. \\\relax
[$a,\,b$] denotes a range of values from $a$ to $b$.}
\end{table}

To study the dependence of the radius valley on planetary orbital period, we followed the same approach as in \citet{vanEylen2018} and \citet{ho2023}, but focused on M-dwarf hosts ($M_{\star}\lesssim0.6\,M_{\odot}$). This complements the stellar mass range spanned by the F, G, and K type stars investigated by \citet{ho2023}. In detail, we first clustered our M-dwarf exoplanets into two different groups, according to their location with respect to the radius valley (above or below), by performing a Gaussian mixture model selection \citep[e.g.][]{Huang2017,fruewirth2018}. To this end, we employed the \textsc{Python} \textit{sklearn} \texttt{GridSearchCV()} class, which allows specifying four different covariance types to define the clustering. After rescaling the period $P$ by a factor of five to avoid misclassification \citep{ho2023}, we fit the selection model within the $\log{R_p}$-$\log{P}$ plane, and we finally selected the model inferred from the spherical covariance type, which has the lowest associated BIC.

After this, we followed a support vector machine (SVM) procedure \citep[e.g.][]{cortes1995,benhur2002} implemented via the \textit{sklearn} \texttt{SVC()} class. After we set a linear kernel and a penalty parameter $C=10$ \citep[see][]{vanEylen2018}, the \texttt{fit} method of \texttt{SVC()} was able to compute the best-fit line separating the two groups of exoplanets in any desired space of covariates. In particular, we obtained
\begin{equation}
\log{R_{p,\mathrm{valley}}}=-0.065_{-0.013}^{+0.024}\log{P}+0.344_{-0.018}^{+0.008}\,,
\label{eq:RpvalleyP}
\end{equation}
where the uncertainties (at the 1\,$\sigma$ level) were computed by bootstrapping the \texttt{Msample} 10\,000 times and repeating the algorithm outlined above.

When compared to the outcome obtained by \citet{ho2023} ($\Rslope_{\mathrm{H23}}=-0.11\pm0.02$), the slope value we obtained differs by almost a factor of two (tension at the 2\,$\sigma$ level), which may suggest that formation and evolution mechanisms enter with different weights in the case of exoplanets orbiting M dwarfs or FGK stars.
Instead, when we performed a homogeneous comparison with other works targeting the $R_{p,\mathrm{valley}}$ slope of exoplanets around low-mass stars, our $\Rslope$ value is consistent within 1\,$\sigma$ with the estimate from L22 ($\Rslope_{\mathrm{L22}}=-0.02\pm0.05$), it is milder than the slope found by \citet{vanEylen2021} ($\Rslope_{\mathrm{V21}}=-0.11_{-0.04}^{+0.05}$), but still consistent at the $\sim$\,1\,$\sigma$ level, and it differs from the outcome of \citet[][$\Rslope_{C20}=+0.058\pm0.022$]{cloutier2020valley}. 
The sample of \citet{cloutier2020valley} also comprises planets orbiting K dwarfs (with a spectral type later than K3.5V, i.e. $M_{\star}\lesssim0.8\,M_{\odot}$), and the reason for the difference in slope may be that the $R_p$ precision for half of the planets they analysed is lower than our 8\% threshold (the 99$^\mathrm{th}$-quantile of their $R_p$ relative uncertainties is $\sim$\,26\%). Instead, both \citet{vanEylen2021} and L22 focused on planets orbiting M dwarfs alone, and the difference with our $\Rslope$ value decreases as the selection threshold for the sample is set to a lower $R_p$ uncertainty (below $20\%$ for \citet{vanEylen2018} and below 8\% for L22). Only the sample by L22 reaches the same precision level as our \texttt{Msample} (because we adopted the same selection criteria), but our sample contains 30\% more planets (45 versus 34 planets).

\begin{figure}
    \centering
    \includegraphics[width=\columnwidth]{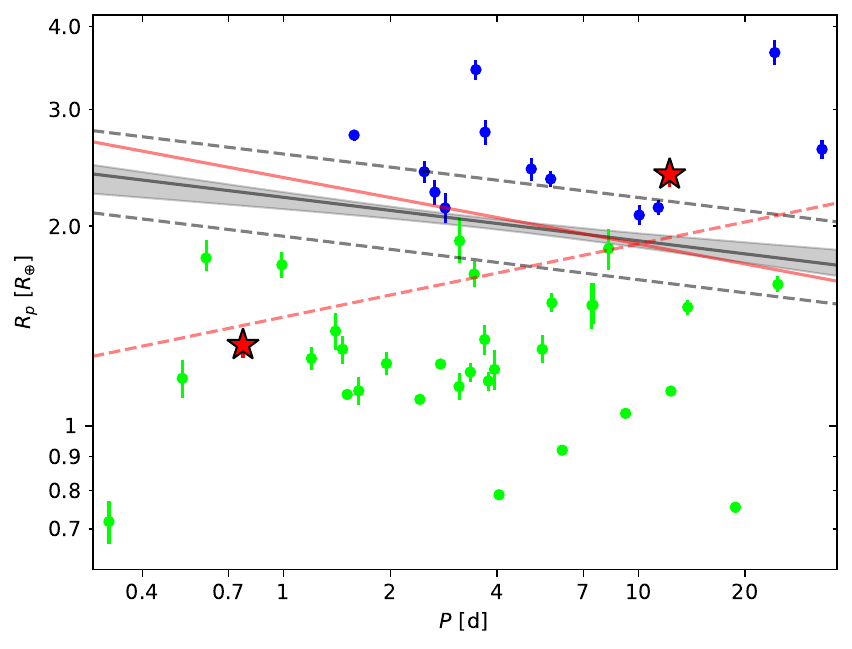}
    \caption{$R_p$ vs $P$ distribution representing the planets in our \texttt{Msample}. Planets classified above and below the radius valley are shown in blue and green, respectively, while the red star-shaped markers are for TOI-732\,b and c. The radius valley inferred via the SVM-based method is marked by the solid grey line with the shaded region highlighting the 1\,$\sigma$ limits of the best-fit line. The two parallel dashed grey lines are the median boundaries passing through the supporting vectors that determine the location of the solid line. Finally, the red lines represent the theoretically expected $R_{p,\mathrm{valley}}$ boundary in case of a thermally-driven mass-loss scenario (solid line as computed from \citet{affolter2023}; negative slope) and the $R_p$ upper limit of planets borned rocky in a gas-empty disc (dashed line as computed from \citet{lopez2018}; positive slope).}
    \label{fig:RvalleyP}
\end{figure}

A visual synthesis of our results is given in Fig.~\ref{fig:RvalleyP}, where the best-fit line marking the radius valley (solid grey line) is compared with the theoretical slopes expected from a thermally driven mass-loss model \citep[solid red line;][]{affolter2023} and a gas-empty formation model \citep[dashed red line;][]{lopez2018}. When compared with the theoretical slope expected from a mixed scenario, where both photo-evaporation and core-powered mass loss are at play \citep[that is $-0.10$;][]{affolter2023}, the negative $\Rslope$ slope we computed (i.e. $-0.065_{-0.013}^{+0.024}$) is shallower by a factor of $\sim$\,1.5 (tension at the $\sim$\,3\,sigma level). Slopes milder than $-0.10$ possibly tending towards positive values indicate a stronger impact of gas-poor formation according to \citet{lopez2018,lee2021}. Therefore, we may conclude that although thermally driven mechanisms appear to be statistically prevalent, the currently observed properties of some of the planets orbiting M dwarfs may be caused by late formation in gas-depleted discs. This scenario has indeed been proposed for a few M-dwarf systems, such as TOI-1634 \citep{cloutier2021}, where the composition of the close-in USP TOI-1634\,b is inconsistent with that of the Earth, or LHS\,1903 \citep{wilson2023}, where the outermost planet at $P$\,$\sim$\,29.3\,d lacks any gaseous envelope, in contrast to some of the inner planets. An alternative scenario explaining our $\Rslope$ findings is investigated in Sect.~\ref{sec:densityValley}.

As the strength of core-powered mass-loss experienced by a planet scales proportionally to $R_p T_{\mathrm{eq}}^4$ \citep{gupta2019}, the colour-coding in Fig.~\ref{fig:RvalleyPcbarCorePow} is an attempt of investigating the impact of core-powered mass loss in shaping the radius valley. However, the colour gradient from the bottom right to top left just reflects the increase in $R_p T_{\mathrm{eq}}^4$ at greater radii and lower orbital period (hotter planets).
The radii of billion-year-old planets dominating the \texttt{Msample} are thought to have significantly shrunk during their evolution due to planetary cooling and evaporation, which effect is correlated to the strength of the atmospheric escape \citep{LopezFortney2014,chen2016,kubyshkina2022,affolter2023}.
Thus, the present-day radii cannot unambiguously define the strength of core-powered mass loss because they are not indicative of the escape rates during the early evolution phases, when core-powered mass loss can dominate.
In addition, the equilibrium temperature strongly correlates with the (poorly constrained) amount of XUV radiation received by the planet because both $T_{\mathrm{eq}}$ and XUV radiation scale with the planet distance. Thus, the $T_{\mathrm{eq}}$ dependence does not allow us to distinguish the inputs from the core-powered and XUV-driven escape mechanisms sufficiently well.

\begin{figure}
    \centering
    \includegraphics[width=\columnwidth]{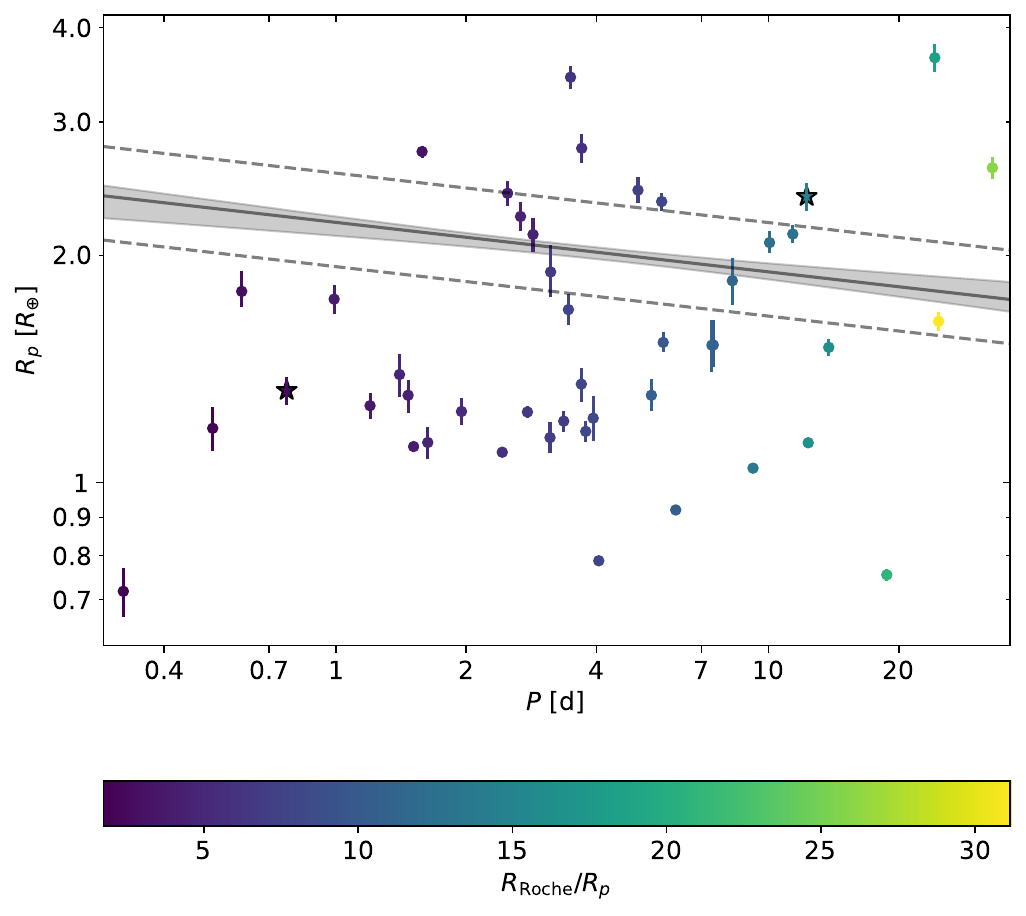} \\
    \includegraphics[width=\columnwidth]{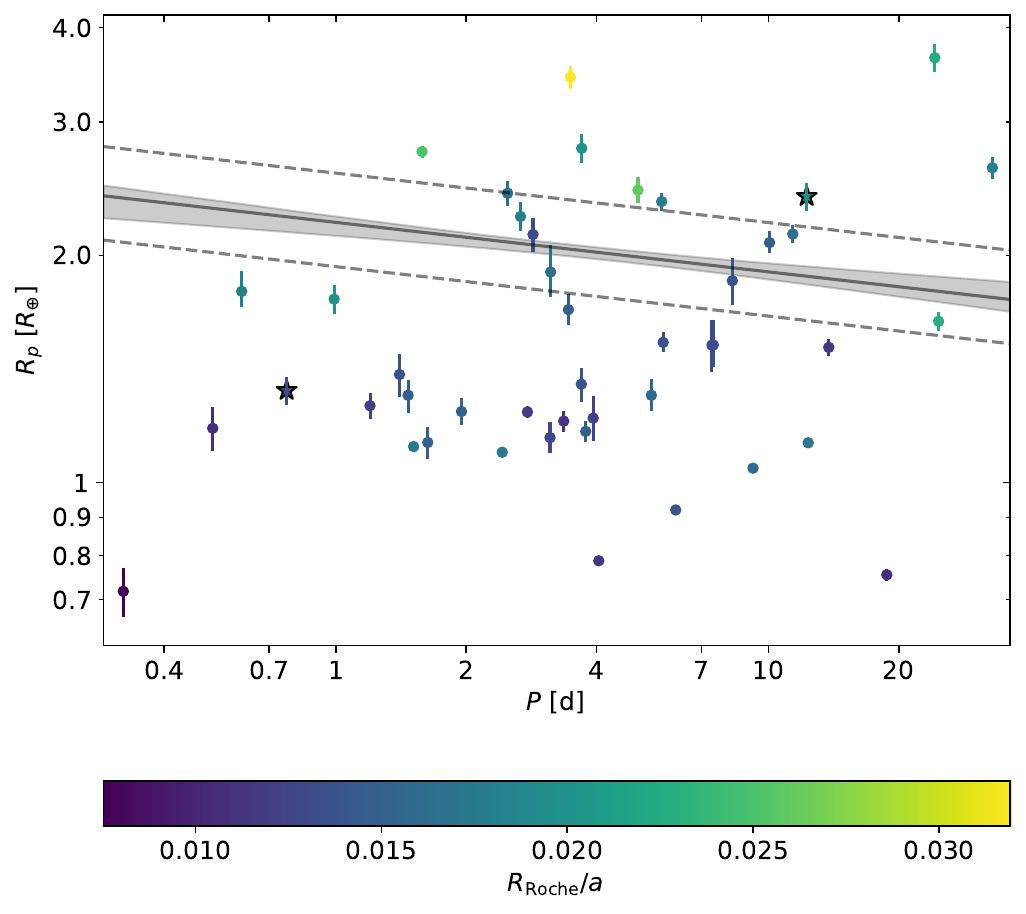}
    \caption{Same as Fig.~\ref{fig:RvalleyP}, but the markers are colour-coded against the Roche lobe of each planet normalised to the planetary radius (\textit{Top}) or to the orbital semi-major axis (\textit{Bottom}).}
    \label{fig:RvalleyPcbar}
\end{figure}

On the other hand, we know from hydrodynamic modeling that core-powered mass loss dominates the atmospheric escape completely if the atmospheric density is sufficiently high in the upper atmospheric layers to prevent the penetration of XUV radiation inside the planetary Roche lobe \citep{kubyshkina2018,kubyshkina2023}. This situation occurs most likely for planets with low masses and small Roche radii (comprising a few $R_p$ at young ages); of these two parameters, the Roche radius carries more information than the planetary mass alone.
Along this line, the two panels of Fig.~\ref{fig:RvalleyPcbar} still represent the planets of our \texttt{Msample} in the $R_p$-$P$ plane, but with a specific focus on the role of the core-powered mass-loss mechanism by tracing the size of the planetary Roche radius. We computed the Roche-lobe radius \citep{eggleton1983} of each planet
\begin{equation}
R_{\mathrm{Roche}}=\frac{0.49q^{\frac{2}{3}}}{0.6q^{\frac{2}{3}} + \ln{(1+q^{\frac{1}{3}})}} a,\quad \mathrm{being}\; q\equiv\frac{M_p}{M_{\star}},
\label{eq:Rroche}
\end{equation}
as a measure of the region within which a possible atmospheric envelope is bounded to the planet. The larger $R_{\mathrm{Roche}}$, the less effective the core-powered atmospheric escape. After normalisation to $R_p$ (top panel), $R_{\mathrm{Roche}}$ still maintains the linear dependence upon the semi-major axis $a$, and indeed, $R_{\mathrm{Roche}}/R_p$ increases with the orbital period. On the one hand, while this trend is expected, this panel emphasises on the other hand, that planets on long-period orbits are less subject to core-powered mass loss. Therefore, a rocky planet (i.e. without a low mean-molecular weight envelope) farther away from its host is more likely to be born in a gas-depleted environment.

The bottom panel of Fig.~\ref{fig:RvalleyPcbar} is similar to the top panel, but this time, the colour-coding follows $R_{\mathrm{Roche}}$ normalised to $a$. In this way, we removed the linear dependence of the Roche radius on $a$, which means that $R_{\mathrm{Roche}}$ depends solely on the $M_p/M_{\star}$ ratio. Now, $R_{\mathrm{Roche}}/a$ increases as $R_p$ increases, with the highest $R_{\mathrm{Roche}}/a$ values clustering above the radius valley. The larger Roche radius of these planets enabled them to keep their atmospheric envelope, and they therefore appear to be more puffy than the planets below the radius valley.

\subsection{Dependence of the radius valley on stellar mass}
\begin{figure}
    \centering
    \includegraphics[width=\columnwidth]{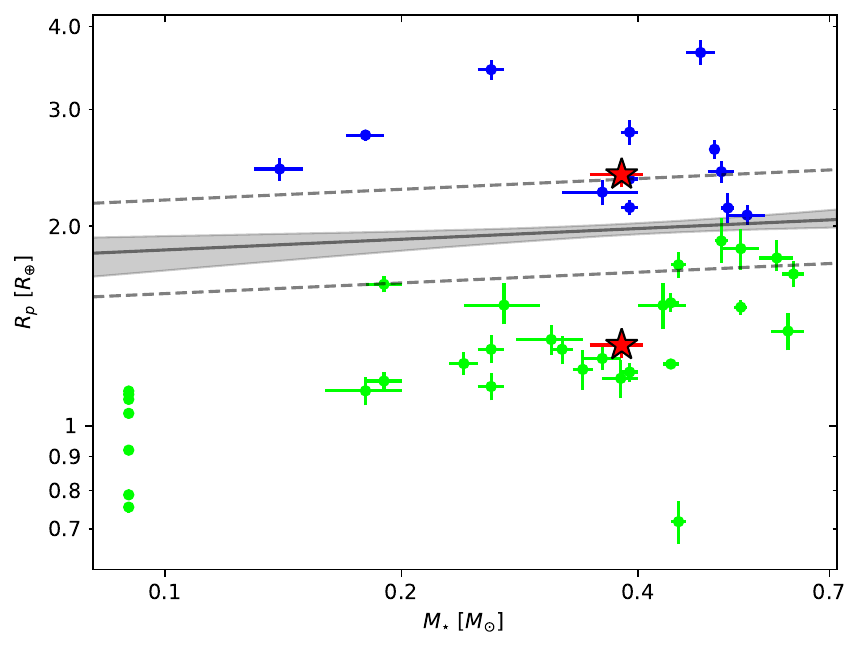}
    \caption{Same as Fig.~\ref{fig:RvalleyP}, but this time, $R_p$ is plotted against stellar mass.}
    \label{fig:RvalleyMs}
\end{figure}

Considering the increasing interest in exploring the trend between the radius valley and the spectral type of the host star \citep[e.g.][]{wu2019,gupta2020,rogers2021,ho2023,berger2023}, we repeated the SVM analysis described above, but assuming the covariate pair $(R_p, M_{\star})$, and we derived
\begin{equation}
    \log{R_{p,\mathrm{valley}}}=+0.054_{-0.034}^{+0.049}\log{M_{\star}}+0.319_{-0.016}^{+0.022}
    \label{eq:RpValleyMs}
\end{equation}
(see Fig.~\ref{fig:RvalleyMs}). Estimates of the radius valley slope $\RMslope$ that are based on observational data as found in the literature \citep{berger2020,petigura2022,ho2023} are mainly the results of works focusing on FGK stars, which lead to steeper slopes (although the accompanying uncertainties are about 40\% or higher). The only homogeneous comparison currently available is with the work by L22, who found $\RMslope=+0.08\pm0.12$ (consistent with our estimate), which may again suggest that planets orbiting M dwarfs differ from those orbiting FGK stars in the context of the radius valley.

However, from a theoretical perspective, it is hard to draw firm conclusions about the mechanisms underlying the formation and evolution of exoplanets when studying the radius valley within the $R_p$-$M_{\star}$ space. \citet{rogers2021} cautioned that the $R_{p,\mathrm{valley}}$-$M_{\star}$ slope shows several degeneracies. They theoretically derived that the expected slope does not only depend on $M_{\star}$, but also on the incident bolometric flux $S$, and it can be expressed as
\begin{equation}
    \left.\frac{\mathrm{d}\log{R_{p,\mathrm{valley}}}}{\mathrm{d}\log{M_{\star}}}\right\vert_{\mathrm{th}}\approx \alpha\left(\zeta - \frac{2}{3} \right)+\beta\,,
    \label{eq:RMslopeth}
\end{equation}
where $\alpha\equiv\partial\log{R_{p,\mathrm{valley}}}/\partial\log{S}$ and $\beta\equiv\partial\log{R_{p,\mathrm{valley}}}/\partial\log{M_{\star}}$ are predicted, depending on the scenario at play (either photo-evaporation or core-powered mass loss), while $\zeta$ is the exponent entering the mass-luminosity relation, that is, $L_{\star}\propto M_{\star}^{\zeta}$. Because $\zeta\gg\alpha$ and $\zeta\gg |\beta|$ \citep{rogers2021}, the slope value is mainly controlled by $\zeta$, which needs to be properly estimated according to the stellar spectral type. 
\citet{cuntz2018} proposed the following expression for $\zeta=\zeta(M_{\star})$ for low-mass stars:
\begin{equation}
    \zeta=-141.7M_{\star}^4+232.4M_{\star}^3-129.1M_{\star}^2+33.29M+0.215
    \label{eq:LMexp}
\end{equation}
and averaging out that function over our mass range of interest, we obtained $\zeta_{M}=4.0$. Plugging in the $(\alpha,\beta)$ predictions by \citet{rogers2021} along with $\zeta_{M}$ in Eq.~(\ref{eq:RMslopeth}), we computed $\RMslope_{\mathrm{th}}$\,$\approx$\,+0.23 and +0.27 for the photo-evaporation and core-powered mass-loss models, respectively. The difference with our observationally inferred estimate may suggest that other mechanisms shape the observed properties of planets orbiting M dwarfs (e.g. a significant role of gas-poor formation, for which \citet{lee2021} theoretically predicted a $\RMslope$ down to $+0.11$). However, \citet{wu2019} first remarked that the specific scaling relation between the planetary core mass and stellar mass further influences Eq.~(\ref{eq:RMslopeth}), and \citet{rogers2021} indeed verified that the $\RMslope$ may be considerably altered when these scalings are accounted for.

\subsection{Density valley}\label{sec:densityValley}
\begin{figure}
    \centering
    \includegraphics[width=\columnwidth]{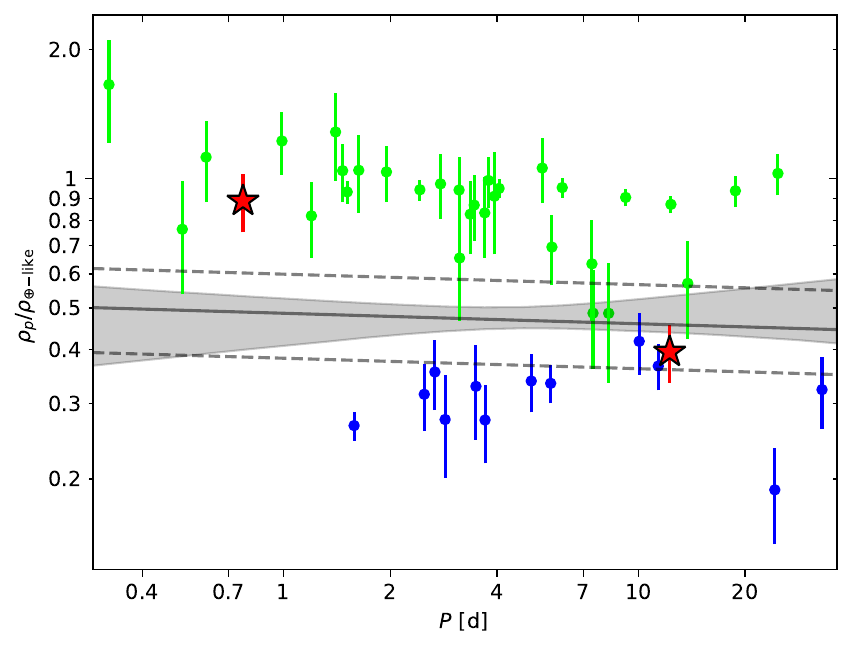}
    \caption{Normalised density as a function of orbital period. The normalised density is the mean density of the planet divided by the density the planet would have if it had an Earth-like composition (same variable as introduced in L22 to display the density valley). The grey line and its shaded area indicate the density valley with its corresponding error, as in Fig.\ref{fig:RvalleyP}.}
    \label{fig:rhovalleyP}
\end{figure}

Finally, as L22 concluded that the demographics of exoplanets can be better visualised by considering the density valley, we repeated the SVM analysis in the $\hat{\rho}$-$P$ space, where the normalised density $\hat{\rho}\equiv\frac{\rho_p}{\rho_{\oplus\cdot\mathrm{like}}}$ and $\rho_{\oplus\cdot\mathrm{like}}$ is the density that a planet of given mass would have if it had an Earth-like composition. As done by L22, we followed \citet{zeng2019}, who computed that an Earth-like planet of mass $M_p$ has a radius $R_{\oplus\cdot\mathrm{like}}=M_p^{\frac{1}{3.7}}$, where both the mass and the radius are expressed in Earth units. Therefore, according to \citet{zeng2019}, the density of an Earth-like planet scales as (Earth units)
\begin{equation}
    \rho_{\oplus\cdot\mathrm{like}}=M_p^{\frac{0.7}{3.7}},
    \label{eq:rhoEarthLike}
\end{equation}
which is the normalisation factor to derive $\hat{\rho}$ from $\rho_p$. The density valley is shown in Fig.~\ref{fig:rhovalleyP} along with the SVM-based best-fit line,
\begin{equation}
    \log{\hat{\rho}_\mathrm{valley}}=-0.02_{-0.04}^{+0.12}\log{P}-0.313_{-0.076}^{+0.034}.
    \label{eq:rhopValleyP}
\end{equation} 
Fig.~\ref{fig:rhovalleyP} confirms that the normalised density $\hat{\rho}$ separates two different populations of exoplanets, as first pointed out by L22. Our quantitative characterisation of the valley yields a slope $\Rhoslope=-0.02_{-0.04}^{+0.12}$, which is well consistent with zero, similar to $\Rhoslope_{\mathrm{L22}}=+0.02\pm0.04$ estimated by L22. 

The agreement of both our $\Rslope$ and $\Rhoslope$ outcomes with the results from L22 may also suggest that the L22 interpretation of planet demographics may be followed. In detail, L22 identified that planets with $R_p\lesssim 1.6\,R_{\oplus}$ are rocky, planets with $R_p\gtrsim 2.3\,R_{\oplus}$ are puffy sub-Neptunes, and planets with intermediate radii are water worlds, that is, planets with the same mass content of condensed water and rocks. L22 interpreted the density gap as a division between rocky planets and water worlds, which also agrees with the conclusions by \citet{venturini2020}, who find that the radius gap separates dry from wet planets.

At lower stellar mass, the minimum mass for a planet to undergo type I migration decreases \citep[e.g.][]{burn2021}. As a result, water worlds are more common around M dwarfs, and their abundance shapes the topology of the radius valley, which is then determined by the favoured inward migration of water worlds rather than by atmospheric loss processes (Venturini et al. in prep.). The migration causes an overlap between rocky planets and water worlds within the mass-radius and $R_p$-$P$ space. Hence, the radius valley is partially filled (as also found by L22) and its slope becomes shallower than expected from thermally-driven atmospheric mass-loss mechanisms.

\section{Conclusions}\label{sec:conclusions}
The M4\,V star TOI-732 hosts two transiting planets, namely a close-in USP planet at $P_b$\,$\sim$\,0.77 d and an outer one at $P_c$\,$\sim$\,12.25 d. They straddle the radius valley and have $R_b$\,$\sim$\,1.3\,$R_{\oplus}$ and $R_c$\,$\sim$\,2.4\,$R_{\oplus}$.
The system has been analysed by C20, N20, and L22, but by collecting 25 CHEOPS LCs and benefiting from a further still unpublished TESS sector, we were able to double the number of space-based observations for a total of $\sim$\,140 transit events observed with both ground- and space-based facilities. Furthermore, in addition to the 127 RV data points already available in the literature, we obtained 38 RV observations with the MAROON-X spectrograph.

We jointly analysed all the available LCs and RV time series using the MCMCI routine by \citet{bonfanti2020}, reaching a transit depth precision of 4.4\% (resp. 4.0\%) and an RV semi-amplitude precision of 6.2\% (3.8\%) for TOI-732\,b (TOI-732\,c). Even with respect to the most recent parameters available in the literature, we were able to improve the precision on the transit and RV observables up to a factor $\sim$\,2.4, with a remarkably positive impact on the mean densities of both planets. We estimated $\rho_b=5.8_{-0.8}^{+1.0}$ g\,cm$^{-3}$ and $\rho_c=3.24_{-0.43}^{+0.55}$ g\,cm$^{-3}$ (hence $\sim$\,15\% uncertainty for both), and only $\sim$\,20\% of the currently known exoplanets around M dwarf are known with a comparable or better precision according to the NASA Exoplanet Archive.

Based on the internal structure modelling we performed, TOI-732\,b probably does not host any gaseous envelope, but it is fully compatible with a rocky composition. Instead, TOI-732\,c is compatible with having a volatile layer, with our interior structure model yielding a H-He envelope mass $M_{\mathrm{gas,c}}=0.02^{+0.05}_{-0.02}$ $M_{\oplus}$, which corresponds to a thickness of $R_{\mathrm{gas,c}}=0.40_{-0.27}^{+0.24}\,R_{\oplus}$. However, based on the \citet{aguichine2021} models, the mass and radius values of TOI-732\,c are also compatible with an Earth-like core surrounded by a steam water layer. From the physical parameters of the planets, we then infer that the inner planet is a super-Earth, while the outer planet is a sub-Neptune. This constitutes a quite common system architecture in the exoplanet field.

We finally built a sample of well-characterised M-dwarf exoplanets (the \texttt{Msample}) with $R_p<4\,R_{\oplus}$ and whose radii and masses are known to better than 8\% and 25\%, respectively. After this, we investigated the slopes of the radius valley as a function of the planet orbital periods and of the host stellar mass because theoretical models predict different trends depending on the mechanisms that have underlain planet formation and evolution. Following an SVM approach \citep[e.g.][]{cortes1995}, we determined a $\Rslope=-0.065_{-0.013}^{+0.024}$, differing by $\sim$\,2$\sigma$ from the $\Rslope_{\mathrm{H23}}=-0.11\pm0.02$ slope derived by \citet{ho2023} when targeting FGK stars, which may imply that formation and evolution mechanisms are at play with different weights in FGK and M-dwarf exoplanet systems. 

Theoretical predictions would associate a $\Rslope_{\mathrm{TD}}=-0.10$ with a thermally driven mass-loss scenario \citep{affolter2023}, while \citet{lopez2018} computed an upper limit for the radius valley slope ($\Rslope_{\mathrm{L18}}=+0.11$)  derived from gas-empty planet formation models. As our result falls in between, with a negative slope, we may argue that thermally driven mass-loss events can explain the evolution of the majority of M-dwarf exoplanets, but some of the planets in our \texttt{Msample} may be compatible with the gas-poor formation scenario. This type of formation mechanism has recently been invoked to justify the physical properties of some exoplanets hosted by M dwarfs, such as the cases of TOI-1634\,b \citep{cloutier2021} or LHS\,1903\,e \citep{wilson2023}.

An alternative explanation for the observed radius valley topology instead relies on the abundance of water worlds around low-mass stars. In particular, the favoured inward migration of water worlds suggested by simulations \citep{venturini2020,burn2021} would cause a partial filling of the radius valley (L22; Venturini et al., in prep.). The radius-valley slope would then become flatter compared to what is theoretically expected if only thermally driven mass-loss mechanisms were at play; this agrees with our $\Rslope$ estimate.
Following L22, we further confirm the presence of a density valley that better separates rocky and water-rich exoplanets. By repeating the SVM analysis in the $(\hat{\rho}, P)$ plane, we computed a slope of $\Rhoslope=-0.02_{-0.04}^{+0.12}$, which agrees well with the $\Rhoslope_{\mathrm{L22}}=+0.02\pm0.04$ value found by L22. Therefore, the interpretation of L22 in terms of planet demographics and formation can be adopted here as well. In summary, when comparing theoretical predictions of $\Rslope$ values (see Tab.~\ref{tab:slopeTheory}) with our findings, our $\Rslope=-0.065_{-0.013}^{+0.024}$ estimate can be justified by invoking further mechanisms (e.g. gas-poor formation or inward migration) in addition to thermally driven mass-loss phenomena.

\begin{acknowledgements}
We thank the anonymous referee for the valuable comments, which improve the quality of the manuscript.
CHEOPS is an ESA mission in partnership with Switzerland with important contributions to the payload and the ground segment from Austria, Belgium, France, Germany, Hungary, Italy, Portugal, Spain, Sweden, and the United Kingdom. The CHEOPS Consortium would like to gratefully acknowledge the support received by all the agencies, offices, universities, and industries involved. Their flexibility and willingness to explore new approaches were essential to the success of this mission. CHEOPS data analysed in this article will be made available in the CHEOPS mission archive (\url{https://cheops.unige.ch/archive_browser/}). 
TWi acknowledges support from the UKSA and the University of Warwick. 
This work has been carried out within the framework of the NCCR PlanetS supported by the Swiss National Science Foundation under grants 51NF40\_182901 and 51NF40\_205606. 
ABr was supported by the SNSA. 
S.G.S. acknowledge support from FCT through FCT contract nr. CEECIND/00826/2018 and POPH/FSE (EC). 
ML acknowledges support of the Swiss National Science Foundation under grant number PCEFP2\_194576. 
ASi and CBr acknowledge support from the Swiss Space Office through the ESA PRODEX program. 
This work was also partially supported by a grant from the Simons Foundation (PI Queloz, grant number 327127). 
YAl acknowledges support from the Swiss National Science Foundation (SNSF) under grant 200020\_192038. 
NCSa acknowledges funding by the European Union (ERC, FIERCE, 101052347). Views and opinions expressed are however those of the author(s) only and do not necessarily reflect those of the European Union or the European Research Council. Neither the European Union nor the granting authority can be held responsible for them. 
V.V.G. is an F.R.S-FNRS Research Associate. 
RAl, DBa, EPa, and IRi acknowledge financial support from the Agencia Estatal de Investigación of the Ministerio de Ciencia e Innovación MCIN/AEI/10.13039/501100011033 and the ERDF “A way of making Europe” through projects PID2019-107061GB-C61, PID2019-107061GB-C66, PID2021-125627OB-C31, and PID2021-125627OB-C32, from the Centre of Excellence “Severo Ochoa'' award to the Instituto de Astrofísica de Canarias (CEX2019-000920-S), from the Centre of Excellence “María de Maeztu” award to the Institut de Ciències de l’Espai (CEX2020-001058-M), and from the Generalitat de Catalunya/CERCA programme. 
S.C.C.B. acknowledges support from FCT through FCT contracts nr. IF/01312/2014/CP1215/CT0004. 
LBo, VNa, IPa, GPi, RRa, GSc, and TZi acknowledge support from CHEOPS ASI-INAF agreement n. 2019-29-HH.0. 
ACC acknowledges support from STFC consolidated grant numbers ST/R000824/1 and ST/V000861/1, and UKSA grant number ST/R003203/1. 
P.E.C. is funded by the Austrian Science Fund (FWF) Erwin Schroedinger Fellowship, program J4595-N. 
This project was supported by the CNES. 
The Belgian participation to CHEOPS has been supported by the Belgian Federal Science Policy Office (BELSPO) in the framework of the PRODEX Program, and by the University of Liège through an ARC grant for Concerted Research Actions financed by the Wallonia-Brussels Federation. 
L.D. is an F.R.S.-FNRS Postdoctoral Researcher. 
This work was supported by FCT - Funda\c{c}\~{a}o para a Ci\^{e}ncia e a Tecnologia through national funds and by FEDER through COMPETE2020 through the research grants UIDB/04434/2020, UIDP/04434/2020, 2022.06962.PTDC. 
O.D.S.D. is supported in the form of work contract (DL 57/2016/CP1364/CT0004) funded by national funds through FCT. 
B.-O. D. acknowledges support from the Swiss State Secretariat for Education, Research and Innovation (SERI) under contract number MB22.00046. 
This project has received funding from the European Research Council (ERC) under the European Union’s Horizon 2020 research and innovation programme (project {\sc Four Aces}. 
grant agreement No 724427). It has also been carried out in the frame of the National Centre for Competence in Research PlanetS supported by the Swiss National Science Foundation (SNSF). DE acknowledges financial support from the Swiss National Science Foundation for project 200021\_200726. 
MF and CMP gratefully acknowledge the support of the Swedish National Space Agency (DNR 65/19, 174/18). 
DG gratefully acknowledges financial support from the CRT foundation under Grant No. 2018.2323 ``Gaseousor rocky? Unveiling the nature of small worlds''. 
M.G. is an F.R.S.-FNRS Senior Research Associate. 
MNG is the ESA CHEOPS Project Scientist and Mission Representative, and as such also responsible for the Guest Observers (GO) Programme. MNG does not relay proprietary information between the GO and Guaranteed Time Observation (GTO) Programmes, and does not decide on the definition and target selection of the GTO Programme. 
CHe acknowledges support from the European Union H2020-MSCA-ITN-2019 under Grant Agreement no. 860470 (CHAMELEON). 
SH gratefully acknowledges CNES funding through the grant 837319. 
KGI is the ESA CHEOPS Project Scientist and is responsible for the ESA CHEOPS Guest Observers Programme. She does not participate in, or contribute to, the definition of the Guaranteed Time Programme of the CHEOPS mission through which observations described in this paper have been taken, nor to any aspect of target selection for the programme. 
K.W.F.L. was supported by Deutsche Forschungsgemeinschaft grants RA714/14-1 within the DFG Schwerpunkt SPP 1992, Exploring the Diversity of Extrasolar Planets. 
This work was granted access to the HPC resources of MesoPSL financed by the Region Ile de France and the project Equip@Meso (reference ANR-10-EQPX-29-01) of the programme Investissements d'Avenir supervised by the Agence Nationale pour la Recherche. 
PM acknowledges support from STFC research grant number ST/M001040/1. 
GS acknowledges support provided by NASA through the NASA Hubble Fellowship grant HST-HF2-51519.001-A awarded by the Space Telescope Science Institute, which is operated by the Association of Universities for Research in Astronomy, Inc., for NASA, under contract NAS5-26555. 
GyMSz acknowledges the support of the Hungarian National Research, Development and Innovation Office (NKFIH) grant K-125015, a a PRODEX Experiment Agreement No. 4000137122, the Lend\"ulet LP2018-7/2021 grant of the Hungarian Academy of Science and the support of the city of Szombathely. 
NAW acknowledges UKSA grant ST/R004838/1.
R.L. acknowledges funding from University of La Laguna through the Margarita Salas Fellowship from the Spanish Ministry of Universities ref. UNI/551/2021-May 26, and under the EU Next Generation funds. 
This work has made use of data from the European Space Agency (ESA) mission
{\it Gaia} (\url{https://www.cosmos.esa.int/gaia}), processed by the {\it Gaia}
Data Processing and Analysis Consortium (DPAC,
\url{https://www.cosmos.esa.int/web/gaia/dpac/consortium}). Funding for the DPAC
has been provided by national institutions, in particular the institutions
participating in the {\it Gaia} Multilateral Agreement.
This article is based on observations made with the MuSCAT2 instrument, developed by ABC, at Telescopio Carlos S\'{a}nchez operated on the island of Tenerife by the IAC in the Spanish Observatorio del Teide. This work is partly financed by the Spanish Ministry of Economics and Competitiveness through grants PGC2018-098153-B-C31.
This work is partly supported by JSPS KAKENHI Grant Numbers JP17H04574, JP18H01265, and JP18H05439, Grant-in-Aid for JSPS Fellows Grant Number JP20J21872, JST PRESTO Grant Number JPMJPR1775, and a University Research Support Grant from the National Astronomical Observatory of Japan (NAOJ).
This material is based upon work supported by the National Science Foundation Graduate Research Fellowship under Grant No. DGE 1746045.
\end{acknowledgements}

%
%

\bibliographystyle{aa}
\bibliography{biblio}

\begin{thebibliography}{211}
\expandafter\ifx\csname natexlab\endcsname\relax\def\natexlab#1{#1}\fi

\bibitem[{{Abdurro'uf} {et~al.}(2022){Abdurro'uf}, {Accetta}, {Aerts}, {Silva
  Aguirre}, {Ahumada}, {Ajgaonkar}, {Filiz Ak}, {Alam}, {Allende Prieto},
  {Almeida}, \& et~al.}]{Abdurrouf-22}
{Abdurro'uf}, {Accetta}, K., {Aerts}, C., {et~al.} 2022, \apjs, 259, 35

\bibitem[{{Adibekyan} {et~al.}(2021){Adibekyan}, {Dorn}, {Sousa}, {Santos},
  {Bitsch}, {Israelian}, {Mordasini}, {Barros}, {Delgado Mena}, {Demangeon},
  {Faria}, {Figueira}, {Hakobyan}, {Oshagh}, {Soares}, {Kunitomo}, {Takeda},
  {Jofr{\'e}}, {Petrucci}, \& {Martioli}}]{adibekyan2021}
{Adibekyan}, V., {Dorn}, C., {Sousa}, S.~G., {et~al.} 2021, Science, 374, 330

\bibitem[{{Adibekyan} {et~al.}(2012){Adibekyan}, {Sousa}, {Santos}, {Delgado
  Mena}, {Gonz{\'a}lez Hern{\'a}ndez}, {Israelian}, {Mayor}, \&
  {Khachatryan}}]{Adibekyan-12b}
{Adibekyan}, V.~Z., {Sousa}, S.~G., {Santos}, N.~C., {et~al.} 2012, \aap, 545,
  A32

\bibitem[{{Affolter} {et~al.}(2023){Affolter}, {Mordasini}, {Oza},
  {Kubyshkina}, \& {Fossati}}]{affolter2023}
{Affolter}, L., {Mordasini}, C., {Oza}, A.~V., {Kubyshkina}, D., \& {Fossati},
  L. 2023, \aap, 676, A119

\bibitem[{{Agol} {et~al.}(2021){Agol}, {Dorn}, {Grimm}, {Turbet}, {Ducrot},
  {Delrez}, {Gillon}, {Demory}, {Burdanov}, {Barkaoui}, {Benkhaldoun},
  {Bolmont}, {Burgasser}, {Carey}, {de Wit}, {Fabrycky}, {Foreman-Mackey},
  {Haldemann}, {Hernandez}, {Ingalls}, {Jehin}, {Langford}, {Leconte},
  {Lederer}, {Luger}, {Malhotra}, {Meadows}, {Morris}, {Pozuelos}, {Queloz},
  {Raymond}, {Selsis}, {Sestovic}, {Triaud}, \& {Van Grootel}}]{agol2021}
{Agol}, E., {Dorn}, C., {Grimm}, S.~L., {et~al.} 2021, \psj, 2, 1

\bibitem[{{Aguichine} {et~al.}(2021){Aguichine}, {Mousis}, {Deleuil}, \&
  {Marcq}}]{aguichine2021}
{Aguichine}, A., {Mousis}, O., {Deleuil}, M., \& {Marcq}, E. 2021, \apj, 914,
  84

\bibitem[{{Alibert}(2017)}]{Alibert17}
{Alibert}, Y. 2017, \aap, 606, A69

\bibitem[{{Alibert} {et~al.}(2013){Alibert}, {Carron}, {Fortier}, {Pfyffer},
  {Benz}, {Mordasini}, \& {Swoboda}}]{Alibert2013}
{Alibert}, Y., {Carron}, F., {Fortier}, A., {et~al.} 2013, \aap, 558, A109

\bibitem[{{Allard}(2014)}]{Allard2014}
{Allard}, F. 2014, in Exploring the Formation and Evolution of Planetary
  Systems, ed. M.~{Booth}, B.~C. {Matthews}, \& J.~R. {Graham}, Vol. 299,
  271--272

\bibitem[{{Almeida-Fernandes} \& {Rocha-Pinto}(2018)}]{AlmeidaFernandes2018}
{Almeida-Fernandes}, F. \& {Rocha-Pinto}, H.~J. 2018, \mnras, 476, 184

\bibitem[{{Almenara} {et~al.}(2022){Almenara}, {Bonfils}, {Otegi}, {Attia},
  {Turbet}, {Astudillo-Defru}, {Collins}, {Polanski}, {Bourrier}, {Hellier},
  {Ziegler}, {Bouchy}, {Briceno}, {Charbonneau}, {Cointepas}, {Collins},
  {Crossfield}, {Delfosse}, {Diaz}, {Dorn}, {Doty}, {Forveille}, {Gaisn{\'e}},
  {Gan}, {Helled}, {Hesse}, {Jenkins}, {Jensen}, {Latham}, {Law}, {Mann},
  {Mao}, {McLean}, {Murgas}, {Myers}, {Seager}, {Shporer}, {Tan}, {Twicken}, \&
  {Winn}}]{almenara2022}
{Almenara}, J.~M., {Bonfils}, X., {Otegi}, J.~F., {et~al.} 2022, \aap, 665, A91

\bibitem[{{Anglada-Escud{\'e}} \& {Butler}(2012)}]{angladaEscude2012}
{Anglada-Escud{\'e}}, G. \& {Butler}, R.~P. 2012, \apjs, 200, 15

\bibitem[{{Antoniadis-Karnavas} {et~al.}(2020){Antoniadis-Karnavas}, {Sousa},
  {Delgado-Mena}, {Santos}, {Teixeira}, \& {Neves}}]{Antoniadis-2020}
{Antoniadis-Karnavas}, A., {Sousa}, S.~G., {Delgado-Mena}, E., {et~al.} 2020,
  \aap, 636, A9

\bibitem[{{Astudillo-Defru} {et~al.}(2020){Astudillo-Defru}, {Cloutier},
  {Wang}, {Teske}, {Brahm}, {Hellier}, {Ricker}, {Vanderspek}, {Latham},
  {Seager}, {Winn}, {Jenkins}, {Collins}, {Stassun}, {Ziegler}, {Almenara},
  {Anderson}, {Artigau}, {Bonfils}, {Bouchy}, {Brice{\~n}o}, {Butler},
  {Charbonneau}, {Conti}, {Crane}, {Crossfield}, {Davies}, {Delfosse},
  {D{\'\i}az}, {Doyon}, {Dragomir}, {Eastman}, {Espinoza}, {Essack}, {Feng},
  {Figueira}, {Forveille}, {Gan}, {Glidden}, {Guerrero}, {Hart}, {Henning},
  {Horch}, {Isopi}, {Jenkins}, {Jord{\'a}n}, {Kielkopf}, {Law}, {Lovis},
  {Mallia}, {Mann}, {de Medeiros}, {Melo}, {Mennickent}, {Mignon}, {Murgas},
  {Nusdeo}, {Pepe}, {Relles}, {Rose}, {Santos}, {S{\'e}gransan}, {Shectman},
  {Shporer}, {Smith}, {Torres}, {Udry}, {Villasenor}, {Winters}, \&
  {Zhou}}]{astudilloDefru2020}
{Astudillo-Defru}, N., {Cloutier}, R., {Wang}, S.~X., {et~al.} 2020, \aap, 636,
  A58

\bibitem[{{Barkaoui} {et~al.}(2019){Barkaoui}, {Burdanov}, {Hellier}, {Gillon},
  {Smalley}, {Maxted}, {Lendl}, {Triaud}, {Anderson}, {McCormac}, {Jehin},
  {Almleaky}, {Armstrong}, {Benkhaldoun}, {Bouchy}, {Brown}, {Cameron},
  {Daassou}, {Delrez}, {Ducrot}, {Foxell}, {Murray}, {Nielsen}, {Pepe},
  {Pollacco}, {Pozuelos}, {Queloz}, {Segransan}, {Udry}, {Thompson}, \&
  {West}}]{barkaoui2019}
{Barkaoui}, K., {Burdanov}, A., {Hellier}, C., {et~al.} 2019, \aj, 157, 43

\bibitem[{{Barkaoui} {et~al.}(2023){Barkaoui}, {Timmermans}, {Soubkiou},
  {Rackham}, {Burgasser}, {Chouqar}, {Pozuelos}, {Collins}, {Howell}, {Simcoe},
  {Melis}, {Stassun}, {Tregloan-Reed}, {Cointepas}, {Gillon}, {Bonfils},
  {Furlan}, {Gnilka}, {Almenara}, {Alonso}, {Benkhaldoun}, {Bonavita},
  {Bouchy}, {Burdanov}, {Chinchilla}, {Davoudi}, {Delrez}, {Demangeon},
  {Dominik}, {Demory}, {de Wit}, {Dransfield}, {Ducrot}, {Fukui}, {Hinse},
  {Hooton}, {Jehin}, {Jenkins}, {J{\o}rgensen}, {Latham}, {Garcia},
  {Carrazco-Gaxiola}, {Ghachoui}, {G{\'o}mez Maqueo Chew}, {G{\"u}nther},
  {McCormac}, {Murgas}, {Murray}, {Narita}, {Niraula}, {Pedersen}, {Queloz},
  {Rebolo-L{\'o}pez}, {Ricker}, {Sabin}, {Sajadian}, {Schanche}, {Schwarz},
  {Seager}, {Sebastian}, {Sefako}, {Sohy}, {Southworth}, {Srdoc}, {Thompson},
  {Triaud}, {Vanderspek}, {Wells}, {Winn}, \&
  {Z{\'u}{\~n}iga-Fern{\'a}ndez}}]{barkaoui2023}
{Barkaoui}, K., {Timmermans}, M., {Soubkiou}, A., {et~al.} 2023, \aap, 677, A38

\bibitem[{{Barnes}(2017)}]{barnes2017}
{Barnes}, R. 2017, Celestial Mechanics and Dynamical Astronomy, 129, 509

\bibitem[{{Bastian} {et~al.}(2010){Bastian}, {Covey}, \& {Meyer}}]{bastian2010}
{Bastian}, N., {Covey}, K.~R., \& {Meyer}, M.~R. 2010, \araa, 48, 339

\bibitem[{{Bean} {et~al.}(2010){Bean}, {Seifahrt}, {Hartman}, {Nilsson},
  {Wiedemann}, {Reiners}, {Dreizler}, \& {Henry}}]{Bean10}
{Bean}, J.~L., {Seifahrt}, A., {Hartman}, H., {et~al.} 2010, \apj, 713, 410

\bibitem[{Ben-Hur {et~al.}(2002)Ben-Hur, Horn, Siegelmann, \&
  Vapnik}]{benhur2002}
Ben-Hur, A., Horn, D., Siegelmann, H.~T., \& Vapnik, V. 2002, J. Mach. Learn.
  Res., 2, 125–137

\bibitem[{{Benedict} {et~al.}(2016){Benedict}, {Henry}, {Franz}, {McArthur},
  {Wasserman}, {Jao}, {Cargile}, {Dieterich}, {Bradley}, {Nelan}, \&
  {Whipple}}]{benedict2016}
{Benedict}, G.~F., {Henry}, T.~J., {Franz}, O.~G., {et~al.} 2016, \aj, 152, 141

\bibitem[{{Benneke} {et~al.}(2019){Benneke}, {Wong}, {Piaulet}, {Knutson},
  {Lothringer}, {Morley}, {Crossfield}, {Gao}, {Greene}, {Dressing},
  {Dragomir}, {Howard}, {McCullough}, {Kempton}, {Fortney}, \&
  {Fraine}}]{benneke2019}
{Benneke}, B., {Wong}, I., {Piaulet}, C., {et~al.} 2019, \apjl, 887, L14

\bibitem[{{Benz} {et~al.}(2021){Benz}, {Broeg}, {Fortier}, {Rando}, {Beck},
  {Beck}, {Queloz}, {Ehrenreich}, {Maxted}, {Isaak}, {Billot}, {Alibert},
  {Alonso}, {Ant{\'o}nio}, {Asquier}, {Bandy}, {B{\'a}rczy}, {Barrado},
  {Barros}, {Baumjohann}, {Bekkelien}, {Bergomi}, {Biondi}, {Bonfils},
  {Borsato}, {Brandeker}, {Busch}, {Cabrera}, {Cessa}, {Charnoz}, {Chazelas},
  {Collier Cameron}, {Corral Van Damme}, {Cortes}, {Davies}, {Deleuil},
  {Deline}, {Delrez}, {Demangeon}, {Demory}, {Erikson}, {Farinato}, {Fossati},
  {Fridlund}, {Futyan}, {Gandolfi}, {Garcia Munoz}, {Gillon}, {Guterman},
  {Gutierrez}, {Hasiba}, {Heng}, {Hernandez}, {Hoyer}, {Kiss}, {Kovacs},
  {Kuntzer}, {Laskar}, {Lecavelier des Etangs}, {Lendl}, {L{\'o}pez}, {Lora},
  {Lovis}, {L{\"u}ftinger}, {Magrin}, {Malvasio}, {Marafatto}, {Michaelis}, {de
  Miguel}, {Modrego}, {Munari}, {Nascimbeni}, {Olofsson}, {Ottacher},
  {Ottensamer}, {Pagano}, {Palacios}, {Pall{\'e}}, {Peter}, {Piazza}, {Piotto},
  {Pizarro}, {Pollaco}, {Ragazzoni}, {Ratti}, {Rauer}, {Ribas}, {Rieder},
  {Rohlfs}, {Safa}, {Salatti}, {Santos}, {Scandariato}, {S{\'e}gransan},
  {Simon}, {Smith}, {Sordet}, {Sousa}, {Steller}, {Szab{\'o}}, {Szoke},
  {Thomas}, {Tschentscher}, {Udry}, {Van Grootel}, {Viotto}, {Walter},
  {Walton}, {Wildi}, \& {Wolter}}]{benz2021}
{Benz}, W., {Broeg}, C., {Fortier}, A., {et~al.} 2021, Experimental Astronomy,
  51, 109

\bibitem[{{Berger} {et~al.}(2020){Berger}, {Huber}, {Gaidos}, {van Saders}, \&
  {Weiss}}]{berger2020}
{Berger}, T.~A., {Huber}, D., {Gaidos}, E., {van Saders}, J.~L., \& {Weiss},
  L.~M. 2020, \aj, 160, 108

\bibitem[{{Berger} {et~al.}(2023){Berger}, {Schlieder}, {Huber}, \&
  {Barclay}}]{berger2023}
{Berger}, T.~A., {Schlieder}, J.~E., {Huber}, D., \& {Barclay}, T. 2023, arXiv
  e-prints, arXiv:2302.00009

\bibitem[{{Blackwell} \& {Shallis}(1977)}]{Blackwell1977}
{Blackwell}, D.~E. \& {Shallis}, M.~J. 1977, \mnras, 180, 177

\bibitem[{{Bluhm} {et~al.}(2020){Bluhm}, {Luque}, {Espinoza}, {Pall{\'e}},
  {Caballero}, {Dreizler}, {Livingston}, {Mathur}, {Quirrenbach}, {Stock}, {Van
  Eylen}, {Nowak}, {L{\'o}pez}, {Csizmadia}, {Zapatero Osorio}, {Sch{\"o}fer},
  {Lillo-Box}, {Oshagh}, {Gonz{\'a}lez-{\'A}lvarez}, {Amado}, {Barrado},
  {B{\'e}jar}, {Cale}, {Chaturvedi}, {Cifuentes}, {Cochran}, {Collins},
  {Collins}, {Cort{\'e}s-Contreras}, {D{\'\i}ez Alonso}, {El Mufti},
  {Ercolino}, {Fridlund}, {Gaidos}, {Garc{\'\i}a}, {Georgieva},
  {Gonz{\'a}lez-Cuesta}, {Guerra}, {Hatzes}, {Henning}, {Herrero}, {Hidalgo},
  {Isopi}, {Jeffers}, {Jenkins}, {Jensen}, {K{\'a}bath}, {Kaminski}, {Kemmer},
  {Korth}, {Kossakowski}, {K{\"u}rster}, {Lafarga}, {Mallia}, {Montes},
  {Morales}, {Morales-Calder{\'o}n}, {Murgas}, {Narita}, {Passegger}, {Pedraz},
  {Persson}, {Plavchan}, {Rauer}, {Redfield}, {Reffert}, {Reiners}, {Ribas},
  {Ricker}, {Rodr{\'\i}guez-L{\'o}pez}, {Santos}, {Seager}, {Schlecker},
  {Schweitzer}, {Shan}, {Soto}, {Subjak}, {Tal-Or}, {Trifonov}, {Vanaverbeke},
  {Vanderspek}, {Wittrock}, {Zechmeister}, \& {Zohrabi}}]{bluhm2020}
{Bluhm}, P., {Luque}, R., {Espinoza}, N., {et~al.} 2020, \aap, 639, A132

\bibitem[{{Bonfanti} {et~al.}(2021){Bonfanti}, {Delrez}, {Hooton}, {Wilson},
  {Fossati}, {Alibert}, {Hoyer}, {Mustill}, {Osborn}, {Adibekyan}, {Gandolfi},
  {Salmon}, {Sousa}, {Tuson}, {Van Grootel}, {Cabrera}, {Nascimbeni}, {Maxted},
  {Barros}, {Billot}, {Bonfils}, {Borsato}, {Broeg}, {Davies}, {Deleuil},
  {Demangeon}, {Fridlund}, {Lacedelli}, {Lendl}, {Persson}, {Santos},
  {Scandariato}, {Szab{\'o}}, {Collier Cameron}, {Udry}, {Benz}, {Beck},
  {Ehrenreich}, {Fortier}, {Isaak}, {Queloz}, {Alonso}, {Asquier}, {Bandy},
  {B{\'a}rczy}, {Barrado}, {Barrag{\'a}n}, {Baumjohann}, {Beck}, {Bekkelien},
  {Bergomi}, {Brandeker}, {Busch}, {Cessa}, {Charnoz}, {Chazelas}, {Corral Van
  Damme}, {Demory}, {Erikson}, {Farinato}, {Futyan}, {Garcia Mu{\~n}oz},
  {Gillon}, {Guedel}, {Guterman}, {Hasiba}, {Heng}, {Hernandez}, {Kiss},
  {Kuntzer}, {Laskar}, {Lecavelier des Etangs}, {Lovis}, {Magrin}, {Malvasio},
  {Marafatto}, {Michaelis}, {Munari}, {Olofsson}, {Ottacher}, {Ottensamer},
  {Pagano}, {Pall{\'e}}, {Peter}, {Piazza}, {Piotto}, {Pollacco}, {Ragazzoni},
  {Rando}, {Ratti}, {Rauer}, {Ribas}, {Rieder}, {Rohlfs}, {Safa}, {Salatti},
  {S{\'e}gransan}, {Simon}, {Smith}, {Sordet}, {Steller}, {Thomas},
  {Tschentscher}, {Van Eylen}, {Viotto}, {Walter}, {Walton}, {Wildi}, \&
  {Wolter}}]{bonfanti2021}
{Bonfanti}, A., {Delrez}, L., {Hooton}, M.~J., {et~al.} 2021, \aap, 646, A157

\bibitem[{{Bonfanti} \& {Gillon}(2020)}]{bonfanti2020}
{Bonfanti}, A. \& {Gillon}, M. 2020, \aap, 635, A6

\bibitem[{{Bonfanti} {et~al.}(2016){Bonfanti}, {Ortolani}, \&
  {Nascimbeni}}]{bonfanti2016}
{Bonfanti}, A., {Ortolani}, S., \& {Nascimbeni}, V. 2016, \aap, 585, A5

\bibitem[{{Bonfanti} {et~al.}(2015){Bonfanti}, {Ortolani}, {Piotto}, \&
  {Nascimbeni}}]{bonfanti2015}
{Bonfanti}, A., {Ortolani}, S., {Piotto}, G., \& {Nascimbeni}, V. 2015, \aap,
  575, A18

\bibitem[{{Bonfils} {et~al.}(2018){Bonfils}, {Almenara}, {Cloutier},
  {W{\"u}nsche}, {Astudillo-Defru}, {Berta-Thompson}, {Bouchy}, {Charbonneau},
  {Delfosse}, {D{\'\i}az}, {Dittmann}, {Doyon}, {Forveille}, {Irwin}, {Lovis},
  {Mayor}, {Menou}, {Murgas}, {Newton}, {Pepe}, {Santos}, \&
  {Udry}}]{bonfils2018}
{Bonfils}, X., {Almenara}, J.~M., {Cloutier}, R., {et~al.} 2018, \aap, 618,
  A142

\bibitem[{{Bonfils} {et~al.}(2005){Bonfils}, {Forveille}, {Delfosse}, {Udry},
  {Mayor}, {Perrier}, {Bouchy}, {Pepe}, {Queloz}, \& {Bertaux}}]{bonfils2005}
{Bonfils}, X., {Forveille}, T., {Delfosse}, X., {et~al.} 2005, \aap, 443, L15

\bibitem[{{Brandeker} {et~al.}(2022){Brandeker}, {Heng}, {Lendl}, {Patel},
  {Morris}, {Broeg}, {Guterman}, {Beck}, {Maxted}, {Demangeon}, {Delrez},
  {Demory}, {Kitzmann}, {Santos}, {Singh}, {Alibert}, {Alonso}, {Anglada},
  {B{\'a}rczy}, {Barrado y Navascues}, {Barros}, {Baumjohann}, {Beck}, {Benz},
  {Billot}, {Bonfils}, {Bruno}, {Cabrera}, {Charnoz}, {Collier Cameron},
  {Corral van Damme}, {Csizmadia}, {Davies}, {Deleuil}, {Deline}, {Ehrenreich},
  {Erikson}, {Farinato}, {Fortier}, {Fossati}, {Fridlund}, {Gandolfi},
  {Gillon}, {G{\"u}del}, {Hoyer}, {Isaak}, {Kiss}, {Laskar}, {Lecavelier des
  Etangs}, {Lovis}, {Luntzer}, {Magrin}, {Nascimbeni}, {Olofsson},
  {Ottensamer}, {Pagano}, {Pall{\'e}}, {Peter}, {Piotto}, {Pollacco}, {Queloz},
  {Ragazzoni}, {Rando}, {Rauer}, {Ribas}, {Scandariato}, {S{\'e}gransan},
  {Simon}, {Smith}, {Sousa}, {Steller}, {Szab{\'o}}, {Thomas}, {Udry}, {Van
  Grootel}, {Walton}, \& {Wolter}}]{brandeker2022}
{Brandeker}, A., {Heng}, K., {Lendl}, M., {et~al.} 2022, \aap, 659, L4

\bibitem[{{Brown} {et~al.}(2013){Brown}, {Baliber}, {Bianco}, {Bowman},
  {Burleson}, {Conway}, {Crellin}, {Depagne}, {De Vera}, {Dilday}, {Dragomir},
  {Dubberley}, {Eastman}, {Elphick}, {Falarski}, {Foale}, {Ford}, {Fulton},
  {Garza}, {Gomez}, {Graham}, {Greene}, {Haldeman}, {Hawkins}, {Haworth},
  {Haynes}, {Hidas}, {Hjelstrom}, {Howell}, {Hygelund}, {Lister}, {Lobdill},
  {Martinez}, {Mullins}, {Norbury}, {Parrent}, {Paulson}, {Petry}, {Pickles},
  {Posner}, {Rosing}, {Ross}, {Sand}, {Saunders}, {Shobbrook}, {Shporer},
  {Street}, {Thomas}, {Tsapras}, {Tufts}, {Valenti}, {Vander Horst}, {Walker},
  {White}, \& {Willis}}]{brown2013}
{Brown}, T.~M., {Baliber}, N., {Bianco}, F.~B., {et~al.} 2013, \pasp, 125, 1031

\bibitem[{{Burn} {et~al.}(2021){Burn}, {Schlecker}, {Mordasini}, {Emsenhuber},
  {Alibert}, {Henning}, {Klahr}, \& {Benz}}]{burn2021}
{Burn}, R., {Schlecker}, M., {Mordasini}, C., {et~al.} 2021, \aap, 656, A72

\bibitem[{{Burt} {et~al.}(2021){Burt}, {Dragomir}, {Molli{\`e}re},
  {Youngblood}, {Garc{\'\i}a Mu{\~n}oz}, {McCann}, {Kreidberg}, {Huang},
  {Collins}, {Eastman}, {Abe}, {Almenara}, {Crossfield}, {Ziegler},
  {Rodriguez}, {Mamajek}, {Stassun}, {Halverson}, {Villanueva}, {Butler},
  {Wang}, {Schwarz}, {Ricker}, {Vanderspek}, {Latham}, {Seager}, {Winn},
  {Jenkins}, {Agabi}, {Bonfils}, {Ciardi}, {Cointepas}, {Crane}, {Crouzet},
  {Dransfield}, {Feng}, {Furlan}, {Guillot}, {Gupta}, {Howell}, {Jensen},
  {Law}, {Mann}, {Marie-Sainte}, {Matson}, {Matthews}, {M{\'e}karnia},
  {Pepper}, {Scott}, {Shectman}, {Schlieder}, {Schmider}, {Stevens}, {Teske},
  {Triaud}, {Charbonneau}, {Berta-Thompson}, {Burke}, {Daylan}, {Barclay},
  {Wohler}, \& {Brasseur}}]{burt2021}
{Burt}, J.~A., {Dragomir}, D., {Molli{\`e}re}, P., {et~al.} 2021, \aj, 162, 87

\bibitem[{{Butler} {et~al.}(2004){Butler}, {Vogt}, {Marcy}, {Fischer},
  {Wright}, {Henry}, {Laughlin}, \& {Lissauer}}]{butler2004}
{Butler}, R.~P., {Vogt}, S.~S., {Marcy}, G.~W., {et~al.} 2004, \apj, 617, 580

\bibitem[{{Ca{\~n}as} {et~al.}(2023){Ca{\~n}as}, {Kanodia}, {Libby-Roberts},
  {Lin}, {Schutte}, {Powers}, {Jones}, {Monson}, {Wang}, {Stef{\'a}nsson},
  {Cochran}, {Robertson}, {Mahadevan}, {Kowalski}, {Wisniewski}, {Parker},
  {Larsen}, {Chapman}, {Kobulnicky}, {Gupta}, {Everett}, {Penprase}, {Zeimann},
  {Beard}, {Bender}, {Col{\'o}n}, {Diddams}, {Fredrick}, {Halverson}, {Ninan},
  {Ramsey}, {Roy}, \& {Schwab}}]{canas2023}
{Ca{\~n}as}, C.~I., {Kanodia}, S., {Libby-Roberts}, J., {et~al.} 2023, \aj,
  166, 30

\bibitem[{{Cale} {et~al.}(2019){Cale}, {Plavchan}, {LeBrun}, {Gagn{\'e}},
  {Gao}, {Tanner}, {Beichman}, {Xuesong Wang}, {Gaidos}, {Teske}, {Ciardi},
  {Vasisht}, {Kane}, \& {von Braun}}]{cale2019}
{Cale}, B., {Plavchan}, P., {LeBrun}, D., {et~al.} 2019, \aj, 158, 170

\bibitem[{{Casagrande} {et~al.}(2011){Casagrande}, {Sch{\"o}nrich}, {Asplund},
  {Cassisi}, {Ram{\'\i}rez}, {Mel{\'e}ndez}, {Bensby}, \&
  {Feltzing}}]{Casagrande2011}
{Casagrande}, L., {Sch{\"o}nrich}, R., {Asplund}, M., {et~al.} 2011, \aap, 530,
  A138

\bibitem[{{Castelli} \& {Kurucz}(2003)}]{Castelli2003}
{Castelli}, F. \& {Kurucz}, R.~L. 2003, in IAU Symposium, Vol. 210, Modelling
  of Stellar Atmospheres, ed. N.~{Piskunov}, W.~W. {Weiss}, \& D.~F. {Gray},
  A20

\bibitem[{{Castro-Gonz{\'a}lez} {et~al.}(2023){Castro-Gonz{\'a}lez},
  {Demangeon}, {Lillo-Box}, {Lovis}, {Lavie}, {Adibekyan}, {Acu{\~n}a},
  {Deleuil}, {Aguichine}, {Zapatero Osorio}, {Tabernero}, {Davoult}, {Alibert},
  {Santos}, {Sousa}, {Antoniadis-Karnavas}, {Borsa}, {Winn}, {Allende Prieto},
  {Figueira}, {Jenkins}, {Sozzetti}, {Damasso}, {Silva}, {Astudillo-Defru},
  {Barros}, {Bonfils}, {Cristiani}, {Di Marcantonio}, {Gonz{\'a}lez
  Hern{\'a}ndez}, {Curto}, {Martins}, {Nunes}, {Palle}, {Pepe}, {Seager}, \&
  {Su{\'a}rez Mascare{\~n}o}}]{castroGonzalez2023}
{Castro-Gonz{\'a}lez}, A., {Demangeon}, O.~D.~S., {Lillo-Box}, J., {et~al.}
  2023, \aap, 675, A52

\bibitem[{{Charbonneau} {et~al.}(2008){Charbonneau}, {Irwin}, {Nutzman}, \&
  {Falco}}]{charbonneau2008}
{Charbonneau}, D., {Irwin}, J., {Nutzman}, P., \& {Falco}, E.~E. 2008, in
  American Astronomical Society Meeting Abstracts, Vol. 212, American
  Astronomical Society Meeting Abstracts \#212, 44.02

\bibitem[{{Chaturvedi} {et~al.}(2022){Chaturvedi}, {Bluhm}, {Nagel}, {Hatzes},
  {Morello}, {Brady}, {Korth}, {Molaverdikhani}, {Kossakowski}, {Caballero},
  {Guenther}, {Pall{\'e}}, {Espinoza}, {Seifahrt}, {Lodieu}, {Cifuentes},
  {Furlan}, {Amado}, {Barclay}, {Bean}, {B{\'e}jar}, {Bergond}, {Boyle},
  {Ciardi}, {Collins}, {Collins}, {Esparza-Borges}, {Fukui}, {Gnilka}, {Goeke},
  {Guerra}, {Henning}, {Herrero}, {Howell}, {Jeffers}, {Jenkins}, {Jensen},
  {Kasper}, {Kodama}, {Latham}, {L{\'o}pez-Gonz{\'a}lez}, {Luque}, {Montes},
  {Morales}, {Mori}, {Murgas}, {Narita}, {Nowak}, {Parviainen}, {Passegger},
  {Quirrenbach}, {Reffert}, {Reiners}, {Ribas}, {Ricker}, {Rodriguez},
  {Rodr{\'\i}guez-L{\'o}pez}, {Schlecker}, {Schwarz}, {Schweitzer}, {Seager},
  {Stef{\'a}nsson}, {Stockdale}, {Tal-Or}, {Twicken}, {Vanaverbeke}, {Wang},
  {Watanabe}, {Winn}, \& {Zechmeister}}]{chaturvedi2022}
{Chaturvedi}, P., {Bluhm}, P., {Nagel}, E., {et~al.} 2022, \aap, 666, A155

\bibitem[{{Chen} \& {Rogers}(2016)}]{chen2016}
{Chen}, H. \& {Rogers}, L.~A. 2016, \apj, 831, 180

\bibitem[{{Cherubim} {et~al.}(2023){Cherubim}, {Cloutier}, {Charbonneau},
  {Stockdale}, {Stassun}, {Schwarz}, {Safonov}, {Mortier}, {Lewin}, {Latham},
  {Horne}, {Haywood}, {Gonzales}, {Goliguzova}, {Collins}, {Ciardi}, {Bieryla},
  {Belinski}, {Wohler}, {Watson}, {Vanderspek}, {Udry}, {Sozzetti},
  {S{\'e}gransan}, {Sasselov}, {Ricker}, {Rice}, {Poretti}, {Piotto}, {Pepe},
  {Molinari}, {Micela}, {Mayor}, {Lovis}, {L{\'o}pez-Morales}, {Jenkins},
  {Essack}, {Dumusque}, {Doty}, {Col{\'o}n}, {Cameron}, \&
  {Buchhave}}]{cherubim2023}
{Cherubim}, C., {Cloutier}, R., {Charbonneau}, D., {et~al.} 2023, \aj, 165, 167

\bibitem[{{Childs} {et~al.}(2022){Childs}, {Martin}, \& {Livio}}]{childs2022}
{Childs}, A.~C., {Martin}, R.~G., \& {Livio}, M. 2022, \apjl, 937, L41

\bibitem[{{Ciardi} {et~al.}(2013){Ciardi}, {Fabrycky}, {Ford}, {Gautier},
  {Howell}, {Lissauer}, {Ragozzine}, \& {Rowe}}]{ciardi2013}
{Ciardi}, D.~R., {Fabrycky}, D.~C., {Ford}, E.~B., {et~al.} 2013, \apj, 763, 41

\bibitem[{{Cloutier} {et~al.}(2019){Cloutier}, {Astudillo-Defru}, {Bonfils},
  {Jenkins}, {Berdi{\~n}as}, {Ricker}, {Vanderspek}, {Latham}, {Seager},
  {Winn}, {Jenkins}, {Almenara}, {Bouchy}, {Delfosse}, {D{\'\i}az},
  {D{\'\i}az}, {Doyon}, {Figueira}, {Forveille}, {Kurtovic}, {Lovis}, {Mayor},
  {Menou}, {Morgan}, {Morris}, {Muirhead}, {Murgas}, {Pepe}, {Santos},
  {S{\'e}gransan}, {Smith}, {Tenenbaum}, {Torres}, {Udry}, {Vezie}, \&
  {Villasenor}}]{cloutier2019}
{Cloutier}, R., {Astudillo-Defru}, N., {Bonfils}, X., {et~al.} 2019, \aap, 629,
  A111

\bibitem[{{Cloutier} {et~al.}(2021{\natexlab{a}}){Cloutier}, {Charbonneau},
  {Deming}, {Bonfils}, \& {Astudillo-Defru}}]{cloutier2021GJ1214}
{Cloutier}, R., {Charbonneau}, D., {Deming}, D., {Bonfils}, X., \&
  {Astudillo-Defru}, N. 2021{\natexlab{a}}, \aj, 162, 174

\bibitem[{{Cloutier} {et~al.}(2021{\natexlab{b}}){Cloutier}, {Charbonneau},
  {Stassun}, {Murgas}, {Mortier}, {Massey}, {Lissauer}, {Latham}, {Irwin},
  {Haywood}, {Guerra}, {Girardin}, {Giacalone}, {Bosch-Cabot}, {Bieryla},
  {Winn}, {Watson}, {Vanderspek}, {Udry}, {Tamura}, {Sozzetti}, {Shporer},
  {S{\'e}gransan}, {Seager}, {Savel}, {Sasselov}, {Rose}, {Ricker}, {Rice},
  {Quintana}, {Quinn}, {Piotto}, {Phillips}, {Pepe}, {Pedani}, {Parviainen},
  {Palle}, {Narita}, {Molinari}, {Micela}, {McDermott}, {Mayor}, {Matson},
  {Martinez Fiorenzano}, {Lovis}, {L{\'o}pez-Morales}, {Kusakabe}, {Jensen},
  {Jenkins}, {Huang}, {Howell}, {Harutyunyan}, {F{\H{u}}r{\'e}sz}, {Fukui},
  {Esquerdo}, {Esparza-Borges}, {Dumusque}, {Dressing}, {Fabrizio}, {Collins},
  {Cameron}, {Christiansen}, {Cecconi}, {Buchhave}, {Boschin}, \&
  {Andreuzzi}}]{cloutier2021}
{Cloutier}, R., {Charbonneau}, D., {Stassun}, K.~G., {et~al.}
  2021{\natexlab{b}}, \aj, 162, 79

\bibitem[{{Cloutier} {et~al.}(2020){Cloutier}, {Eastman}, {Rodriguez},
  {Astudillo-Defru}, {Bonfils}, {Mortier}, {Watson}, {Stalport}, {Pinamonti},
  {Lienhard}, {Harutyunyan}, {Damasso}, {Latham}, {Collins}, {Massey}, {Irwin},
  {Winters}, {Charbonneau}, {Ziegler}, {Matthews}, {Crossfield}, {Kreidberg},
  {Quinn}, {Ricker}, {Vanderspek}, {Seager}, {Winn}, {Jenkins}, {Vezie},
  {Udry}, {Twicken}, {Tenenbaum}, {Sozzetti}, {S{\'e}gransan}, {Schlieder},
  {Sasselov}, {Santos}, {Rice}, {Rackham}, {Poretti}, {Piotto}, {Phillips},
  {Pepe}, {Molinari}, {Mignon}, {Micela}, {Melo}, {de Medeiros}, {Mayor},
  {Matson}, {Martinez Fiorenzano}, {Mann}, {Magazz{\'u}}, {Lovis},
  {L{\'o}pez-Morales}, {Lopez}, {Lissauer}, {L{\'e}pine}, {Law}, {Kielkopf},
  {Johnson}, {Jensen}, {Howell}, {Gonzales}, {Ghedina}, {Forveille},
  {Figueira}, {Dumusque}, {Dressing}, {Doyon}, {D{\'\i}az}, {Fabrizio},
  {Delfosse}, {Cosentino}, {Conti}, {Collins}, {Cameron}, {Ciardi}, {Caldwell},
  {Burke}, {Buchhave}, {Brice{\~n}o}, {Boyd}, {Bouchy}, {Beichman}, {Artigau},
  \& {Almenara}}]{cloutier2020}
{Cloutier}, R., {Eastman}, J.~D., {Rodriguez}, J.~E., {et~al.} 2020, \aj, 160,
  3

\bibitem[{{Cloutier} \& {Menou}(2020)}]{cloutier2020valley}
{Cloutier}, R. \& {Menou}, K. 2020, \aj, 159, 211

\bibitem[{{Cointepas} {et~al.}(2021){Cointepas}, {Almenara}, {Bonfils},
  {Bouchy}, {Astudillo-Defru}, {Murgas}, {Otegi}, {Wyttenbach}, {Anderson},
  {Artigau}, {Canto Martins}, {Charbonneau}, {Collins}, {Collins}, {Correia},
  {Curaba}, {Delboulb{\'e}}, {Delfosse}, {D{\'\i}az}, {Dorn}, {Doyon},
  {Feautrier}, {Figueira}, {Forveille}, {Gaisne}, {Gan}, {Gluck}, {Helled},
  {Hellier}, {Jocou}, {Kern}, {Lafrasse}, {Law}, {Le{\~a}o}, {Lovis},
  {Magnard}, {Mann}, {Maurel}, {de Medeiros}, {Melo}, {Moulin}, {Pepe},
  {Rabou}, {Rochat}, {Rodriguez}, {Roux}, {Santos}, {S{\'e}gransan}, {Stadler},
  {Ting}, {Twicken}, {Udry}, {Waalkes}, {West}, {W{\"u}nsche}, {Ziegler},
  {Ricker}, {Vanderspek}, {Latham}, {Seager}, {Winn}, \&
  {Jenkins}}]{cointepas2021}
{Cointepas}, M., {Almenara}, J.~M., {Bonfils}, X., {et~al.} 2021, \aap, 650,
  A145

\bibitem[{{Cortes} \& {Vapnik}(1995)}]{cortes1995}
{Cortes}, C. \& {Vapnik}, V. 1995, Machine Learning, 20, 273

\bibitem[{{Cosentino} {et~al.}(2000){Cosentino}, {Bonanno}, {Bruno}, {Scuderi},
  {Bonoli}, {Bortoletto}, {D'Alessandro}, \& {Fantinel}}]{cosentino2000}
{Cosentino}, R., {Bonanno}, G., {Bruno}, P., {et~al.} 2000, \memsai, 71, 393

\bibitem[{{Cosentino} {et~al.}(2012){Cosentino}, {Lovis}, {Pepe}, {Collier
  Cameron}, {Latham}, {Molinari}, {Udry}, {Bezawada}, {Black}, {Born},
  {Buchschacher}, {Charbonneau}, {Figueira}, {Fleury}, {Galli}, {Gallie},
  {Gao}, {Ghedina}, {Gonzalez}, {Gonzalez}, {Guerra}, {Henry}, {Horne},
  {Hughes}, {Kelly}, {Lodi}, {Lunney}, {Maire}, {Mayor}, {Micela}, {Ordway},
  {Peacock}, {Phillips}, {Piotto}, {Pollacco}, {Queloz}, {Rice}, {Riverol},
  {Riverol}, {San Juan}, {Sasselov}, {Segransan}, {Sozzetti}, {Sosnowska},
  {Stobie}, {Szentgyorgyi}, {Vick}, \& {Weber}}]{cosentino2012}
{Cosentino}, R., {Lovis}, C., {Pepe}, F., {et~al.} 2012, in Society of
  Photo-Optical Instrumentation Engineers (SPIE) Conference Series, Vol. 8446,
  Ground-based and Airborne Instrumentation for Astronomy IV, ed. I.~S.
  {McLean}, S.~K. {Ramsay}, \& H.~{Takami}, 84461V

\bibitem[{{Crossfield} {et~al.}(2022){Crossfield}, {Malik}, {Hill}, {Kane},
  {Foley}, {Polanski}, {Coria}, {Brande}, {Zhang}, {Wienke}, {Kreidberg},
  {Cowan}, {Dragomir}, {Gorjian}, {Mikal-Evans}, {Benneke}, {Christiansen},
  {Deming}, \& {Morales}}]{crossfield2022}
{Crossfield}, I. J.~M., {Malik}, M., {Hill}, M.~L., {et~al.} 2022, \apjl, 937,
  L17

\bibitem[{{Cuntz} \& {Wang}(2018)}]{cuntz2018}
{Cuntz}, M. \& {Wang}, Z. 2018, Research Notes of the American Astronomical
  Society, 2, 19

\bibitem[{{Cutri} {et~al.}(2003){Cutri}, {Skrutskie}, {van Dyk}, {Beichman},
  {Carpenter}, {Chester}, {Cambresy}, {Evans}, {Fowler}, {Gizis}, {Howard},
  {Huchra}, {Jarrett}, {Kopan}, {Kirkpatrick}, {Light}, {Marsh}, {McCallon},
  {Schneider}, {Stiening}, {Sykes}, {Weinberg}, {Wheaton}, {Wheelock}, \&
  {Zacarias}}]{cutri2003}
{Cutri}, R.~M., {Skrutskie}, M.~F., {van Dyk}, S., {et~al.} 2003, VizieR Online
  Data Catalog, II/246

\bibitem[{{Delfosse} {et~al.}(1998){Delfosse}, {Forveille}, {Mayor}, {Perrier},
  {Naef}, \& {Queloz}}]{delfosse1998}
{Delfosse}, X., {Forveille}, T., {Mayor}, M., {et~al.} 1998, \aap, 338, L67

\bibitem[{{Demangeon} {et~al.}(2021{\natexlab{a}}){Demangeon}, {Zapatero
  Osorio}, {Alibert}, {Barros}, {Adibekyan}, {Tabernero},
  {Antoniadis-Karnavas}, {Camacho}, {Su{\'a}rez Mascare{\~n}o}, {Oshagh},
  {Micela}, {Sousa}, {Lovis}, {Pepe}, {Rebolo}, {Cristiani}, {Santos},
  {Allart}, {Allende Prieto}, {Bossini}, {Bouchy}, {Cabral}, {Damasso}, {Di
  Marcantonio}, {D'Odorico}, {Ehrenreich}, {Faria}, {Figueira}, {G{\'e}nova
  Santos}, {Haldemann}, {Hara}, {Gonz{\'a}lez Hern{\'a}ndez}, {Lavie},
  {Lillo-Box}, {Lo Curto}, {Martins}, {M{\'e}gevand}, {Mehner}, {Molaro},
  {Nunes}, {Pall{\'e}}, {Pasquini}, {Poretti}, {Sozzetti}, \&
  {Udry}}]{Demangeon-21}
{Demangeon}, O.~D.~S., {Zapatero Osorio}, M.~R., {Alibert}, Y., {et~al.}
  2021{\natexlab{a}}, \aap, 653, A41

\bibitem[{{Demangeon} {et~al.}(2021{\natexlab{b}}){Demangeon}, {Zapatero
  Osorio}, {Alibert}, {Barros}, {Adibekyan}, {Tabernero},
  {Antoniadis-Karnavas}, {Camacho}, {Su{\'a}rez Mascare{\~n}o}, {Oshagh},
  {Micela}, {Sousa}, {Lovis}, {Pepe}, {Rebolo}, {Cristiani}, {Santos},
  {Allart}, {Allende Prieto}, {Bossini}, {Bouchy}, {Cabral}, {Damasso}, {Di
  Marcantonio}, {D'Odorico}, {Ehrenreich}, {Faria}, {Figueira}, {G{\'e}nova
  Santos}, {Haldemann}, {Hara}, {Gonz{\'a}lez Hern{\'a}ndez}, {Lavie},
  {Lillo-Box}, {Lo Curto}, {Martins}, {M{\'e}gevand}, {Mehner}, {Molaro},
  {Nunes}, {Pall{\'e}}, {Pasquini}, {Poretti}, {Sozzetti}, \&
  {Udry}}]{demangeon2021}
{Demangeon}, O.~D.~S., {Zapatero Osorio}, M.~R., {Alibert}, Y., {et~al.}
  2021{\natexlab{b}}, \aap, 653, A41

\bibitem[{{Diamond-Lowe} {et~al.}(2022){Diamond-Lowe}, {Kreidberg}, {Harman},
  {Kempton}, {Rogers}, {Joyce}, {Eastman}, {King}, {Kopparapu}, {Youngblood},
  {Kosiarek}, {Livingston}, {Hardegree-Ullman}, \&
  {Crossfield}}]{diamondLowe2022}
{Diamond-Lowe}, H., {Kreidberg}, L., {Harman}, C.~E., {et~al.} 2022, \aj, 164,
  172

\bibitem[{{Dorn} {et~al.}(2017){Dorn}, {Venturini}, {Khan}, {Heng}, {Alibert},
  {Helled}, {Rivoldini}, \& {Benz}}]{Dorn2017}
{Dorn}, C., {Venturini}, J., {Khan}, A., {et~al.} 2017, \aap, 597, A37

\bibitem[{{Dransfield} {et~al.}(2023){Dransfield}, {Timmermans}, {Triaud},
  {D{\'e}vora-Pajares}, {Aganze}, {Barkaoui}, {Burgasser}, {Collins},
  {Cointepas}, {Ducrot}, {G{\"u}nther}, {Howell}, {Murray}, {Niraula},
  {Rackham}, {Sebastian}, {Stassun}, {Z{\'u}{\~n}iga-Fern{\'a}ndez},
  {Almenara}, {Bonfils}, {Bouchy}, {Burke}, {Charbonneau}, {Christiansen},
  {Delrez}, {Gan}, {Garc{\'\i}a}, {Gillon}, {Chew}, {Hesse}, {Hooton}, {Isopi},
  {Jehin}, {Jenkins}, {Latham}, {Mallia}, {Murgas}, {Pedersen}, {Pozuelos},
  {Queloz}, {Rodriguez}, {Schanche}, {Seager}, {Srdoc}, {Stockdale}, {Twicken},
  {Vanderspek}, {Wells}, {Winn}, {de Wit}, \& {Zapparata}}]{dransfield2023}
{Dransfield}, G., {Timmermans}, M., {Triaud}, A. H.~M.~J., {et~al.} 2023,
  \mnras [\eprint[arXiv]{2305.06206}]

\bibitem[{{Eggleton}(1983)}]{eggleton1983}
{Eggleton}, P.~P. 1983, \apj, 268, 368

\bibitem[{{Emsenhuber} {et~al.}(2021){Emsenhuber}, {Mordasini}, {Burn},
  {Alibert}, {Benz}, \& {Asphaug}}]{Emsenhuber21}
{Emsenhuber}, A., {Mordasini}, C., {Burn}, R., {et~al.} 2021, \aap, 656, A70

\bibitem[{{Erkaev} {et~al.}(2007){Erkaev}, {Kulikov}, {Lammer}, {Selsis},
  {Langmayr}, {Jaritz}, \& {Biernat}}]{erkaev2007}
{Erkaev}, N.~V., {Kulikov}, Y.~N., {Lammer}, H., {et~al.} 2007, \aap, 472, 329

\bibitem[{{Espinoza} \& {Jord{\'a}n}(2015)}]{espinoza2015}
{Espinoza}, N. \& {Jord{\'a}n}, A. 2015, \mnras, 450, 1879

\bibitem[{{Essack} {et~al.}(2023){Essack}, {Shporer}, {Burt}, {Seager},
  {Cambioni}, {Lin}, {Collins}, {Mamajek}, {Stassun}, {Ricker}, {Vanderspek},
  {Latham}, {Winn}, {Jenkins}, {Butler}, {Charbonneau}, {Collins}, {Crane},
  {Gan}, {Hellier}, {Howell}, {Irwin}, {Mann}, {Ramadhan}, {Shectman}, {Teske},
  {Yee}, {Mireles}, {Quintana}, {Tenenbaum}, {Torres}, \&
  {Furlan}}]{essack2023}
{Essack}, Z., {Shporer}, A., {Burt}, J.~A., {et~al.} 2023, \aj, 165, 47

\bibitem[{{Foreman-Mackey} {et~al.}(2017){Foreman-Mackey}, {Agol},
  {Ambikasaran}, \& {Angus}}]{foreman2017}
{Foreman-Mackey}, D., {Agol}, E., {Ambikasaran}, S., \& {Angus}, R. 2017, \aj,
  154, 220

\bibitem[{{Fridlund} {et~al.}(2023){Fridlund}, {Georgieva}, {Bonfanti}, \&
  {et~al.}}]{fridlund2023}
{Fridlund}, M., {Georgieva}, I.~Y., {Bonfanti}, A., \& {et~al.} 2023, accepted
  for publication in A\&A

\bibitem[{{Fruhwirth-Schnatter} {et~al.}(2018){Fruhwirth-Schnatter}, {Celeux},
  \& {Robert}}]{fruewirth2018}
{Fruhwirth-Schnatter}, S., {Celeux}, G., \& {Robert}, C.~P. 2018, Handbook of
  Mixture Analysis, 1st edn. (Chapman and Hall/CRC)

\bibitem[{{Fulton} \& {Petigura}(2018)}]{fulton2018}
{Fulton}, B.~J. \& {Petigura}, E.~A. 2018, \aj, 156, 264

\bibitem[{{Fulton} {et~al.}(2017){Fulton}, {Petigura}, {Howard}, {Isaacson},
  {Marcy}, {Cargile}, {Hebb}, {Weiss}, {Johnson}, {Morton}, {Sinukoff},
  {Crossfield}, \& {Hirsch}}]{fulton2017}
{Fulton}, B.~J., {Petigura}, E.~A., {Howard}, A.~W., {et~al.} 2017, \aj, 154,
  109

\bibitem[{{Gaia Collaboration} {et~al.}(2023){Gaia Collaboration}, {Vallenari},
  {Brown}, {Prusti}, {de Bruijne}, {Arenou}, {Babusiaux}, {Biermann},
  {Creevey}, {Ducourant}, {Evans}, {Eyer}, {Guerra}, {Hutton}, {Jordi},
  {Klioner}, {Lammers}, {Lindegren}, {Luri}, {Mignard}, {Panem}, {Pourbaix},
  {Randich}, {Sartoretti}, {Soubiran}, {Tanga}, {Walton}, {Bailer-Jones},
  {Bastian}, {Drimmel}, {Jansen}, {Katz}, {Lattanzi}, {van Leeuwen}, {Bakker},
  {Cacciari}, {Casta{\~n}eda}, {De Angeli}, {Fabricius}, {Fouesneau},
  {Fr{\'e}mat}, {Galluccio}, {Guerrier}, {Heiter}, {Masana}, {Messineo},
  {Mowlavi}, {Nicolas}, {Nienartowicz}, {Pailler}, {Panuzzo}, {Riclet}, {Roux},
  {Seabroke}, {Sordo}, {Th{\'e}venin}, {Gracia-Abril}, {Portell}, {Teyssier},
  {Altmann}, {Andrae}, {Audard}, {Bellas-Velidis}, {Benson}, {Berthier},
  {Blomme}, {Burgess}, {Busonero}, {Busso}, {C{\'a}novas}, {Carry}, {Cellino},
  {Cheek}, {Clementini}, {Damerdji}, {Davidson}, {de Teodoro}, {Nu{\~n}ez
  Campos}, {Delchambre}, {Dell'Oro}, {Esquej}, {Fern{\'a}ndez-Hern{\'a}ndez},
  {Fraile}, {Garabato}, {Garc{\'\i}a-Lario}, {Gosset}, {Haigron}, {Halbwachs},
  {Hambly}, {Harrison}, {Hern{\'a}ndez}, {Hestroffer}, {Hodgkin}, {Holl},
  {Jan{\ss}en}, {Jevardat de Fombelle}, {Jordan}, {Krone-Martins}, {Lanzafame},
  {L{\"o}ffler}, {Marchal}, {Marrese}, {Moitinho}, {Muinonen}, {Osborne},
  {Pancino}, {Pauwels}, {Recio-Blanco}, {Reyl{\'e}}, {Riello}, {Rimoldini},
  {Roegiers}, {Rybizki}, {Sarro}, {Siopis}, {Smith}, {Sozzetti}, {Utrilla},
  {van Leeuwen}, {Abbas}, {{\'A}brah{\'a}m}, {Abreu Aramburu}, {Aerts},
  {Aguado}, {Ajaj}, {Aldea-Montero}, {Altavilla}, {{\'A}lvarez}, {Alves},
  {Anders}, {Anderson}, {Anglada Varela}, {Antoja}, {Baines}, {Baker},
  {Balaguer-N{\'u}{\~n}ez}, {Balbinot}, {Balog}, {Barache}, {Barbato},
  {Barros}, {Barstow}, {Bartolom{\'e}}, {Bassilana}, {Bauchet}, {Becciani},
  {Bellazzini}, {Berihuete}, {Bernet}, {Bertone}, {Bianchi}, {Binnenfeld},
  {Blanco-Cuaresma}, {Blazere}, {Boch}, {Bombrun}, {Bossini}, {Bouquillon},
  {Bragaglia}, {Bramante}, {Breedt}, {Bressan}, {Brouillet}, {Brugaletta},
  {Bucciarelli}, {Burlacu}, {Butkevich}, {Buzzi}, {Caffau}, {Cancelliere},
  {Cantat-Gaudin}, {Carballo}, {Carlucci}, {Carnerero}, {Carrasco},
  {Casamiquela}, {Castellani}, {Castro-Ginard}, {Chaoul}, {Charlot}, {Chemin},
  {Chiaramida}, {Chiavassa}, {Chornay}, {Comoretto}, {Contursi}, {Cooper},
  {Cornez}, {Cowell}, {Crifo}, {Cropper}, {Crosta}, {Crowley}, {Dafonte},
  {Dapergolas}, {David}, {David}, {de Laverny}, {De Luise}, {De March}, {De
  Ridder}, {de Souza}, {de Torres}, {del Peloso}, {del Pozo}, {Delbo},
  {Delgado}, {Delisle}, {Demouchy}, {Dharmawardena}, {Di Matteo}, {Diakite},
  {Diener}, {Distefano}, {Dolding}, {Edvardsson}, {Enke}, {Fabre}, {Fabrizio},
  {Faigler}, {Fedorets}, {Fernique}, {Fienga}, {Figueras}, {Fournier},
  {Fouron}, {Fragkoudi}, {Gai}, {Garcia-Gutierrez}, {Garcia-Reinaldos},
  {Garc{\'\i}a-Torres}, {Garofalo}, {Gavel}, {Gavras}, {Gerlach}, {Geyer},
  {Giacobbe}, {Gilmore}, {Girona}, {Giuffrida}, {Gomel}, {Gomez},
  {Gonz{\'a}lez-N{\'u}{\~n}ez}, {Gonz{\'a}lez-Santamar{\'\i}a},
  {Gonz{\'a}lez-Vidal}, {Granvik}, {Guillout}, {Guiraud},
  {Guti{\'e}rrez-S{\'a}nchez}, {Guy}, {Hatzidimitriou}, {Hauser}, {Haywood},
  {Helmer}, {Helmi}, {Sarmiento}, {Hidalgo}, {Hilger}, {H{\l}adczuk}, {Hobbs},
  {Holland}, {Huckle}, {Jardine}, {Jasniewicz}, {Jean-Antoine Piccolo},
  {Jim{\'e}nez-Arranz}, {Jorissen}, {Juaristi Campillo}, {Julbe}, {Karbevska},
  {Kervella}, {Khanna}, {Kontizas}, {Kordopatis}, {Korn}, {K{\'o}sp{\'a}l},
  {Kostrzewa-Rutkowska}, {Kruszy{\'n}ska}, {Kun}, {Laizeau}, {Lambert},
  {Lanza}, {Lasne}, {Le Campion}, {Lebreton}, {Lebzelter}, {Leccia}, {Leclerc},
  {Lecoeur-Taibi}, {Liao}, {Licata}, {Lindstr{\o}m}, {Lister}, {Livanou},
  {Lobel}, {Lorca}, {Loup}, {Madrero Pardo}, {Magdaleno Romeo}, {Managau},
  {Mann}, {Manteiga}, {Marchant}, {Marconi}, {Marcos}, {Marcos Santos},
  {Mar{\'\i}n Pina}, {Marinoni}, {Marocco}, {Marshall}, {Martin Polo},
  {Mart{\'\i}n-Fleitas}, {Marton}, {Mary}, {Masip}, {Massari},
  {Mastrobuono-Battisti}, {Mazeh}, {McMillan}, {Messina}, {Michalik}, {Millar},
  {Mints}, {Molina}, {Molinaro}, {Moln{\'a}r}, {Monari}, {Mongui{\'o}},
  {Montegriffo}, {Montero}, {Mor}, {Mora}, {Morbidelli}, {Morel}, {Morris},
  {Muraveva}, {Murphy}, {Musella}, {Nagy}, {Noval}, {Oca{\~n}a}, {Ogden},
  {Ordenovic}, {Osinde}, {Pagani}, {Pagano}, {Palaversa}, {Palicio},
  {Pallas-Quintela}, {Panahi}, {Payne-Wardenaar}, {Pe{\~n}alosa Esteller},
  {Penttil{\"a}}, {Pichon}, {Piersimoni}, {Pineau}, {Plachy}, {Plum}, {Poggio},
  {Pr{\v{s}}a}, {Pulone}, {Racero}, {Ragaini}, {Rainer}, {Raiteri}, {Rambaux},
  {Ramos}, {Ramos-Lerate}, {Re Fiorentin}, {Regibo}, {Richards}, {Rios Diaz},
  {Ripepi}, {Riva}, {Rix}, {Rixon}, {Robichon}, {Robin}, {Robin}, {Roelens},
  {Rogues}, {Rohrbasser}, {Romero-G{\'o}mez}, {Rowell}, {Royer}, {Ruz Mieres},
  {Rybicki}, {Sadowski}, {S{\'a}ez N{\'u}{\~n}ez}, {Sagrist{\`a} Sell{\'e}s},
  {Sahlmann}, {Salguero}, {Samaras}, {Sanchez Gimenez}, {Sanna},
  {Santove{\~n}a}, {Sarasso}, {Schultheis}, {Sciacca}, {Segol}, {Segovia},
  {S{\'e}gransan}, {Semeux}, {Shahaf}, {Siddiqui}, {Siebert}, {Siltala},
  {Silvelo}, {Slezak}, {Slezak}, {Smart}, {Snaith}, {Solano}, {Solitro},
  {Souami}, {Souchay}, {Spagna}, {Spina}, {Spoto}, {Steele},
  {Steidelm{\"u}ller}, {Stephenson}, {S{\"u}veges}, {Surdej}, {Szabados},
  {Szegedi-Elek}, {Taris}, {Taylor}, {Teixeira}, {Tolomei}, {Tonello}, {Torra},
  {Torra}, {Torralba Elipe}, {Trabucchi}, {Tsounis}, {Turon}, {Ulla}, {Unger},
  {Vaillant}, {van Dillen}, {van Reeven}, {Vanel}, {Vecchiato}, {Viala},
  {Vicente}, {Voutsinas}, {Weiler}, {Wevers}, {Wyrzykowski}, {Yoldas}, {Yvard},
  {Zhao}, {Zorec}, {Zucker}, \& {Zwitter}}]{GaiaDR3}
{Gaia Collaboration}, {Vallenari}, A., {Brown}, A.~G.~A., {et~al.} 2023, \aap,
  674, A1

\bibitem[{{Gelman} \& {Rubin}(1992)}]{gelman1992}
{Gelman}, A. \& {Rubin}, D.~B. 1992, Statistical Science, 7, 457

\bibitem[{{Ghachoui} {et~al.}(2023){Ghachoui}, {Soubkiou}, {Wells}, {Rackham},
  {Triaud}, {Sebastian}, {Giacalone}, {Stassun}, {Ciardi}, {Collins}, {Liu},
  {G{\'o}mez Maqueo Chew}, {Gillon}, {Benkhaldoun}, {Delrez}, {Eastman},
  {Demangeon}, {Barkaoui}, {Burdanov}, {Demory}, {de Wit}, {Dransfield},
  {Ducrot}, {Garcia}, {G{\'o}mez-Mu{\~n}oz}, {Hooton}, {Jehin}, {Murray},
  {Pedersen}, {Pozuelos}, {Queloz}, {Sabin}, {Schanche}, {Timmermans},
  {Gonzales}, {Dressing}, {Aganze}, {Burgasser}, {Gerasimov}, {Hsu},
  {Theissen}, {Charbonneau}, {Jenkins}, {Latham}, {Ricker}, {Seager},
  {Shporer}, {Twicken}, {Vanderspek}, {Winn}, {Collins}, {Fukui}, {Gan},
  {Narita}, \& {Schwarz}}]{ghachoui2023}
{Ghachoui}, M., {Soubkiou}, A., {Wells}, R.~D., {et~al.} 2023, \aap, 677, A31

\bibitem[{{Gillon} {et~al.}(2007){Gillon}, {Pont}, {Demory}, {Mallmann},
  {Mayor}, {Mazeh}, {Queloz}, {Shporer}, {Udry}, \& {Vuissoz}}]{gillon2007}
{Gillon}, M., {Pont}, F., {Demory}, B.~O., {et~al.} 2007, \aap, 472, L13

\bibitem[{{Ginzburg} {et~al.}(2018){Ginzburg}, {Schlichting}, \&
  {Sari}}]{ginzburg2018}
{Ginzburg}, S., {Schlichting}, H.~E., \& {Sari}, R. 2018, \mnras, 476, 759

\bibitem[{{Gupta} \& {Schlichting}(2019)}]{gupta2019}
{Gupta}, A. \& {Schlichting}, H.~E. 2019, \mnras, 487, 24

\bibitem[{{Gupta} \& {Schlichting}(2020)}]{gupta2020}
{Gupta}, A. \& {Schlichting}, H.~E. 2020, \mnras, 493, 792

\bibitem[{{Hakim} {et~al.}(2018){Hakim}, {Rivoldini}, {Van Hoolst},
  {Cottenier}, {Jaeken}, {Chust}, \& {Steinle-Neumann}}]{Hakim2018}
{Hakim}, K., {Rivoldini}, A., {Van Hoolst}, T., {et~al.} 2018, \icarus, 313, 61

\bibitem[{{Haldemann} {et~al.}(2020){Haldemann}, {Alibert}, {Mordasini}, \&
  {Benz}}]{Haldemann2020}
{Haldemann}, J., {Alibert}, Y., {Mordasini}, C., \& {Benz}, W. 2020, \aap, 643,
  A105

\bibitem[{{Haldemann} {et~al.}(2023){Haldemann}, {Dorn}, {Venturini}, \&
  {et~al.}}]{haldemann2023}
{Haldemann}, J., {Dorn}, C., {Venturini}, J., \& {et~al.} 2023, A\&A in press

\bibitem[{{Hatzes}(2016)}]{hatzes2016}
{Hatzes}, A.~P. 2016, in Astrophysics and Space Science Library, Vol. 428,
  Methods of Detecting Exoplanets: 1st Advanced School on Exoplanetary Science,
  ed. V.~{Bozza}, L.~{Mancini}, \& A.~{Sozzetti}, 3

\bibitem[{{Hirano} {et~al.}(2020){Hirano}, {Gaidos}, {Winn}, {Dai}, {Fukui},
  {Kuzuhara}, {Kotani}, {Tamura}, {Hjorth}, {Albrecht}, {Huber}, {Bolmont},
  {Harakawa}, {Hodapp}, {Ishizuka}, {Jacobson}, {Konishi}, {Kudo}, {Kurokawa},
  {Nishikawa}, {Omiya}, {Serizawa}, {Ueda}, \& {Weiss}}]{hirano2020}
{Hirano}, T., {Gaidos}, E., {Winn}, J.~N., {et~al.} 2020, \apjl, 890, L27

\bibitem[{{Hirano} {et~al.}(2021){Hirano}, {Livingston}, {Fukui}, {Narita},
  {Harakawa}, {Ishikawa}, {Miyakawa}, {Kimura}, {Nakayama}, {Fujita}, {Hori},
  {Stassun}, {Bieryla}, {Cadieux}, {Ciardi}, {Collins}, {Ikoma}, {Vanderburg},
  {Barclay}, {Brasseur}, {de Leon}, {Doty}, {Doyon}, {Esparza-Borges},
  {Esquerdo}, {Furlan}, {Gaidos}, {Gonzales}, {Hodapp}, {Howell}, {Isogai},
  {Jacobson}, {Jenkins}, {Jensen}, {Kawauchi}, {Kotani}, {Kudo}, {Kurita},
  {Kurokawa}, {Kusakabe}, {Kuzuhara}, {Lafreni{\`e}re}, {Latham}, {Massey},
  {Mori}, {Murgas}, {Nishikawa}, {Nishiumi}, {Omiya}, {Paegert}, {Palle},
  {Parviainen}, {Quinn}, {Ricker}, {Schwarz}, {Seager}, {Tamura}, {Tenenbaum},
  {Terada}, {Vanderspek}, {Vievard}, {Watanabe}, \& {Winn}}]{hirano2021}
{Hirano}, T., {Livingston}, J.~H., {Fukui}, A., {et~al.} 2021, \aj, 162, 161

\bibitem[{{Ho} \& {Van Eylen}(2023)}]{ho2023}
{Ho}, C. S.~K. \& {Van Eylen}, V. 2023, \mnras, 519, 4056

\bibitem[{{Hoerl} \& {Kennard}(1970)}]{hoerlKennard1970}
{Hoerl}, A.~E. \& {Kennard}, R.~W. 1970, Technometrics, 12, 55

\bibitem[{{Holman} {et~al.}(2006){Holman}, {Winn}, {Latham}, {O'Donovan},
  {Charbonneau}, {Bakos}, {Esquerdo}, {Hergenrother}, {Everett}, \&
  {P{\'a}l}}]{holman2006}
{Holman}, M.~J., {Winn}, J.~N., {Latham}, D.~W., {et~al.} 2006, \apj, 652, 1715

\bibitem[{{Hoyer} {et~al.}(2020){Hoyer}, {Guterman}, {Demangeon}, {Sousa},
  {Deleuil}, {Meunier}, \& {Benz}}]{hoyer2020}
{Hoyer}, S., {Guterman}, P., {Demangeon}, O., {et~al.} 2020, \aap, 635, A24

\bibitem[{{Huang} {et~al.}(2017){Huang}, {Peng}, \& {Zhang}}]{Huang2017}
{Huang}, T., {Peng}, H., \& {Zhang}, K. 2017, Statistica Sinica, 27, 147

\bibitem[{{Incha} {et~al.}(2023){Incha}, {Vanderburg}, {Jacobs}, {LaCourse},
  {Bieryla}, {Pass}, {Howell}, {Berlind}, {Calkins}, {Esquerdo}, {Latham}, \&
  {Mann}}]{incha2023}
{Incha}, E., {Vanderburg}, A., {Jacobs}, T., {et~al.} 2023, \mnras, 523, 474

\bibitem[{{Jehin} {et~al.}(2011){Jehin}, {Gillon}, {Queloz}, {Magain},
  {Manfroid}, {Chantry}, {Lendl}, {Hutsem{\'e}kers}, \& {Udry}}]{jehin2011}
{Jehin}, E., {Gillon}, M., {Queloz}, D., {et~al.} 2011, The Messenger, 145, 2

\bibitem[{{Jenkins} {et~al.}(2016){Jenkins}, {Twicken}, {McCauliff},
  {Campbell}, {Sanderfer}, {Lung}, {Mansouri-Samani}, {Girouard}, {Tenenbaum},
  {Klaus}, {Smith}, {Caldwell}, {Chacon}, {Henze}, {Heiges}, {Latham},
  {Morgan}, {Swade}, {Rinehart}, \& {Vanderspek}}]{jenkins2016}
{Jenkins}, J.~M., {Twicken}, J.~D., {McCauliff}, S., {et~al.} 2016, in Society
  of Photo-Optical Instrumentation Engineers (SPIE) Conference Series, Vol.
  9913, Software and Cyberinfrastructure for Astronomy IV, 99133E

\bibitem[{{Kaminski} {et~al.}(2018){Kaminski}, {Trifonov}, {Caballero},
  {Quirrenbach}, {Ribas}, {Reiners}, {Amado}, {Zechmeister}, {Dreizler},
  {Perger}, {Tal-Or}, {Bonfils}, {Mayor}, {Astudillo-Defru}, {Bauer},
  {B{\'e}jar}, {Cifuentes}, {Colom{\'e}}, {Cort{\'e}s-Contreras}, {Delfosse},
  {D{\'\i}ez-Alonso}, {Forveille}, {Guenther}, {Hatzes}, {Henning}, {Jeffers},
  {K{\"u}rster}, {Lafarga}, {Luque}, {Mandel}, {Montes}, {Morales},
  {Passegger}, {Pedraz}, {Reffert}, {Sadegi}, {Schweitzer}, {Seifert}, {Stahl},
  \& {Udry}}]{kaminski2018}
{Kaminski}, A., {Trifonov}, T., {Caballero}, J.~A., {et~al.} 2018, \aap, 618,
  A115

\bibitem[{{Kasting} {et~al.}(1993){Kasting}, {Whitmire}, \&
  {Reynolds}}]{kasting1993}
{Kasting}, J.~F., {Whitmire}, D.~P., \& {Reynolds}, R.~T. 1993, \icarus, 101,
  108

\bibitem[{{Kay} {et~al.}(2016){Kay}, {Opher}, \& {Kornbleuth}}]{kay2016}
{Kay}, C., {Opher}, M., \& {Kornbleuth}, M. 2016, \apj, 826, 195

\bibitem[{{Kegerreis} {et~al.}(2020){Kegerreis}, {Eke}, {Catling}, {Massey},
  {Teodoro}, \& {Zahnle}}]{kegerreis2020}
{Kegerreis}, J.~A., {Eke}, V.~R., {Catling}, D.~C., {et~al.} 2020, \apjl, 901,
  L31

\bibitem[{{Kemmer} {et~al.}(2020){Kemmer}, {Stock}, {Kossakowski}, {Kaminski},
  {Molaverdikhani}, {Schlecker}, {Caballero}, {Amado}, {Astudillo-Defru},
  {Bonfils}, {Ciardi}, {Collins}, {Espinoza}, {Fukui}, {Hirano}, {Jenkins},
  {Latham}, {Matthews}, {Narita}, {Pall{\'e}}, {Parviainen}, {Quirrenbach},
  {Reiners}, {Ribas}, {Ricker}, {Schlieder}, {Seager}, {Vanderspek}, {Winn},
  {Almenara}, {B{\'e}jar}, {Bluhm}, {Bouchy}, {Boyd}, {Christiansen},
  {Cifuentes}, {Cloutier}, {Collins}, {Cort{\'e}s-Contreras}, {Crossfield},
  {Crouzet}, {de Leon}, {Della-Rose}, {Delfosse}, {Dreizler}, {Esparza-Borges},
  {Essack}, {Forveille}, {Figueira}, {Galad{\'\i}-Enr{\'\i}quez}, {Gan},
  {Glidden}, {Gonzales}, {Guerra}, {Harakawa}, {Hatzes}, {Henning}, {Herrero},
  {Hodapp}, {Hori}, {Howell}, {Ikoma}, {Isogai}, {Jeffers}, {K{\"u}rster},
  {Kawauchi}, {Kimura}, {Klagyivik}, {Kotani}, {Kurokawa}, {Kusakabe},
  {Kuzuhara}, {Lafarga}, {Livingston}, {Luque}, {Matson}, {Morales}, {Mori},
  {Muirhead}, {Murgas}, {Nishikawa}, {Nishiumi}, {Omiya}, {Reffert},
  {Rodr{\'\i}guez L{\'o}pez}, {Santos}, {Sch{\"o}fer}, {Schwarz}, {Shiao},
  {Tamura}, {Terada}, {Twicken}, {Ueda}, {Vievard}, {Watanabe}, \&
  {Zechmeister}}]{kemmer2020}
{Kemmer}, J., {Stock}, S., {Kossakowski}, D., {et~al.} 2020, \aap, 642, A236

\bibitem[{{Kiman} {et~al.}(2021){Kiman}, {Faherty}, {Cruz}, {Gagn{\'e}},
  {Angus}, {Schmidt}, {Mann}, {Bardalez Gagliuffi}, \& {Rice}}]{kiman2021}
{Kiman}, R., {Faherty}, J.~K., {Cruz}, K.~L., {et~al.} 2021, \aj, 161, 277

\bibitem[{{Kopparapu} {et~al.}(2013){Kopparapu}, {Ramirez}, {Kasting}, {Eymet},
  {Robinson}, {Mahadevan}, {Terrien}, {Domagal-Goldman}, {Meadows}, \&
  {Deshpande}}]{kopparapu2013}
{Kopparapu}, R.~K., {Ramirez}, R., {Kasting}, J.~F., {et~al.} 2013, \apj, 765,
  131

\bibitem[{{Kossakowski} {et~al.}(2021){Kossakowski}, {Kemmer}, {Bluhm},
  {Stock}, {Caballero}, {B{\'e}jar}, {Guill{\'e}n}, {Lodieu}, {Collins},
  {Oshagh}, {Schlecker}, {Espinoza}, {Pall{\'e}}, {Henning}, {Kreidberg},
  {K{\"u}rster}, {Amado}, {Anderson}, {Morales}, {Cartwright}, {Charbonneau},
  {Chaturvedi}, {Cifuentes}, {Conti}, {Cort{\'e}s-Contreras}, {Dreizler},
  {Galad{\'\i}-Enr{\'\i}quez}, {Guerra}, {Hart}, {Hellier}, {Henze}, {Herrero},
  {Jeffers}, {Jenkins}, {Jensen}, {Kaminski}, {Kielkopf}, {Kunimoto},
  {Lafarga}, {Latham}, {Lillo-Box}, {Luque}, {Molaverdikhani}, {Montes},
  {Morello}, {Morgan}, {Nowak}, {Pavlov}, {Perger}, {Quintana}, {Quirrenbach},
  {Reffert}, {Reiners}, {Ricker}, {Ribas}, {Rodr{\'\i}guez L{\'o}pez},
  {Zapatero Osorio}, {Seager}, {Sch{\"o}fer}, {Schweitzer}, {Trifonov},
  {Vanaverbeke}, {Vanderspek}, {West}, {Winn}, \&
  {Zechmeister}}]{kossakowski2021}
{Kossakowski}, D., {Kemmer}, J., {Bluhm}, P., {et~al.} 2021, \aap, 656, A124

\bibitem[{{Kotani} {et~al.}(2018){Kotani}, {Tamura}, {Nishikawa}, {Ueda},
  {Kuzuhara}, {Omiya}, {Hashimoto}, {Ishizuka}, {Hirano}, {Suto}, {Kurokawa},
  {Kokubo}, {Mori}, {Tanaka}, {Kashiwagi}, {Konishi}, {Kudo}, {Sato},
  {Jacobson}, {Hodapp}, {Hall}, {Aoki}, {Usuda}, {Nishiyama}, {Nakajima},
  {Ikeda}, {Yamamuro}, {Morino}, {Baba}, {Hosokawa}, {Ishikawa}, {Narita},
  {Kokubo}, {Hayano}, {Izumiura}, {Kambe}, {Kusakabe}, {Kwon}, {Ikoma}, {Hori},
  {Genda}, {Fukui}, {Fujii}, {Kawahara}, {Olivier}, {Jovanovic}, {Harakawa},
  {Hayashi}, {Hidai}, {Machida}, {Matsuo}, {Nagata}, {Ogihara}, {Takami},
  {Takato}, {Terada}, \& {Oh}}]{kotani2018}
{Kotani}, T., {Tamura}, M., {Nishikawa}, J., {et~al.} 2018, in Society of
  Photo-Optical Instrumentation Engineers (SPIE) Conference Series, Vol. 10702,
  Ground-based and Airborne Instrumentation for Astronomy VII, ed. C.~J.
  {Evans}, L.~{Simard}, \& H.~{Takami}, 1070211

\bibitem[{{Kubyshkina}(2023)}]{kubyshkina2023}
{Kubyshkina}, D. 2023, IAU Symposium, 370, 103

\bibitem[{{Kubyshkina} \& {Fossati}(2022)}]{kubyshkina2022}
{Kubyshkina}, D. \& {Fossati}, L. 2022, \aap, 668, A178

\bibitem[{{Kubyshkina} {et~al.}(2018){Kubyshkina}, {Fossati}, {Erkaev},
  {Cubillos}, {Johnstone}, {Kislyakova}, {Lammer}, {Lendl}, \&
  {Odert}}]{kubyshkina2018}
{Kubyshkina}, D., {Fossati}, L., {Erkaev}, N.~V., {et~al.} 2018, \apjl, 866,
  L18

\bibitem[{{Kurucz}(1993)}]{Kurucz1993}
{Kurucz}, R.~L. 1993, {SYNTHE spectrum synthesis programs and line data}
  (Astrophysics Source Code Library)

\bibitem[{{Lam} {et~al.}(2021){Lam}, {Csizmadia}, {Astudillo-Defru}, {Bonfils},
  {Gandolfi}, {Padovan}, {Esposito}, {Hellier}, {Hirano}, {Livingston},
  {Murgas}, {Smith}, {Collins}, {Mathur}, {Garcia}, {Howell}, {Santos}, {Dai},
  {Ricker}, {Vanderspek}, {Latham}, {Seager}, {Winn}, {Jenkins}, {Albrecht},
  {Almenara}, {Artigau}, {Barrag{\'a}n}, {Bouchy}, {Cabrera}, {Charbonneau},
  {Chaturvedi}, {Chaushev}, {Christiansen}, {Cochran}, {De Meideiros},
  {Delfosse}, {D{\'\i}az}, {Doyon}, {Eigm{\"u}ller}, {Figueira}, {Forveille},
  {Fridlund}, {Gaisn{\'e}}, {Goffo}, {Georgieva}, {Grziwa}, {Guenther},
  {Hatzes}, {Johnson}, {Kab{\'a}th}, {Knudstrup}, {Korth}, {Lewin}, {Lissauer},
  {Lovis}, {Luque}, {Melo}, {Morgan}, {Morris}, {Mayor}, {Narita}, {Osborne},
  {Palle}, {Pepe}, {Persson}, {Quinn}, {Rauer}, {Redfield}, {Schlieder},
  {S{\'e}gransan}, {Serrano}, {Smith}, {{\v{S}}ubjak}, {Twicken}, {Udry}, {Van
  Eylen}, \& {Vezie}}]{lam2021}
{Lam}, K. W.~F., {Csizmadia}, S., {Astudillo-Defru}, N., {et~al.} 2021,
  Science, 374, 1271

\bibitem[{{Lam} {et~al.}(2020){Lam}, {Korth}, {Masuda}, {Csizmadia},
  {Eigm{\"u}ller}, {Stef{\'a}nsson}, {Endl}, {Albrecht}, {Robertson}, {Luque},
  {Livingston}, {Hirano}, {Sobrino}, {Barrag{\'a}n}, {Cabrera}, {Carleo},
  {Chaushev}, {Cochran}, {Dai}, {Leon}, {Deeg}, {Erikson}, {Esposito},
  {Fridlund}, {Fukui}, {Gandolfi}, {Georgieva}, {Cuesta}, {Grziwa}, {Guenther},
  {Hatzes}, {Hidalgo}, {Hjorth}, {Kabath}, {Knudstrup}, {Lund}, {Mahadevan},
  {Mathur}, {Rodr{\'\i}guez}, {Murgas}, {Narita}, {Nespral}, {Niraula},
  {Palle}, {P{\"a}tzold}, {Persson}, {Prieto-Arranz}, {Rauer}, {Redfield},
  {Ribas}, {Skarka}, {Smith}, {Subjak}, \& {Van Eylen}}]{lam2020}
{Lam}, K. W.~F., {Korth}, J., {Masuda}, K., {et~al.} 2020, \aj, 159, 120

\bibitem[{{Lammer} {et~al.}(2003){Lammer}, {Selsis}, {Ribas}, {Guinan},
  {Bauer}, \& {Weiss}}]{lammer2003}
{Lammer}, H., {Selsis}, F., {Ribas}, I., {et~al.} 2003, \apjl, 598, L121

\bibitem[{{Lee} {et~al.}(2014){Lee}, {Chiang}, \& {Ormel}}]{lee2014}
{Lee}, E.~J., {Chiang}, E., \& {Ormel}, C.~W. 2014, \apj, 797, 95

\bibitem[{{Lee} \& {Connors}(2021)}]{lee2021}
{Lee}, E.~J. \& {Connors}, N.~J. 2021, \apj, 908, 32

\bibitem[{{Lee} {et~al.}(2022){Lee}, {Karalis}, \& {Thorngren}}]{lee2022}
{Lee}, E.~J., {Karalis}, A., \& {Thorngren}, D.~P. 2022, \apj, 941, 186

\bibitem[{{Leleu} {et~al.}(2021){Leleu}, {Alibert}, {Hara}, {Hooton}, {Wilson},
  {Robutel}, {Delisle}, {Laskar}, {Hoyer}, {Lovis}, {Bryant}, {Ducrot},
  {Cabrera}, {Delrez}, {Acton}, {Adibekyan}, {Allart}, {Allende Prieto},
  {Alonso}, {Alves}, {Anderson}, {Angerhausen}, {Anglada Escud{\'e}},
  {Asquier}, {Barrado}, {Barros}, {Baumjohann}, {Bayliss}, {Beck}, {Beck},
  {Bekkelien}, {Benz}, {Billot}, {Bonfanti}, {Bonfils}, {Bouchy}, {Bourrier},
  {Bou{\'e}}, {Brandeker}, {Broeg}, {Buder}, {Burdanov}, {Burleigh},
  {B{\'a}rczy}, {Cameron}, {Chamberlain}, {Charnoz}, {Cooke}, {Corral Van
  Damme}, {Correia}, {Cristiani}, {Damasso}, {Davies}, {Deleuil}, {Demangeon},
  {Demory}, {Di Marcantonio}, {Di Persio}, {Dumusque}, {Ehrenreich}, {Erikson},
  {Figueira}, {Fortier}, {Fossati}, {Fridlund}, {Futyan}, {Gandolfi},
  {Garc{\'\i}a Mu{\~n}oz}, {Garcia}, {Gill}, {Gillen}, {Gillon}, {Goad},
  {Gonz{\'a}lez Hern{\'a}ndez}, {Guedel}, {G{\"u}nther}, {Haldemann},
  {Henderson}, {Heng}, {Hogan}, {Isaak}, {Jehin}, {Jenkins}, {Jord{\'a}n},
  {Kiss}, {Kristiansen}, {Lam}, {Lavie}, {Lecavelier des Etangs}, {Lendl},
  {Lillo-Box}, {Lo Curto}, {Magrin}, {Martins}, {Maxted}, {McCormac}, {Mehner},
  {Micela}, {Molaro}, {Moyano}, {Murray}, {Nascimbeni}, {Nunes}, {Olofsson},
  {Osborn}, {Oshagh}, {Ottensamer}, {Pagano}, {Pall{\'e}}, {Pedersen}, {Pepe},
  {Persson}, {Peter}, {Piotto}, {Polenta}, {Pollacco}, {Poretti}, {Pozuelos},
  {Queloz}, {Ragazzoni}, {Rando}, {Ratti}, {Rauer}, {Raynard}, {Rebolo},
  {Reimers}, {Ribas}, {Santos}, {Scandariato}, {Schneider}, {Sebastian},
  {Sestovic}, {Simon}, {Smith}, {Sousa}, {Sozzetti}, {Steller}, {Su{\'a}rez
  Mascare{\~n}o}, {Szab{\'o}}, {S{\'e}gransan}, {Thomas}, {Thompson},
  {Tilbrook}, {Triaud}, {Turner}, {Udry}, {Van Grootel}, {Venus}, {Verrecchia},
  {Vines}, {Walton}, {West}, {Wheatley}, {Wolter}, \& {Zapatero
  Osorio}}]{Leleu2021}
{Leleu}, A., {Alibert}, Y., {Hara}, N.~C., {et~al.} 2021, \aap, 649, A26

\bibitem[{{Lillo-Box} {et~al.}(2020){Lillo-Box}, {Figueira}, {Leleu},
  {Acu{\~n}a}, {Faria}, {Hara}, {Santos}, {Correia}, {Robutel}, {Deleuil},
  {Barrado}, {Sousa}, {Bonfils}, {Mousis}, {Almenara}, {Astudillo-Defru},
  {Marcq}, {Udry}, {Lovis}, \& {Pepe}}]{lilloBox2020}
{Lillo-Box}, J., {Figueira}, P., {Leleu}, A., {et~al.} 2020, \aap, 642, A121

\bibitem[{{Lindegren} {et~al.}(2021){Lindegren}, {Bastian}, {Biermann},
  {Bombrun}, {de Torres}, {Gerlach}, {Geyer}, {Hern{\'a}ndez}, {Hilger},
  {Hobbs}, {Klioner}, {Lammers}, {McMillan}, {Ramos-Lerate},
  {Steidelm{\"u}ller}, {Stephenson}, \& {van Leeuwen}}]{lindegren2021}
{Lindegren}, L., {Bastian}, U., {Biermann}, M., {et~al.} 2021, \aap, 649, A4

\bibitem[{{Lobo} {et~al.}(2023){Lobo}, {Shields}, {Palubski}, \&
  {Wolf}}]{lobo2023}
{Lobo}, A.~H., {Shields}, A.~L., {Palubski}, I.~Z., \& {Wolf}, E. 2023, \apj,
  945, 161

\bibitem[{{Lopez} \& {Fortney}(2013)}]{lopez2013}
{Lopez}, E.~D. \& {Fortney}, J.~J. 2013, \apj, 776, 2

\bibitem[{{Lopez} \& {Fortney}(2014)}]{LopezFortney2014}
{Lopez}, E.~D. \& {Fortney}, J.~J. 2014, \apj, 792, 1

\bibitem[{{Lopez} \& {Rice}(2018)}]{lopez2018}
{Lopez}, E.~D. \& {Rice}, K. 2018, \mnras, 479, 5303

\bibitem[{{Luger} \& {Barnes}(2015)}]{luger2015}
{Luger}, R. \& {Barnes}, R. 2015, Astrobiology, 15, 119

\bibitem[{{Luque} {et~al.}(2022){Luque}, {Fulton}, {Kunimoto}, {Amado},
  {Gorrini}, {Dreizler}, {Hellier}, {Henry}, {Molaverdikhani}, {Morello},
  {Pe{\~n}a-Mo{\~n}ino}, {P{\'e}rez-Torres}, {Pozuelos}, {Shan},
  {Anglada-Escud{\'e}}, {B{\'e}jar}, {Bergond}, {Boyle}, {Caballero},
  {Charbonneau}, {Ciardi}, {Dufoer}, {Espinoza}, {Everett}, {Fischer},
  {Hatzes}, {Henning}, {Hesse}, {Howard}, {Howell}, {Isaacson}, {Jeffers},
  {Jenkins}, {Kane}, {Kemmer}, {Khalafinejad}, {Kidwell}, {Kossakowski},
  {Latham}, {Lillo-Box}, {Lissauer}, {Montes}, {Orell-Miquel}, {Pall{\'e}},
  {Pollacco}, {Quirrenbach}, {Reffert}, {Reiners}, {Ribas}, {Ricker}, {Rogers},
  {Sanz-Forcada}, {Schlecker}, {Schweitzer}, {Seager}, {Shporer}, {Stassun},
  {Stock}, {Tal-Or}, {Ting}, {Trifonov}, {Vanaverbeke}, {Vanderspek},
  {Villase{\~n}or}, {Winn}, {Winters}, \& {Zapatero
  Osorio}}]{luque2022HD260655}
{Luque}, R., {Fulton}, B.~J., {Kunimoto}, M., {et~al.} 2022, \aap, 664, A199

\bibitem[{{Luque} \& {Pall{\'e}}(2022)}]{luque2022}
{Luque}, R. \& {Pall{\'e}}, E. 2022, Science, 377, 1211

\bibitem[{{Luque} {et~al.}(2019){Luque}, {Pall{\'e}}, {Kossakowski},
  {Dreizler}, {Kemmer}, {Espinoza}, {Burt}, {Anglada-Escud{\'e}}, {B{\'e}jar},
  {Caballero}, {Collins}, {Collins}, {Cort{\'e}s-Contreras},
  {D{\'\i}ez-Alonso}, {Feng}, {Hatzes}, {Hellier}, {Henning}, {Jeffers},
  {Kaltenegger}, {K{\"u}rster}, {Madden}, {Molaverdikhani}, {Montes}, {Narita},
  {Nowak}, {Ofir}, {Oshagh}, {Parviainen}, {Quirrenbach}, {Reffert}, {Reiners},
  {Rodr{\'\i}guez-L{\'o}pez}, {Schlecker}, {Stock}, {Trifonov}, {Winn},
  {Zapatero Osorio}, {Zechmeister}, {Amado}, {Anderson}, {Batalha}, {Bauer},
  {Bluhm}, {Burke}, {Butler}, {Caldwell}, {Chen}, {Crane}, {Dragomir},
  {Dressing}, {Dynes}, {Jenkins}, {Kaminski}, {Klahr}, {Kotani}, {Lafarga},
  {Latham}, {Lewin}, {McDermott}, {Monta{\~n}{\'e}s-Rodr{\'\i}guez}, {Morales},
  {Murgas}, {Nagel}, {Pedraz}, {Ribas}, {Ricker}, {Rowden}, {Seager},
  {Shectman}, {Tamura}, {Teske}, {Twicken}, {Vanderspeck}, {Wang}, \&
  {Wohler}}]{luque2019}
{Luque}, R., {Pall{\'e}}, E., {Kossakowski}, D., {et~al.} 2019, \aap, 628, A39

\bibitem[{{Luque} {et~al.}(2021){Luque}, {Serrano}, {Molaverdikhani}, {Nixon},
  {Livingston}, {Guenther}, {Pall{\'e}}, {Madhusudhan}, {Nowak}, {Korth},
  {Cochran}, {Hirano}, {Chaturvedi}, {Goffo}, {Albrecht}, {Barrag{\'a}n},
  {Brice{\~n}o}, {Cabrera}, {Charbonneau}, {Cloutier}, {Collins}, {Collins},
  {Col{\'o}n}, {Crossfield}, {Csizmadia}, {Dai}, {Deeg}, {Esposito},
  {Fridlund}, {Gandolfi}, {Georgieva}, {Glidden}, {Goeke}, {Grziwa}, {Hatzes},
  {Henze}, {Howell}, {Irwin}, {Jenkins}, {Jensen}, {K{\'a}bath}, {Kidwell},
  {Kielkopf}, {Knudstrup}, {Lam}, {Latham}, {Lissauer}, {Mann}, {Matthews},
  {Mireles}, {Narita}, {Paegert}, {Persson}, {Redfield}, {Ricker}, {Rodler},
  {Schlieder}, {Scott}, {Seager}, {{\v{S}}ubjak}, {Tan}, {Ting}, {Vanderspek},
  {Van Eylen}, {Winn}, \& {Ziegler}}]{luque2021}
{Luque}, R., {Serrano}, L.~M., {Molaverdikhani}, K., {et~al.} 2021, \aap, 645,
  A41

\bibitem[{{MacDonald}(2019)}]{macdonald2019}
{MacDonald}, M.~G. 2019, \mnras, 487, 5062

\bibitem[{{Maciel} {et~al.}(2011){Maciel}, {Rodrigues}, \&
  {Costa}}]{Maciel2011}
{Maciel}, W.~J., {Rodrigues}, T.~S., \& {Costa}, R.~D.~D. 2011, \aap, 47, 401

\bibitem[{{Maldonado} {et~al.}(2020){Maldonado}, {Micela}, {Baratella},
  {D'Orazi}, {Affer}, {Biazzo}, {Lanza}, {Maggio}, {Gonz{\'a}lez
  Hern{\'a}ndez}, {Perger}, {Pinamonti}, {Scandariato}, {Sozzetti}, {Locci},
  {Di Maio}, {Bignamini}, {Claudi}, {Molinari}, {Rebolo}, {Ribas},
  {Toledo-Padr{\'o}n}, {Covino}, {Desidera}, {Herrero}, {Morales},
  {Su{\'a}rez-Mascare{\~n}o}, {Pagano}, {Petralia}, {Piotto}, \&
  {Poretti}}]{Maldonado-20}
{Maldonado}, J., {Micela}, G., {Baratella}, M., {et~al.} 2020, \aap, 644, A68

\bibitem[{{Mandel} \& {Agol}(2002)}]{mandel2002}
{Mandel}, K. \& {Agol}, E. 2002, \apjl, 580, L171

\bibitem[{{Marboeuf} {et~al.}(2014){Marboeuf}, {Thiabaud}, {Alibert}, {Cabral},
  \& {Benz}}]{Marboeuf2014}
{Marboeuf}, U., {Thiabaud}, A., {Alibert}, Y., {Cabral}, N., \& {Benz}, W.
  2014, \aap, 570, A36

\bibitem[{{Marcy} {et~al.}(1998){Marcy}, {Butler}, {Vogt}, {Fischer}, \&
  {Lissauer}}]{marcy1998}
{Marcy}, G.~W., {Butler}, R.~P., {Vogt}, S.~S., {Fischer}, D., \& {Lissauer},
  J.~J. 1998, \apjl, 505, L147

\bibitem[{{Marigo} {et~al.}(2017){Marigo}, {Girardi}, {Bressan}, {Rosenfield},
  {Aringer}, {Chen}, {Dussin}, {Nanni}, {Pastorelli}, {Rodrigues}, {Trabucchi},
  {Bladh}, {Dalcanton}, {Groenewegen}, {Montalb{\'a}n}, \& {Wood}}]{marigo2017}
{Marigo}, P., {Girardi}, L., {Bressan}, A., {et~al.} 2017, \apj, 835, 77

\bibitem[{{Martinez} {et~al.}(2019){Martinez}, {Cunha}, {Ghezzi}, \&
  {Smith}}]{martinez2019}
{Martinez}, C.~F., {Cunha}, K., {Ghezzi}, L., \& {Smith}, V.~V. 2019, \apj,
  875, 29

\bibitem[{{Matsumura} {et~al.}(2008){Matsumura}, {Takeda}, \&
  {Rasio}}]{matsumura2008}
{Matsumura}, S., {Takeda}, G., \& {Rasio}, F.~A. 2008, \apjl, 686, L29

\bibitem[{{Mayor} {et~al.}(2003){Mayor}, {Pepe}, {Queloz}, {Bouchy},
  {Rupprecht}, {Lo Curto}, {Avila}, {Benz}, {Bertaux}, {Bonfils}, {Dall},
  {Dekker}, {Delabre}, {Eckert}, {Fleury}, {Gilliotte}, {Gojak}, {Guzman},
  {Kohler}, {Lizon}, {Longinotti}, {Lovis}, {Megevand}, {Pasquini}, {Reyes},
  {Sivan}, {Sosnowska}, {Soto}, {Udry}, {van Kesteren}, {Weber}, \&
  {Weilenmann}}]{mayor2003}
{Mayor}, M., {Pepe}, F., {Queloz}, D., {et~al.} 2003, The Messenger, 114, 20

\bibitem[{{Miguel} {et~al.}(2020){Miguel}, {Cridland}, {Ormel}, {Fortney}, \&
  {Ida}}]{Miguel2020}
{Miguel}, Y., {Cridland}, A., {Ormel}, C.~W., {Fortney}, J.~J., \& {Ida}, S.
  2020, \mnras, 491, 1998

\bibitem[{{Mishra} {et~al.}(2023){Mishra}, {Alibert}, {Udry}, \&
  {Mordasini}}]{mishra2023}
{Mishra}, L., {Alibert}, Y., {Udry}, S., \& {Mordasini}, C. 2023, \aap, 670,
  A68

\bibitem[{{Morris} {et~al.}(2021){Morris}, {Delrez}, {Brandeker}, {Cameron},
  {Simon}, {Futyan}, {Olofsson}, {Hoyer}, {Fortier}, {Demory}, {Lendl},
  {Wilson}, {Oshagh}, {Heng}, {Ehrenreich}, {Sulis}, {Alibert}, {Alonso},
  {Anglada Escud{\'e}}, {Barrado}, {Barros}, {Baumjohann}, {Beck}, {Beck},
  {Bekkelien}, {Benz}, {Bergomi}, {Billot}, {Bonfils}, {Bourrier}, {Broeg},
  {B{\'a}rczy}, {Cabrera}, {Charnoz}, {Davies}, {De Miguel Ferreras},
  {Deleuil}, {Deline}, {Demangeon}, {Erikson}, {Floren}, {Fossati}, {Fridlund},
  {Gandolfi}, {Garc{\'\i}a Mu{\~n}oz}, {Gillon}, {Guedel}, {Guterman}, {Isaak},
  {Kiss}, {Laskar}, {Lecavelier des Etangs}, {Lieder}, {Lovis}, {Magrin},
  {Maxted}, {Nascimbeni}, {Ottensamer}, {Pagano}, {Pall{\'e}}, {Peter},
  {Piotto}, {Pizarro Rubio}, {Pollacco}, {Pozuelos}, {Queloz}, {Ragazzoni},
  {Rando}, {Rauer}, {Ribas}, {Santos}, {Scandariato}, {Smith}, {Sousa},
  {Steller}, {Szab{\'o}}, {S{\'e}gransan}, {Thomas}, {Udry}, {Ulmer}, {Van
  Grootel}, \& {Walton}}]{morris2021}
{Morris}, B.~M., {Delrez}, L., {Brandeker}, A., {et~al.} 2021, \aap, 653, A173

\bibitem[{{Narita} {et~al.}(2019){Narita}, {Fukui}, {Kusakabe}, {Watanabe},
  {Palle}, {Parviainen}, {Monta{\~n}{\'e}s-Rodr{\'\i}guez}, {Murgas},
  {Monelli}, {Aguiar}, {Perez Prieto}, {Oscoz}, {de Leon}, {Mori}, {Tamura},
  {Yamamuro}, {B{\'e}jar}, {Crouzet}, {Hidalgo}, {Klagyivik}, {Luque}, \&
  {Nishiumi}}]{narita2019}
{Narita}, N., {Fukui}, A., {Kusakabe}, N., {et~al.} 2019, Journal of
  Astronomical Telescopes, Instruments, and Systems, 5, 015001

\bibitem[{{Nordstr{\"o}m} {et~al.}(2004){Nordstr{\"o}m}, {Mayor}, {Andersen},
  {Holmberg}, {Pont}, {J{\o}rgensen}, {Olsen}, {Udry}, \&
  {Mowlavi}}]{Nordstrom2004}
{Nordstr{\"o}m}, B., {Mayor}, M., {Andersen}, J., {et~al.} 2004, \aap, 418, 989

\bibitem[{{Nowak} {et~al.}(2020){Nowak}, {Luque}, {Parviainen}, {Pall{\'e}},
  {Molaverdikhani}, {B{\'e}jar}, {Lillo-Box}, {Rodr{\'\i}guez-L{\'o}pez},
  {Caballero}, {Zechmeister}, {Passegger}, {Cifuentes}, {Schweitzer}, {Narita},
  {Cale}, {Espinoza}, {Murgas}, {Hidalgo}, {Zapatero Osorio}, {Pozuelos},
  {Aceituno}, {Amado}, {Barkaoui}, {Barrado}, {Bauer}, {Benkhaldoun},
  {Caldwell}, {Casasayas Barris}, {Chaturvedi}, {Chen}, {Collins}, {Collins},
  {Cort{\'e}s-Contreras}, {Crossfield}, {de Le{\'o}n}, {D{\'\i}ez Alonso},
  {Dreizler}, {El Mufti}, {Esparza-Borges}, {Essack}, {Fukui}, {Gaidos},
  {Gillon}, {Gonzales}, {Guerra}, {Hatzes}, {Henning}, {Herrero}, {Hesse},
  {Hirano}, {Howell}, {Jeffers}, {Jehin}, {Jenkins}, {Kaminski}, {Kemmer},
  {Kielkopf}, {Kossakowski}, {Kotani}, {K{\"u}rster}, {Lafarga}, {Latham},
  {Law}, {Lissauer}, {Lodieu}, {Madrigal-Aguado}, {Mann}, {Massey}, {Matson},
  {Matthews}, {Monta{\~n}{\'e}s-Rodr{\'\i}guez}, {Montes}, {Morales}, {Mori},
  {Nagel}, {Oshagh}, {Pedraz}, {Plavchan}, {Pollacco}, {Quirrenbach},
  {Reffert}, {Reiners}, {Ribas}, {Ricker}, {Rose}, {Schlecker}, {Schlieder},
  {Seager}, {Stangret}, {Stock}, {Tamura}, {Tanner}, {Teske}, {Trifonov},
  {Twicken}, {Vanderspek}, {Watanabe}, {Wittrock}, {Ziegler}, \&
  {Zohrabi}}]{nowak2020}
{Nowak}, G., {Luque}, R., {Parviainen}, H., {et~al.} 2020, \aap, 642, A173

\bibitem[{{Ojha} {et~al.}(2022){Ojha}, {Troncone}, {Buffo}, {Journaux}, \&
  {McDonald}}]{ojha2022}
{Ojha}, L., {Troncone}, B., {Buffo}, J., {Journaux}, B., \& {McDonald}, G.
  2022, Nature Communications, 13, 7521

\bibitem[{{Oliva}(2006)}]{oliva2006}
{Oliva}, E. 2006, Memorie della Societa Astronomica Italiana Supplementi, 9,
  409

\bibitem[{{Owen} \& {Wu}(2013)}]{owen2013}
{Owen}, J.~E. \& {Wu}, Y. 2013, \apj, 775, 105

\bibitem[{{Owen} \& {Wu}(2017)}]{owen2017}
{Owen}, J.~E. \& {Wu}, Y. 2017, \apj, 847, 29

\bibitem[{{Parviainen} {et~al.}(2019){Parviainen}, {Tingley}, {Deeg}, {Palle},
  {Alonso}, {Montanes Rodriguez}, {Murgas}, {Narita}, {Fukui}, {Watanabe},
  {Kusakabe}, {Tamura}, {Nishiumi}, {Prieto-Arranz}, {Klagyivik}, {B{\'e}jar},
  {Crouzet}, {Mori}, {Hidalgo Soto}, {Casasayas Barris}, \&
  {Luque}}]{Parviainen2019}
{Parviainen}, H., {Tingley}, B., {Deeg}, H.~J., {et~al.} 2019, \aap, 630, A89

\bibitem[{{Pass} {et~al.}(2022){Pass}, {Charbonneau}, {Irwin}, \&
  {Winters}}]{pass2022}
{Pass}, E.~K., {Charbonneau}, D., {Irwin}, J.~M., \& {Winters}, J.~G. 2022,
  \apj, 936, 109

\bibitem[{Pedregosa {et~al.}(2011)Pedregosa, Varoquaux, Gramfort, Michel,
  Thirion, Grisel, Blondel, Prettenhofer, Weiss, Dubourg, Vanderplas, Passos,
  Cournapeau, Brucher, Perrot, \& Duchesnay}]{sklearn2011}
Pedregosa, F., Varoquaux, G., Gramfort, A., {et~al.} 2011, Journal of Machine
  Learning Research, 12, 2825

\bibitem[{{Pepe} {et~al.}(2021){Pepe}, {Cristiani}, {Rebolo}, {Santos},
  {Dekker}, {Cabral}, {Di Marcantonio}, {Figueira}, {Lo Curto}, {Lovis},
  {Mayor}, {M{\'e}gevand}, {Molaro}, {Riva}, {Zapatero Osorio}, {Amate},
  {Manescau}, {Pasquini}, {Zerbi}, {Adibekyan}, {Abreu}, {Affolter}, {Alibert},
  {Aliverti}, {Allart}, {Allende Prieto}, {{\'A}lvarez}, {Alves}, {Avila},
  {Baldini}, {Bandy}, {Barros}, {Benz}, {Bianco}, {Borsa}, {Bourrier},
  {Bouchy}, {Broeg}, {Calderone}, {Cirami}, {Coelho}, {Conconi}, {Coretti},
  {Cumani}, {Cupani}, {D'Odorico}, {Damasso}, {Deiries}, {Delabre},
  {Demangeon}, {Dumusque}, {Ehrenreich}, {Faria}, {Fragoso}, {Genolet},
  {Genoni}, {G{\'e}nova Santos}, {Gonz{\'a}lez Hern{\'a}ndez}, {Hughes},
  {Iwert}, {Kerber}, {Knudstrup}, {Landoni}, {Lavie}, {Lillo-Box}, {Lizon},
  {Maire}, {Martins}, {Mehner}, {Micela}, {Modigliani}, {Monteiro}, {Monteiro},
  {Moschetti}, {Murphy}, {Nunes}, {Oggioni}, {Oliveira}, {Oshagh}, {Pall{\'e}},
  {Pariani}, {Poretti}, {Rasilla}, {Rebord{\~a}o}, {Redaelli}, {Santana
  Tschudi}, {Santin}, {Santos}, {S{\'e}gransan}, {Schmidt}, {Segovia},
  {Sosnowska}, {Sozzetti}, {Sousa}, {Span{\`o}}, {Su{\'a}rez Mascare{\~n}o},
  {Tabernero}, {Tenegi}, {Udry}, \& {Zanutta}}]{pepe2021}
{Pepe}, F., {Cristiani}, S., {Rebolo}, R., {et~al.} 2021, \aap, 645, A96

\bibitem[{{Peterson} {et~al.}(2023){Peterson}, {Benneke}, {Collins}, {Piaulet},
  {Crossfield}, {Ali-Dib}, {Christiansen}, {Gagn{\'e}}, {Faherty}, {Kite},
  {Dressing}, {Charbonneau}, {Murgas}, {Cointepas}, {Almenara}, {Bonfils},
  {Kane}, {Werner}, {Gorjian}, {Roy}, {Shporer}, {Pozuelos}, {Socia},
  {Cloutier}, {Dietrich}, {Irwin}, {Weiss}, {Waalkes}, {Berta-Thomson},
  {Evans}, {Apai}, {Parviainen}, {Pall{\'e}}, {Narita}, {Howard}, {Dragomir},
  {Barkaoui}, {Gillon}, {Jehin}, {Ducrot}, {Benkhaldoun}, {Fukui}, {Mori},
  {Nishiumi}, {Kawauchi}, {Ricker}, {Latham}, {Winn}, {Seager}, {Isaacson},
  {Bixel}, {Gibbs}, {Jenkins}, {Smith}, {Chavez}, {Rackham}, {Henning},
  {Gabor}, {Chen}, {Espinoza}, {Jensen}, {Collins}, {Schwarz}, {Conti}, {Wang},
  {Kielkopf}, {Mao}, {Horne}, {Sefako}, {Quinn}, {Moldovan}, {Fausnaugh},
  {F{\.z}{\.z}r{\'e}sz}, \& {Barclay}}]{peterson2023}
{Peterson}, M.~S., {Benneke}, B., {Collins}, K., {et~al.} 2023, \nat, 617, 701

\bibitem[{{Petigura} {et~al.}(2022){Petigura}, {Rogers}, {Isaacson}, {Owen},
  {Kraus}, {Winn}, {MacDougall}, {Howard}, {Fulton}, {Kosiarek}, {Weiss},
  {Behmard}, \& {Blunt}}]{petigura2022}
{Petigura}, E.~A., {Rogers}, J.~G., {Isaacson}, H., {et~al.} 2022, \aj, 163,
  179

\bibitem[{{Piaulet} {et~al.}(2023){Piaulet}, {Benneke}, {Almenara}, {Dragomir},
  {Knutson}, {Thorngren}, {Peterson}, {Crossfield}, {Kempton}, {Kubyshkina},
  {Howard}, {Angus}, {Isaacson}, {Weiss}, {Beichman}, {Fortney}, {Fossati},
  {Lammer}, {McCullough}, {Morley}, \& {Wong}}]{piaulet2023}
{Piaulet}, C., {Benneke}, B., {Almenara}, J.~M., {et~al.} 2023, Nature
  Astronomy, 7, 206

\bibitem[{{Pont} {et~al.}(2006){Pont}, {Zucker}, \& {Queloz}}]{pont2006}
{Pont}, F., {Zucker}, S., \& {Queloz}, D. 2006, \mnras, 373, 231

\bibitem[{{Powers} {et~al.}(2023){Powers}, {Libby-Roberts}, {Lin}, {Ca{\~n}as},
  {Kanodia}, {Mahadevan}, {Ninan}, {Stef{\'a}nsson}, {Gupta}, {Jones},
  {Kobulnicky}, {Monson}, {Parker}, {Swaby}, {Bender}, {Cochran}, {Hebb},
  {Metcalf}, {Robertson}, {Schwab}, {Wisniewski}, \& {Wright}}]{powers2023}
{Powers}, L.~C., {Libby-Roberts}, J., {Lin}, A. S.~J., {et~al.} 2023, \aj, 166,
  44

\bibitem[{{Pozuelos} {et~al.}(2023){Pozuelos}, {Timmermans}, {Rackham},
  {Garcia}, {Burgasser}, {Kane}, {G{\"u}nther}, {Stassun}, {Van Grootel},
  {D{\'e}vora-Pajares}, {Luque}, {Edwards}, {Niraula}, {Schanche}, {Wells},
  {Ducrot}, {Howell}, {Sebastian}, {Barkaoui}, {Waalkes}, {Cadieux}, {Doyon},
  {Boyle}, {Dietrich}, {Burdanov}, {Delrez}, {Demory}, {de Wit}, {Dransfield},
  {Gillon}, {G{\'o}mez Maqueo Chew}, {Hooton}, {Jehin}, {Murray}, {Pedersen},
  {Queloz}, {Thompson}, {Triaud}, {Z{\'u}{\~n}iga-Fern{\'a}ndez}, {Collins},
  {Fausnaugh}, {Hedges}, {Hesse}, {Jenkins}, {Kunimoto}, {Latham}, {Shporer},
  {Ting}, {Torres}, {Amado}, {Rod{\'o}n}, {Rodr{\'\i}guez-L{\'o}pez},
  {Su{\'a}rez}, {Alonso}, {Benkhaldoun}, {Berta-Thompson}, {Chinchilla},
  {Ghachoui}, {G{\'o}mez-Mu{\~n}oz}, {Rebolo}, {Sabin}, {Schroffenegger},
  {Furlan}, {Gnilka}, {Lester}, {Scott}, {Aganze}, {Gerasimov}, {Hsu},
  {Theissen}, {Apai}, {Chen}, {Gabor}, {Henning}, \& {Mancini}}]{pozuelos2023}
{Pozuelos}, F.~J., {Timmermans}, M., {Rackham}, B.~V., {et~al.} 2023, \aap,
  672, A70

\bibitem[{{Quirrenbach} {et~al.}(2014){Quirrenbach}, {Amado}, {Caballero},
  {Mundt}, {Reiners}, {Ribas}, {Seifert}, {Abril}, {Aceituno},
  {Alonso-Floriano}, {Ammler-von Eiff}, {Antona Jim{\'e}nez},
  {Anwand-Heerwart}, {Azzaro}, {Bauer}, {Barrado}, {Becerril}, {B{\'e}jar},
  {Ben{\'\i}tez}, {Berdi{\~n}as}, {C{\'a}rdenas}, {Casal}, {Claret},
  {Colom{\'e}}, {Cort{\'e}s-Contreras}, {Czesla}, {Doellinger}, {Dreizler},
  {Feiz}, {Fern{\'a}ndez}, {Galad{\'\i}}, {G{\'a}lvez-Ortiz},
  {Garc{\'\i}a-Piquer}, {Garc{\'\i}a-Vargas}, {Garrido}, {Gesa}, {G{\'o}mez
  Galera}, {Gonz{\'a}lez {\'A}lvarez}, {Gonz{\'a}lez Hern{\'a}ndez},
  {Gr{\"o}zinger}, {Gu{\`a}rdia}, {Guenther}, {de Guindos},
  {Guti{\'e}rrez-Soto}, {Hagen}, {Hatzes}, {Hauschildt}, {Helmling}, {Henning},
  {Hermann}, {Hern{\'a}ndez Casta{\~n}o}, {Herrero}, {Hidalgo}, {Holgado},
  {Huber}, {Huber}, {Jeffers}, {Joergens}, {de Juan}, {Kehr}, {Klein},
  {K{\"u}rster}, {Lamert}, {Lalitha}, {Laun}, {Lemke}, {Lenzen}, {L{\'o}pez del
  Fresno}, {L{\'o}pez Mart{\'\i}}, {L{\'o}pez-Santiago}, {Mall}, {Mandel},
  {Mart{\'\i}n}, {Mart{\'\i}n-Ruiz}, {Mart{\'\i}nez-Rodr{\'\i}guez}, {Marvin},
  {Mathar}, {Mirabet}, {Montes}, {Morales Mu{\~n}oz}, {Moya}, {Naranjo},
  {Ofir}, {Oreiro}, {Pall{\'e}}, {Panduro}, {Passegger}, {P{\'e}rez-Calpena},
  {P{\'e}rez Medialdea}, {Perger}, {Pluto}, {Ram{\'o}n}, {Rebolo}, {Redondo},
  {Reffert}, {Reinhardt}, {Rhode}, {Rix}, {Rodler}, {Rodr{\'\i}guez},
  {Rodr{\'\i}guez-L{\'o}pez}, {Rodr{\'\i}guez-P{\'e}rez}, {Rohloff}, {Rosich},
  {S{\'a}nchez-Blanco}, {S{\'a}nchez Carrasco}, {Sanz-Forcada}, {Sarmiento},
  {Sch{\"a}fer}, {Schiller}, {Schmidt}, {Schmitt}, {Solano}, {Stahl}, {Storz},
  {St{\"u}rmer}, {Su{\'a}rez}, {Ulbrich}, {Veredas}, {Wagner}, {Winkler},
  {Zapatero Osorio}, {Zechmeister}, {Abell{\'a}n de Paco},
  {Anglada-Escud{\'e}}, {del Burgo}, {Klutsch}, {Lizon}, {L{\'o}pez-Morales},
  {Morales}, {Perryman}, {Tulloch}, \& {Xu}}]{quirrenback2014}
{Quirrenbach}, A., {Amado}, P.~J., {Caballero}, J.~A., {et~al.} 2014, in
  Society of Photo-Optical Instrumentation Engineers (SPIE) Conference Series,
  Vol. 9147, Ground-based and Airborne Instrumentation for Astronomy V, ed.
  S.~K. {Ramsay}, I.~S. {McLean}, \& H.~{Takami}, 91471F

\bibitem[{{Quirrenbach} {et~al.}(2018){Quirrenbach}, {Amado}, {Ribas},
  {Reiners}, {Caballero}, {Seifert}, {Aceituno}, {Azzaro}, {Baroch}, {Barrado},
  {Bauer}, {Becerril}, {B{\`e}jar}, {Ben{\'\i}tez}, {Brinkm{\"o}ller}, {Cardona
  Guill{\'e}n}, {Cifuentes}, {Colom{\'e}}, {Cort{\'e}s-Contreras}, {Czesla},
  {Dreizler}, {Fr{\"o}lich}, {Fuhrmeister}, {Galad{\'\i}-Enr{\'\i}quez},
  {Gonz{\'a}lez Hern{\'a}ndez}, {Gonz{\'a}lez Peinado}, {Guenther}, {de
  Guindos}, {Hagen}, {Hatzes}, {Hauschildt}, {Helmling}, {Henning}, {Herbort},
  {Hern{\'a}ndez Casta{\~n}o}, {Herrero}, {Hintz}, {Jeffers}, {Johnson}, {de
  Juan}, {Kaminski}, {Klahr}, {K{\"u}rster}, {Lafarga}, {Sairam}, {Lamp{\'o}n},
  {Lara}, {Launhardt}, {L{\'o}pez del Fresno}, {L{\'o}pez-Puertas}, {Luque},
  {Mandel}, {Marfil}, {Mart{\'\i}n}, {Mart{\'\i}n-Ruiz}, {Mathar}, {Montes},
  {Morales}, {Nagel}, {Nortmann}, {Nowak}, {Pall{\'e}}, {Passegger}, {Pavlov},
  {Pedraz}, {P{\'e}rez-Medialdea}, {Perger}, {Rebolo}, {Reffert},
  {Rodr{\'\i}guez}, {Rodr{\'\i}guez L{\'o}pez}, {Rosich}, {Sabotta}, {Sadegi},
  {Salz}, {S{\'a}nchez-L{\'o}pez}, {Sanz-Forcada}, {Sarkis}, {Sch{\"a}fer},
  {Schiller}, {Schmitt}, {Sch{\"o}fer}, {Schweitzer}, {Shulyak}, {Solano},
  {Stahl}, {Tala Pinto}, {Trifonov}, {Zapatero Osorio}, {Yan}, {Zechmeister},
  {Abell{\'a}n}, {Abril}, {Alonso-Floriano}, {Ammler-von Eiff},
  {Anglada-Escud{\'e}}, {Anwand-Heerwart}, {Arroyo-Torres}, {Berdi{\~n}as},
  {Bergondy}, {Bl{\"u}mcke}, {del Burgo}, {Cano}, {Carro}, {C{\'a}rdenas},
  {Casal}, {Claret}, {D{\'\i}ez-Alonso}, {Doellinger}, {Dorda}, {Feiz},
  {Fern{\'a}ndez}, {Ferro}, {Gaisn{\'e}}, {Gallardo}, {G{\'a}lvez-Ortiz},
  {Garc{\'\i}a-Piquer}, {Garc{\'\i}a-Vargas}, {Garrido}, {Gesa}, {G{\'o}mez
  Galera}, {Gonz{\'a}lez-{\'A}lvarez}, {Gonz{\'a}lez-Cuesta}, {Grohnert},
  {Gr{\"o}zinger}, {Gu{\`a}rdia}, {Guijarro}, {Hedrosa}, {Hermann}, {Hermelo},
  {Hern{\'a}ndez Arab{\'\i}}, {Hern{\'a}ndez Hernando}, {Hidalgo}, {Holgado},
  {Huber}, {Huber}, {Huke}, {Kehr}, {Kim}, {Klein}, {Kl{\"u}ter}, {Klutsch},
  {Labarga}, {Labiche}, {Lamert}, {Laun}, {L{\'a}zaro}, {Lemke}, {Lenzen},
  {Llamas}, {Lizon}, {Lodieu}, {L{\'o}pez Gonz{\'a}lez}, {L{\'o}pez-Morales},
  {L{\'o}pez Salas}, {L{\'o}pez-Santiago}, {Mag{\'a}n Madinabeitia}, {Mall},
  {Mancini}, {Mar{\'\i}n Molina}, {Mart{\'\i}nez-Rodr{\'\i}guez}, {Maroto
  Fern{\'a}ndez}, {Marvin}, {Mirabet}, {Moreno-Raya}, {Moya}, {Mundt},
  {Naranjo}, {Panduro}, {Pascual}, {P{\'e}rez-Calpena}, {Perryman}, {Pluto},
  {Ram{\'o}n}, {Redondo}, {Reinhart}, {Rhode}, {Rix}, {Rodler}, {Rohloff},
  {S{\'a}nchez-Blanco}, {S{\'a}nchez Carrasco}, {Sarmiento}, {Schmidt},
  {Storz}, {Strachan}, {St{\"u}rmer}, {Su{\'a}rez}, {Tabernero}, {Tal-Or},
  {Tulloch}, {Ulbrich}, {Veredas}, {Vico Linares}, {Vidal-Dasilva},
  {Vilardell}, {Wagner}, {Winkler}, {Wolthoff}, {Xu}, \&
  {Zhao}}]{quirrenback2018}
{Quirrenbach}, A., {Amado}, P.~J., {Ribas}, I., {et~al.} 2018, in Society of
  Photo-Optical Instrumentation Engineers (SPIE) Conference Series, Vol. 10702,
  Ground-based and Airborne Instrumentation for Astronomy VII, ed. C.~J.
  {Evans}, L.~{Simard}, \& H.~{Takami}, 107020W

\bibitem[{Rasmussen \& Williams(2005)}]{rasmussen2005}
Rasmussen, C.~E. \& Williams, C. K.~I. 2005, Gaussian Processes for Machine
  Learning (Adaptive Computation and Machine Learning) (The MIT Press)

\bibitem[{{Rayner} {et~al.}(2022){Rayner}, {Tokunaga}, {Jaffe}, {Bond},
  {Bonnet}, {Ching}, {Connelley}, {Cushing}, {Kokubun}, {Lockhart}, {Vacca}, \&
  {Warmbier}}]{rayner2022}
{Rayner}, J., {Tokunaga}, A., {Jaffe}, D., {et~al.} 2022, \pasp, 134, 015002

\bibitem[{{Rayner} {et~al.}(2016){Rayner}, {Tokunaga}, {Jaffe}, {Bonnet},
  {Ching}, {Connelley}, {Kokubun}, {Lockhart}, \& {Warmbier}}]{rayner2016}
{Rayner}, J., {Tokunaga}, A., {Jaffe}, D., {et~al.} 2016, in Society of
  Photo-Optical Instrumentation Engineers (SPIE) Conference Series, Vol. 9908,
  Ground-based and Airborne Instrumentation for Astronomy VI, ed. C.~J.
  {Evans}, L.~{Simard}, \& H.~{Takami}, 990884

\bibitem[{{Reddy} {et~al.}(2006){Reddy}, {Lambert}, \& {Allende
  Prieto}}]{Reddy-06}
{Reddy}, B.~E., {Lambert}, D.~L., \& {Allende Prieto}, C. 2006, \mnras, 367,
  1329

\bibitem[{{Reiners} {et~al.}(2009){Reiners}, {Basri}, \&
  {Browning}}]{reiners2009}
{Reiners}, A., {Basri}, G., \& {Browning}, M. 2009, \apj, 692, 538

\bibitem[{{Ricker} {et~al.}(2015){Ricker}, {Winn}, {Vanderspek}, {Latham},
  {Bakos}, {Bean}, {Berta-Thompson}, {Brown}, {Buchhave}, {Butler}, {Butler},
  {Chaplin}, {Charbonneau}, {Christensen-Dalsgaard}, {Clampin}, {Deming},
  {Doty}, {De Lee}, {Dressing}, {Dunham}, {Endl}, {Fressin}, {Ge}, {Henning},
  {Holman}, {Howard}, {Ida}, {Jenkins}, {Jernigan}, {Johnson}, {Kaltenegger},
  {Kawai}, {Kjeldsen}, {Laughlin}, {Levine}, {Lin}, {Lissauer}, {MacQueen},
  {Marcy}, {McCullough}, {Morton}, {Narita}, {Paegert}, {Palle}, {Pepe},
  {Pepper}, {Quirrenbach}, {Rinehart}, {Sasselov}, {Sato}, {Seager},
  {Sozzetti}, {Stassun}, {Sullivan}, {Szentgyorgyi}, {Torres}, {Udry}, \&
  {Villasenor}}]{ricker2015}
{Ricker}, G.~R., {Winn}, J.~N., {Vanderspek}, R., {et~al.} 2015, Journal of
  Astronomical Telescopes, Instruments, and Systems, 1, 014003

\bibitem[{{Rogers} {et~al.}(2021){Rogers}, {Gupta}, {Owen}, \&
  {Schlichting}}]{rogers2021}
{Rogers}, J.~G., {Gupta}, A., {Owen}, J.~E., \& {Schlichting}, H.~E. 2021,
  \mnras, 508, 5886

\bibitem[{{Saar} \& {Linsky}(1985)}]{saar1985}
{Saar}, S.~H. \& {Linsky}, J.~L. 1985, \apjl, 299, L47

\bibitem[{{Salmon} {et~al.}(2021){Salmon}, {Van Grootel}, {Buldgen}, {Dupret},
  \& {Eggenberger}}]{salmon2021}
{Salmon}, S.~J.~A.~J., {Van Grootel}, V., {Buldgen}, G., {Dupret}, M.~A., \&
  {Eggenberger}, P. 2021, \aap, 646, A7

\bibitem[{{Schanche} {et~al.}(2020){Schanche}, {H{\'e}brard}, {Collier
  Cameron}, {Dalal}, {Smalley}, {Wilson}, {Boisse}, {Bouchy}, {Brown},
  {Demangeon}, {Haswell}, {Hellier}, {Kolb}, {Lopez}, {Maxted}, {Pollacco},
  {West}, \& {Wheatley}}]{Schanche2020}
{Schanche}, N., {H{\'e}brard}, G., {Collier Cameron}, A., {et~al.} 2020,
  \mnras, 499, 428

\bibitem[{{Schmidt} {et~al.}(2015){Schmidt}, {Hawley}, {West}, {Bochanski},
  {Davenport}, {Ge}, \& {Schneider}}]{Schmidt-2015}
{Schmidt}, S.~J., {Hawley}, S.~L., {West}, A.~A., {et~al.} 2015, \aj, 149, 158

\bibitem[{{Scholz} {et~al.}(2005){Scholz}, {Meusinger}, \&
  {Jahrei{\ss}}}]{scholz2005}
{Scholz}, R.~D., {Meusinger}, H., \& {Jahrei{\ss}}, H. 2005, \aap, 442, 211

\bibitem[{{Sch{\"o}nrich} {et~al.}(2010){Sch{\"o}nrich}, {Binney}, \&
  {Dehnen}}]{Schonrich-10}
{Sch{\"o}nrich}, R., {Binney}, J., \& {Dehnen}, W. 2010, \mnras, 403, 1829

\bibitem[{{Schwarz}(1978)}]{schwarz1978}
{Schwarz}, G. 1978, Annals of Statistics, 6, 461

\bibitem[{{Schweitzer} {et~al.}(2019){Schweitzer}, {Passegger}, {Cifuentes},
  {B{\'e}jar}, {Cort{\'e}s-Contreras}, {Caballero}, {del Burgo}, {Czesla},
  {K{\"u}rster}, {Montes}, {Zapatero Osorio}, {Ribas}, {Reiners},
  {Quirrenbach}, {Amado}, {Aceituno}, {Anglada-Escud{\'e}}, {Bauer},
  {Dreizler}, {Jeffers}, {Guenther}, {Henning}, {Kaminski}, {Lafarga},
  {Marfil}, {Morales}, {Schmitt}, {Seifert}, {Solano}, {Tabernero}, \&
  {Zechmeister}}]{schweitzer2019}
{Schweitzer}, A., {Passegger}, V.~M., {Cifuentes}, C., {et~al.} 2019, \aap,
  625, A68

\bibitem[{{Scuflaire} {et~al.}(2008){Scuflaire}, {Th{\'e}ado}, {Montalb{\'a}n},
  {Miglio}, {Bourge}, {Godart}, {Thoul}, \& {Noels}}]{scuflaire2008}
{Scuflaire}, R., {Th{\'e}ado}, S., {Montalb{\'a}n}, J., {et~al.} 2008, \apss,
  316, 83

\bibitem[{{Seifahrt} {et~al.}(2022){Seifahrt}, {Bean}, {Kasper}, {St{\"u}rmer},
  {Brady}, {Liu}, {Zechmeister}, {Stef{\'a}nsson}, {Montet}, {White}, {Tapia},
  {Mocnik}, {Xu}, \& {Schwab}}]{Seifahrt22}
{Seifahrt}, A., {Bean}, J.~L., {Kasper}, D., {et~al.} 2022, in Society of
  Photo-Optical Instrumentation Engineers (SPIE) Conference Series, Vol. 12184,
  Ground-based and Airborne Instrumentation for Astronomy IX, ed. C.~J.
  {Evans}, J.~J. {Bryant}, \& K.~{Motohara}, 121841G

\bibitem[{{Seifahrt} {et~al.}(2018){Seifahrt}, {St{\"u}rmer}, {Bean}, \&
  {Schwab}}]{Seifahrt18}
{Seifahrt}, A., {St{\"u}rmer}, J., {Bean}, J.~L., \& {Schwab}, C. 2018, in
  Society of Photo-Optical Instrumentation Engineers (SPIE) Conference Series,
  Vol. 10702, Ground-based and Airborne Instrumentation for Astronomy VII, ed.
  C.~J. {Evans}, L.~{Simard}, \& H.~{Takami}, 107026D

\bibitem[{{Sergeev} {et~al.}(2020){Sergeev}, {Lambert}, {Mayne}, {Boutle},
  {Manners}, \& {Kohary}}]{sergeev2020}
{Sergeev}, D.~E., {Lambert}, F.~H., {Mayne}, N.~J., {et~al.} 2020, \apj, 894,
  84

\bibitem[{{Shulyak} {et~al.}(2019){Shulyak}, {Reiners}, {Nagel}, {Tal-Or},
  {Caballero}, {Zechmeister}, {B{\'e}jar}, {Cort{\'e}s-Contreras}, {Martin},
  {Kaminski}, {Ribas}, {Quirrenbach}, {Amado}, {Anglada-Escud{\'e}}, {Bauer},
  {Dreizler}, {Guenther}, {Henning}, {Jeffers}, {K{\"u}rster}, {Lafarga},
  {Montes}, {Morales}, \& {Pedraz}}]{shulyak2019}
{Shulyak}, D., {Reiners}, A., {Nagel}, E., {et~al.} 2019, \aap, 626, A86

\bibitem[{{Skrutskie} {et~al.}(2006){Skrutskie}, {Cutri}, {Stiening},
  {Weinberg}, {Schneider}, {Carpenter}, {Beichman}, {Capps}, {Chester},
  {Elias}, {Huchra}, {Liebert}, {Lonsdale}, {Monet}, {Price}, {Seitzer},
  {Jarrett}, {Kirkpatrick}, {Gizis}, {Howard}, {Evans}, {Fowler}, {Fullmer},
  {Hurt}, {Light}, {Kopan}, {Marsh}, {McCallon}, {Tam}, {Van Dyk}, \&
  {Wheelock}}]{Skrutskie2006}
{Skrutskie}, M.~F., {Cutri}, R.~M., {Stiening}, R., {et~al.} 2006, \aj, 131,
  1163

\bibitem[{{Sotin} {et~al.}(2007){Sotin}, {Grasset}, \& {Mocquet}}]{Sotin2007}
{Sotin}, C., {Grasset}, O., \& {Mocquet}, A. 2007, \icarus, 191, 337

\bibitem[{{Soto} {et~al.}(2021){Soto}, {Anglada-Escud{\'e}}, {Dreizler},
  {Molaverdikhani}, {Kemmer}, {Rodr{\'\i}guez-L{\'o}pez}, {Lillo-Box},
  {Pall{\'e}}, {Espinoza}, {Caballero}, {Quirrenbach}, {Ribas}, {Reiners},
  {Narita}, {Hirano}, {Amado}, {B{\'e}jar}, {Bluhm}, {Burke}, {Caldwell},
  {Charbonneau}, {Cloutier}, {Collins}, {Cort{\'e}s-Contreras}, {Girardin},
  {Guerra}, {Harakawa}, {Hatzes}, {Irwin}, {Jenkins}, {Jensen}, {Kawauchi},
  {Kotani}, {Kudo}, {Kunimoto}, {Kuzuhara}, {Latham}, {Montes}, {Morales},
  {Mori}, {Nelson}, {Omiya}, {Pedraz}, {Passegger}, {Rackham}, {Rudat},
  {Schlieder}, {Sch{\"o}fer}, {Schweitzer}, {Selezneva}, {Stockdale}, {Tamura},
  {Trifonov}, {Vanderspek}, \& {Watanabe}}]{soto2021}
{Soto}, M.~G., {Anglada-Escud{\'e}}, G., {Dreizler}, S., {et~al.} 2021, \aap,
  649, A144

\bibitem[{{Sousa} {et~al.}(2021){Sousa}, {Adibekyan}, {Delgado-Mena}, {Santos},
  {Rojas-Ayala}, {Soares}, {Legoinha}, {Ulmer-Moll}, {Camacho}, {Barros},
  {Demangeon}, {Hoyer}, {Israelian}, {Mortier}, {Tsantaki}, \&
  {Monteiro}}]{Sousa-2021}
{Sousa}, S.~G., {Adibekyan}, V., {Delgado-Mena}, E., {et~al.} 2021, \aap, 656,
  A53

\bibitem[{{Stefansson} {et~al.}(2020){Stefansson}, {Mahadevan}, {Maney},
  {Ninan}, {Robertson}, {Rajagopal}, {Haase}, {Allen}, {Ford}, {Winn},
  {Wolfgang}, {Dawson}, {Wisniewski}, {Bender}, {Ca{\~n}as}, {Cochran},
  {Diddams}, {Fredrick}, {Halverson}, {Hearty}, {Hebb}, {Kanodia}, {Levi},
  {Metcalf}, {Monson}, {Ramsey}, {Roy}, {Schwab}, {Terrien}, \&
  {Wright}}]{stefansson2020}
{Stefansson}, G., {Mahadevan}, S., {Maney}, M., {et~al.} 2020, \aj, 160, 192

\bibitem[{{Thiabaud} {et~al.}(2014){Thiabaud}, {Marboeuf}, {Alibert}, {Cabral},
  {Leya}, \& {Mezger}}]{Thiabaud2014}
{Thiabaud}, A., {Marboeuf}, U., {Alibert}, Y., {et~al.} 2014, \aap, 562, A27

\bibitem[{{Thiabaud} {et~al.}(2015){Thiabaud}, {Marboeuf}, {Alibert}, {Leya},
  \& {Mezger}}]{Thiabaud2015}
{Thiabaud}, A., {Marboeuf}, U., {Alibert}, Y., {Leya}, I., \& {Mezger}, K.
  2015, \aap, 574, A138

\bibitem[{{Tilley} {et~al.}(2019){Tilley}, {Segura}, {Meadows}, {Hawley}, \&
  {Davenport}}]{tilley2019}
{Tilley}, M.~A., {Segura}, A., {Meadows}, V., {Hawley}, S., \& {Davenport}, J.
  2019, Astrobiology, 19, 64

\bibitem[{{Tody}(1986)}]{tody1986}
{Tody}, D. 1986, in Society of Photo-Optical Instrumentation Engineers (SPIE)
  Conference Series, Vol. 627, Instrumentation in astronomy VI, ed. D.~L.
  {Crawford}, 733

\bibitem[{{Tody}(1993)}]{tody1993}
{Tody}, D. 1993, in Astronomical Society of the Pacific Conference Series,
  Vol.~52, Astronomical Data Analysis Software and Systems II, ed. R.~J.
  {Hanisch}, R.~J.~V. {Brissenden}, \& J.~{Barnes}, 173

\bibitem[{{Trifonov} {et~al.}(2021){Trifonov}, {Caballero}, {Morales},
  {Seifahrt}, {Ribas}, {Reiners}, {Bean}, {Luque}, {Parviainen}, {Pall{\'e}},
  {Stock}, {Zechmeister}, {Amado}, {Anglada-Escud{\'e}}, {Azzaro}, {Barclay},
  {B{\'e}jar}, {Bluhm}, {Casasayas-Barris}, {Cifuentes}, {Collins}, {Collins},
  {Cort{\'e}s-Contreras}, {de Leon}, {Dreizler}, {Dressing}, {Esparza-Borges},
  {Espinoza}, {Fausnaugh}, {Fukui}, {Hatzes}, {Hellier}, {Henning}, {Henze},
  {Herrero}, {Jeffers}, {Jenkins}, {Jensen}, {Kaminski}, {Kasper},
  {Kossakowski}, {K{\"u}rster}, {Lafarga}, {Latham}, {Mann}, {Molaverdikhani},
  {Montes}, {Montet}, {Murgas}, {Narita}, {Oshagh}, {Passegger}, {Pollacco},
  {Quinn}, {Quirrenbach}, {Ricker}, {Rodr{\'\i}guez L{\'o}pez}, {Sanz-Forcada},
  {Schwarz}, {Schweitzer}, {Seager}, {Shporer}, {Stangret}, {St{\"u}rmer},
  {Tan}, {Tenenbaum}, {Twicken}, {Vanderspek}, \& {Winn}}]{trifonov2021}
{Trifonov}, T., {Caballero}, J.~A., {Morales}, J.~C., {et~al.} 2021, Science,
  371, 1038

\bibitem[{{Trifonov} {et~al.}(2018){Trifonov}, {K{\"u}rster}, {Zechmeister},
  {Tal-Or}, {Caballero}, {Quirrenbach}, {Amado}, {Ribas}, {Reiners}, {Reffert},
  {Dreizler}, {Hatzes}, {Kaminski}, {Launhardt}, {Henning}, {Montes},
  {B{\'e}jar}, {Mundt}, {Pavlov}, {Schmitt}, {Seifert}, {Morales}, {Nowak},
  {Jeffers}, {Rodr{\'\i}guez-L{\'o}pez}, {del Burgo}, {Anglada-Escud{\'e}},
  {L{\'o}pez-Santiago}, {Mathar}, {Ammler-von Eiff}, {Guenther}, {Barrado},
  {Gonz{\'a}lez Hern{\'a}ndez}, {Mancini}, {St{\"u}rmer}, {Abril}, {Aceituno},
  {Alonso-Floriano}, {Antona}, {Anwand-Heerwart}, {Arroyo-Torres}, {Azzaro},
  {Baroch}, {Bauer}, {Becerril}, {Ben{\'\i}tez}, {Berdi{\~n}as}, {Bergond},
  {Bl{\"u}mcke}, {Brinkm{\"o}ller}, {Cano}, {C{\'a}rdenas V{\'a}zquez},
  {Casal}, {Cifuentes}, {Claret}, {Colom{\'e}}, {Cort{\'e}s-Contreras},
  {Czesla}, {D{\'\i}ez-Alonso}, {Feiz}, {Fern{\'a}ndez}, {Ferro},
  {Fuhrmeister}, {Galad{\'\i}-Enr{\'\i}quez}, {Garcia-Piquer}, {Garc{\'\i}a
  Vargas}, {Gesa}, {G{\'o}mez Galera}, {Gonz{\'a}lez-Peinado}, {Gr{\"o}zinger},
  {Grohnert}, {Gu{\`a}rdia}, {Guijarro}, {de Guindos}, {Guti{\'e}rrez-Soto},
  {Hagen}, {Hauschildt}, {Hedrosa}, {Helmling}, {Hermelo}, {Hern{\'a}ndez
  Arab{\'\i}}, {Hern{\'a}ndez Casta{\~n}o}, {Hern{\'a}ndez Hernando},
  {Herrero}, {Huber}, {Huke}, {Johnson}, {de Juan}, {Kim}, {Klein},
  {Kl{\"u}ter}, {Klutsch}, {Lafarga}, {Lamp{\'o}n}, {Lara}, {Laun}, {Lemke},
  {Lenzen}, {L{\'o}pez del Fresno}, {L{\'o}pez-Gonz{\'a}lez},
  {L{\'o}pez-Puertas}, {L{\'o}pez Salas}, {Luque}, {Mag{\'a}n Madinabeitia},
  {Mall}, {Mandel}, {Marfil}, {Mar{\'\i}n Molina}, {Maroto Fern{\'a}ndez},
  {Mart{\'\i}n}, {Mart{\'\i}n-Ruiz}, {Marvin}, {Mirabet}, {Moya},
  {Moreno-Raya}, {Nagel}, {Naranjo}, {Nortmann}, {Ofir}, {Oreiro}, {Pall{\'e}},
  {Panduro}, {Pascual}, {Passegger}, {Pedraz}, {P{\'e}rez-Calpena}, {P{\'e}rez
  Medialdea}, {Perger}, {Perryman}, {Pluto}, {Rabaza}, {Ram{\'o}n}, {Rebolo},
  {Redondo}, {Reinhardt}, {Rhode}, {Rix}, {Rodler}, {Rodr{\'\i}guez},
  {Rodr{\'\i}guez Trinidad}, {Rohloff}, {Rosich}, {Sadegi},
  {S{\'a}nchez-Blanco}, {S{\'a}nchez Carrasco}, {S{\'a}nchez-L{\'o}pez},
  {Sanz-Forcada}, {Sarkis}, {Sarmiento}, {Sch{\"a}fer}, {Schiller},
  {Sch{\"o}fer}, {Schweitzer}, {Solano}, {Stahl}, {Strachan}, {Su{\'a}rez},
  {Tabernero}, {Tala}, {Tulloch}, {Veredas}, {Vico Linares}, {Vilardell},
  {Wagner}, {Winkler}, {Wolthoff}, {Xu}, {Yan}, \& {Zapatero
  Osorio}}]{trifonov2018}
{Trifonov}, T., {K{\"u}rster}, M., {Zechmeister}, M., {et~al.} 2018, \aap, 609,
  A117

\bibitem[{{Udry} {et~al.}(2007){Udry}, {Bonfils}, {Delfosse}, {Forveille},
  {Mayor}, {Perrier}, {Bouchy}, {Lovis}, {Pepe}, {Queloz}, \&
  {Bertaux}}]{udry2007}
{Udry}, S., {Bonfils}, X., {Delfosse}, X., {et~al.} 2007, \aap, 469, L43

\bibitem[{{Van Eylen} {et~al.}(2018){Van Eylen}, {Agentoft}, {Lundkvist},
  {Kjeldsen}, {Owen}, {Fulton}, {Petigura}, \& {Snellen}}]{vanEylen2018}
{Van Eylen}, V., {Agentoft}, C., {Lundkvist}, M.~S., {et~al.} 2018, \mnras,
  479, 4786

\bibitem[{{Van Eylen} {et~al.}(2021){Van Eylen}, {Astudillo-Defru}, {Bonfils},
  {Livingston}, {Hirano}, {Luque}, {Lam}, {Justesen}, {Winn}, {Gandolfi},
  {Nowak}, {Palle}, {Albrecht}, {Dai}, {Campos Estrada}, {Owen},
  {Foreman-Mackey}, {Fridlund}, {Korth}, {Mathur}, {Forveille}, {Mikal-Evans},
  {Osborne}, {Ho}, {Almenara}, {Artigau}, {Barrag{\'a}n}, {Barros}, {Bouchy},
  {Cabrera}, {Caldwell}, {Charbonneau}, {Chaturvedi}, {Cochran}, {Csizmadia},
  {Damasso}, {Delfosse}, {De Medeiros}, {D{\'\i}az}, {Doyon}, {Esposito},
  {F{\H{u}}r{\'e}sz}, {Figueira}, {Georgieva}, {Goffo}, {Grziwa}, {Guenther},
  {Hatzes}, {Jenkins}, {Kabath}, {Knudstrup}, {Latham}, {Lavie}, {Lovis},
  {Mennickent}, {Mullally}, {Murgas}, {Narita}, {Pepe}, {Persson}, {Redfield},
  {Ricker}, {Santos}, {Seager}, {Serrano}, {Smith}, {Su{\'a}rez Mascare{\~n}o},
  {Subjak}, {Twicken}, {Udry}, {Vanderspek}, \& {Zapatero
  Osorio}}]{vanEylen2021}
{Van Eylen}, V., {Astudillo-Defru}, N., {Bonfils}, X., {et~al.} 2021, \mnras,
  507, 2154

\bibitem[{{Venturini} {et~al.}(2020){Venturini}, {Guilera}, {Haldemann},
  {Ronco}, \& {Mordasini}}]{venturini2020}
{Venturini}, J., {Guilera}, O.~M., {Haldemann}, J., {Ronco}, M.~P., \&
  {Mordasini}, C. 2020, \aap, 643, L1

\bibitem[{{Watson} {et~al.}(1981){Watson}, {Donahue}, \& {Walker}}]{watson1981}
{Watson}, A.~J., {Donahue}, T.~M., \& {Walker}, J.~C.~G. 1981, \icarus, 48, 150

\bibitem[{{Weiss} {et~al.}(2018){Weiss}, {Marcy}, {Petigura}, {Fulton},
  {Howard}, {Winn}, {Isaacson}, {Morton}, {Hirsch}, {Sinukoff}, {Cumming},
  {Hebb}, \& {Cargile}}]{weiss2018}
{Weiss}, L.~M., {Marcy}, G.~W., {Petigura}, E.~A., {et~al.} 2018, \aj, 155, 48

\bibitem[{{West} {et~al.}(2011){West}, {Morgan}, {Bochanski}, {Andersen},
  {Bell}, {Kowalski}, {Davenport}, {Hawley}, {Schmidt}, {Bernat}, {Hilton},
  {Muirhead}, {Covey}, {Rojas-Ayala}, {Schlawin}, {Gooding}, {Schluns},
  {Dhital}, {Pineda}, \& {Jones}}]{West-2011}
{West}, A.~A., {Morgan}, D.~P., {Bochanski}, J.~J., {et~al.} 2011, \aj, 141, 97

\bibitem[{{Wielen}(1977)}]{Wielen1977}
{Wielen}, R. 1977, \aap, 60, 263

\bibitem[{{Wilson} {et~al.}(2023){Wilson}, {Simpson}, {Collier Cameron}, \&
  {et~al.}}]{wilson2023}
{Wilson}, T.~G., {Simpson}, A.~M., {Collier Cameron}, A., \& {et~al.} 2023,
  submitted to Science

\bibitem[{{Winn}(2010)}]{winn2010}
{Winn}, J.~N. 2010, in Exoplanets, ed. S.~{Seager} (University of Arizona
  Press, Tucson), 55--77

\bibitem[{{Winters} {et~al.}(2022){Winters}, {Cloutier}, {Medina}, {Irwin},
  {Charbonneau}, {Astudillo-Defru}, {Bonfils}, {Howard}, {Isaacson}, {Bean},
  {Seifahrt}, {Teske}, {Eastman}, {Twicken}, {Collins}, {Jensen}, {Quinn},
  {Payne}, {Kristiansen}, {Spencer}, {Vanderburg}, {Zechmeister}, {Weiss},
  {Wang}, {Wang}, {Udry}, {Terentev}, {St{\"u}rmer}, {Stef{\'a}nsson},
  {Shporer}, {Shectman}, {Sefako}, {Schwengeler}, {Schwarz}, {Scarsdale},
  {Rubenzahl}, {Roy}, {Rosenthal}, {Robertson}, {Petigura}, {Pepe},
  {Omohundro}, {Murphy}, {Murgas}, {Mo{\v{c}}nik}, {Montet}, {Mennickent},
  {Mayo}, {Massey}, {Lubin}, {Lovis}, {Lewin}, {Kasper}, {Kane}, {Jenkins},
  {Huber}, {Horne}, {Hill}, {Gorrini}, {Giacalone}, {Fulton}, {Forveille},
  {Figueira}, {Fetherolf}, {Dressing}, {D{\'\i}az}, {Delfosse}, {Dalba}, {Dai},
  {Cort{\'e}s}, {Crossfield}, {Crane}, {Conti}, {Collins}, {Chontos}, {Butler},
  {Brown}, {Brady}, {Behmard}, {Beard}, {Batalha}, \& {Almenara}}]{winters2022}
{Winters}, J.~G., {Cloutier}, R., {Medina}, A.~A., {et~al.} 2022, \aj, 163, 168

\bibitem[{{Wright} {et~al.}(2010){Wright}, {Eisenhardt}, {Mainzer}, {Ressler},
  {Cutri}, {Jarrett}, {Kirkpatrick}, {Padgett}, {McMillan}, {Skrutskie},
  {Stanford}, {Cohen}, {Walker}, {Mather}, {Leisawitz}, {Gautier}, {McLean},
  {Benford}, {Lonsdale}, {Blain}, {Mendez}, {Irace}, {Duval}, {Liu}, {Royer},
  {Heinrichsen}, {Howard}, {Shannon}, {Kendall}, {Walsh}, {Larsen}, {Cardon},
  {Schick}, {Schwalm}, {Abid}, {Fabinsky}, {Naes}, \& {Tsai}}]{Wright2010}
{Wright}, E.~L., {Eisenhardt}, P. R.~M., {Mainzer}, A.~K., {et~al.} 2010, \aj,
  140, 1868

\bibitem[{{Wu}(2019)}]{wu2019}
{Wu}, Y. 2019, \apj, 874, 91

\bibitem[{{Wyatt} {et~al.}(2020){Wyatt}, {Kral}, \& {Sinclair}}]{wyatt2020}
{Wyatt}, M.~C., {Kral}, Q., \& {Sinclair}, C.~A. 2020, \mnras, 491, 782

\bibitem[{{Zacharias} {et~al.}(2012){Zacharias}, {Finch}, {Girard}, {Henden},
  {Bartlett}, {Monet}, \& {Zacharias}}]{zacharias2012}
{Zacharias}, N., {Finch}, C.~T., {Girard}, T.~M., {et~al.} 2012, VizieR Online
  Data Catalog, I/322A

\bibitem[{{Zechmeister} {et~al.}(2018){Zechmeister}, {Reiners}, {Amado},
  {Azzaro}, {Bauer}, {B{\'e}jar}, {Caballero}, {Guenther}, {Hagen}, {Jeffers},
  {Kaminski}, {K{\"u}rster}, {Launhardt}, {Montes}, {Morales}, {Quirrenbach},
  {Reffert}, {Ribas}, {Seifert}, {Tal-Or}, \& {Wolthoff}}]{zechmeister2018}
{Zechmeister}, M., {Reiners}, A., {Amado}, P.~J., {et~al.} 2018, \aap, 609, A12

\bibitem[{{Zechmeister} {et~al.}(2020){Zechmeister}, {Reiners}, {Amado},
  {Azzaro}, {Bauer}, {B{\'e}jar}, {Caballero}, {Guenther}, {Hagen}, {Jeffers},
  {Kaminski}, {K{\"u}rster}, {Launhardt}, {Montes}, {Morales}, {Quirrenbach},
  {Reffert}, {Ribas}, {Seifert}, \& {Tal-Or}}]{Zechmeister20}
{Zechmeister}, M., {Reiners}, A., {Amado}, P.~J., {et~al.} 2020, {SERVAL:
  SpEctrum Radial Velocity AnaLyser}, Astrophysics Source Code Library, record
  ascl:2006.011

\bibitem[{{Zeng} {et~al.}(2019){Zeng}, {Jacobsen}, {Sasselov}, {Petaev},
  {Vanderburg}, {Lopez-Morales}, {Perez-Mercader}, {Mattsson}, {Li}, {Heising},
  {Bonomo}, {Damasso}, {Berger}, {Cao}, {Levi}, \& {Wordsworth}}]{zeng2019}
{Zeng}, L., {Jacobsen}, S.~B., {Sasselov}, D.~D., {et~al.} 2019, Proceedings of
  the National Academy of Science, 116, 9723

\end{thebibliography}

\begin{appendix}

\section{CHEOPS raw light curves}\label{sec:rawCHEOPS}

\begin{minipage}{\textwidth}
\centering
\includegraphics[width=0.495\textwidth]{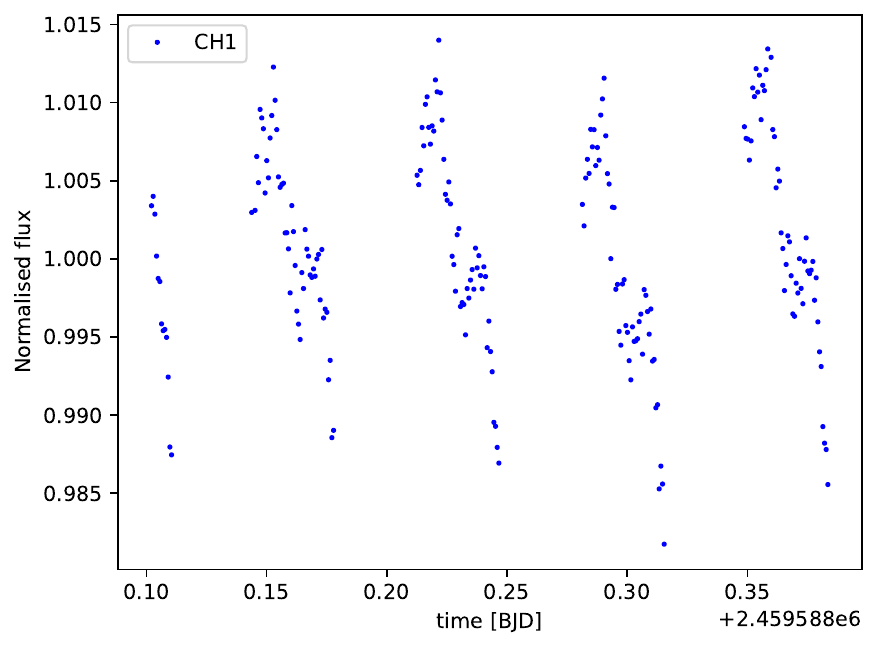}
\includegraphics[width=0.495\textwidth]{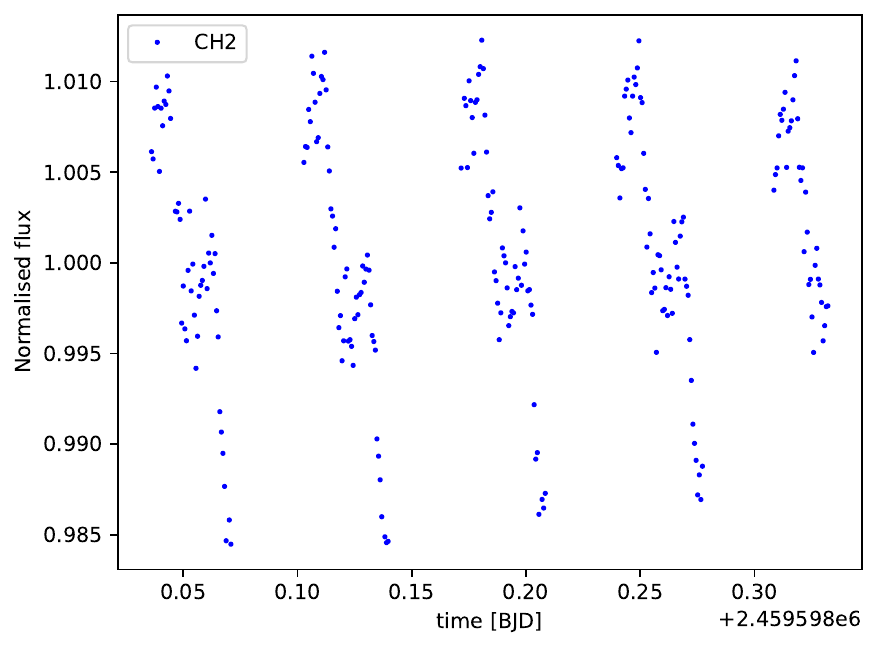} \\
\includegraphics[width=0.495\textwidth]{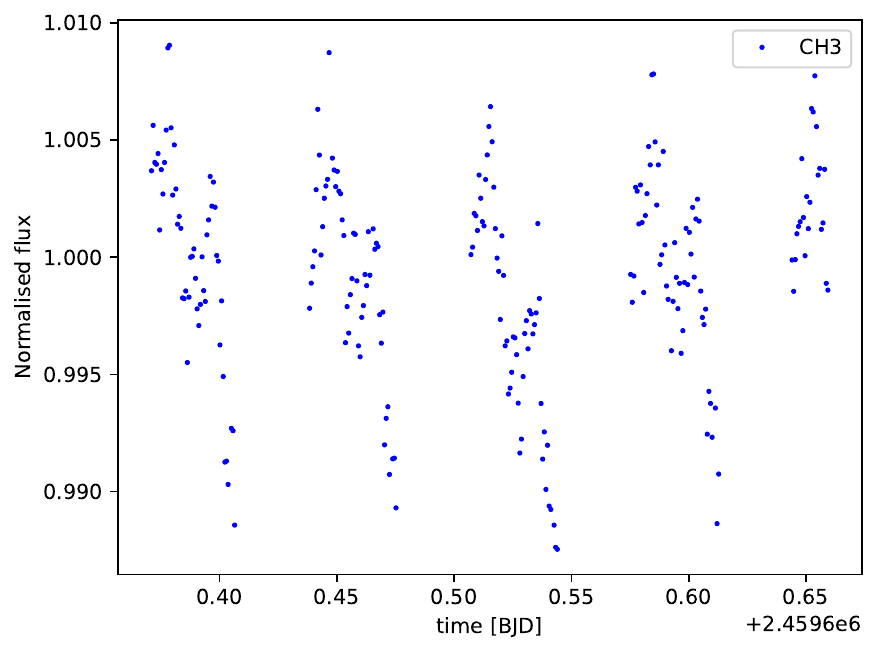}
\includegraphics[width=0.495\textwidth]{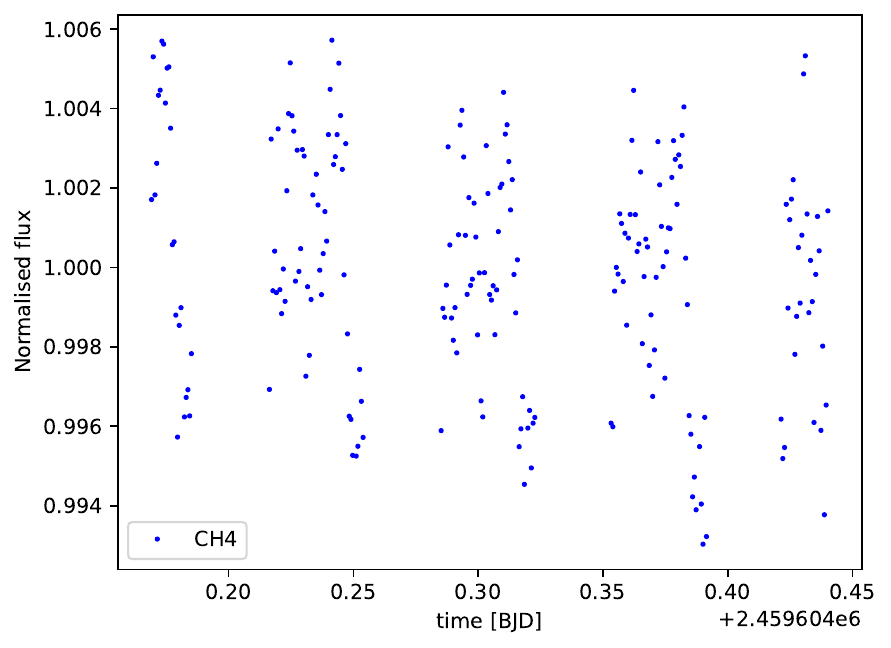} \\
\includegraphics[width=0.495\textwidth]{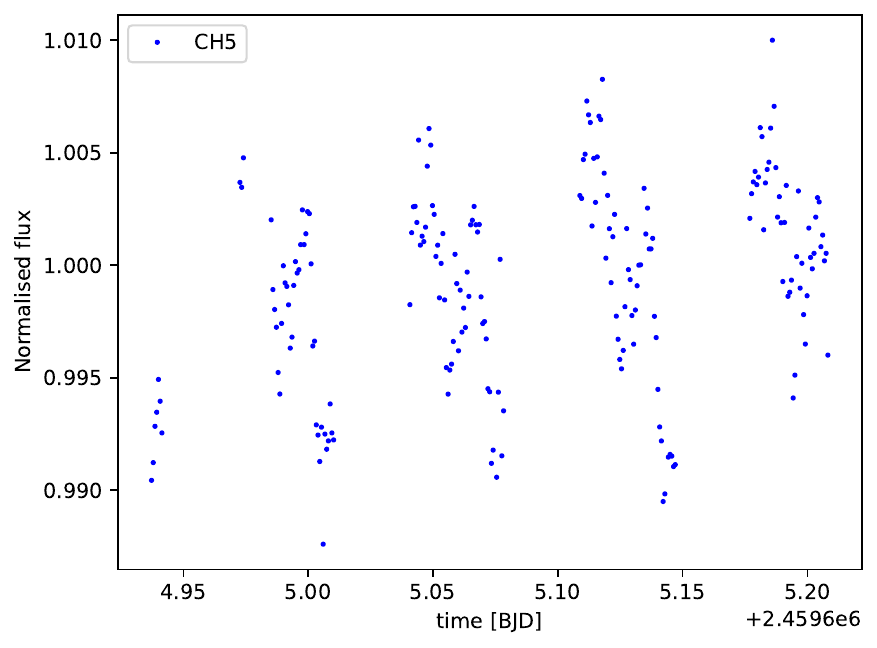}
\includegraphics[width=0.495\textwidth]{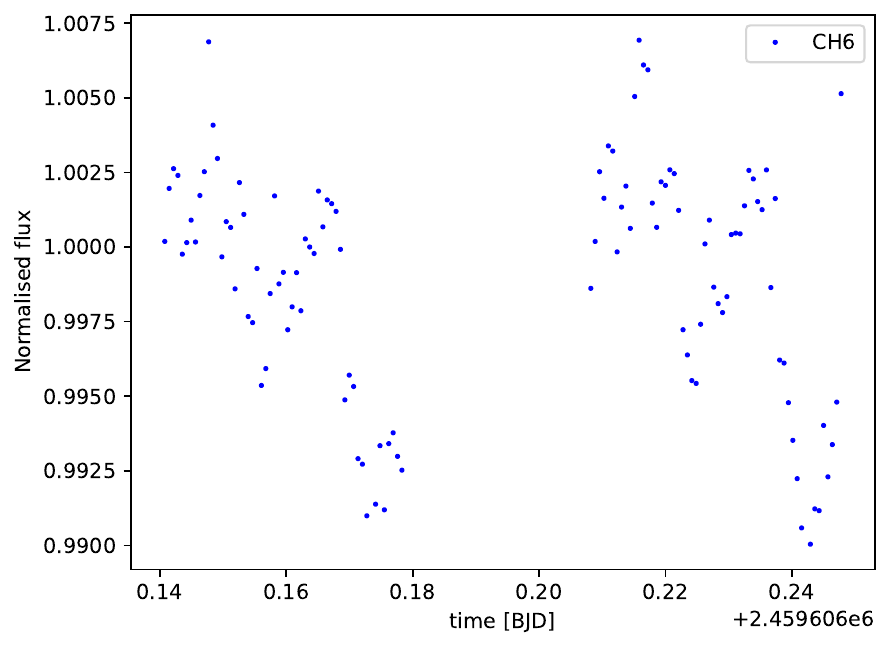}
\captionof{figure}{Raw CHEOPS LCs in chronological order of observations from CH~1 up to CH~6, as presented in Table~\ref{tab:cheopsLog}. The main systematic affecting the LCs is due to the highly variable flux pattern correlating with the spacecraft roll angle.}
\label{fig:rawCheopsLC1-6}
\end{minipage}

\clearpage

\begin{figure*}
\centering
\includegraphics[width=0.495\textwidth]{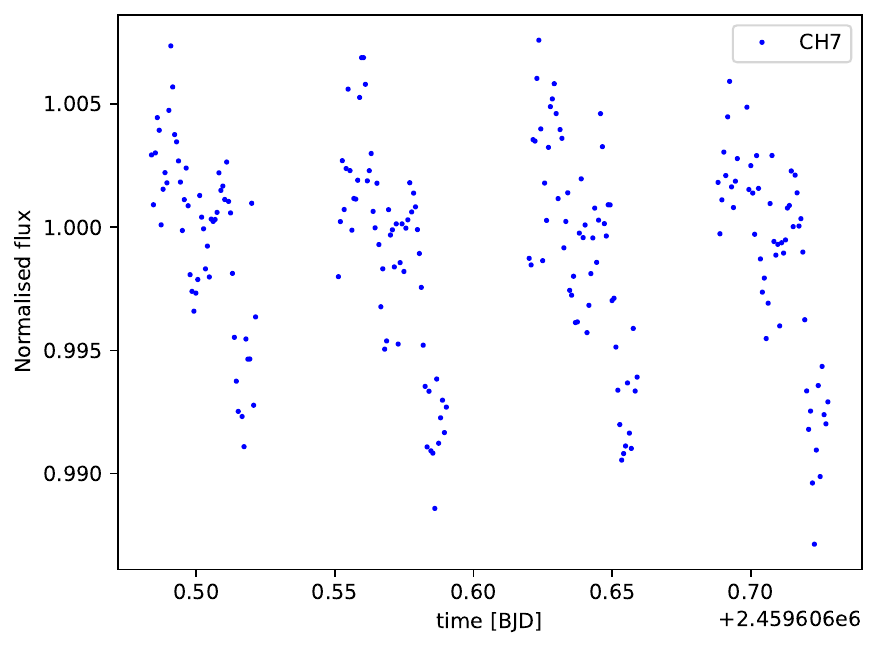}
\includegraphics[width=0.495\textwidth]{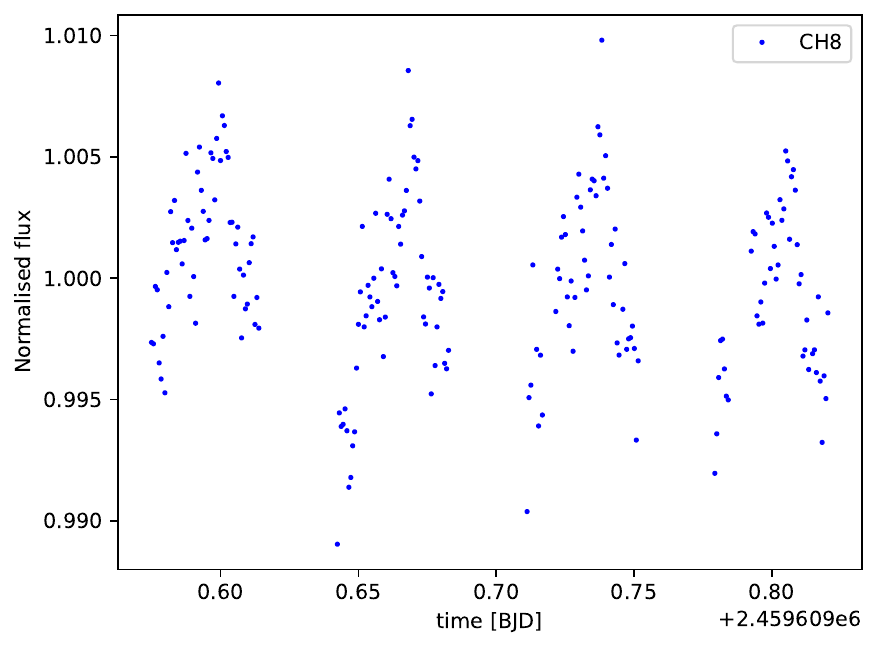} \\
\includegraphics[width=0.495\textwidth]{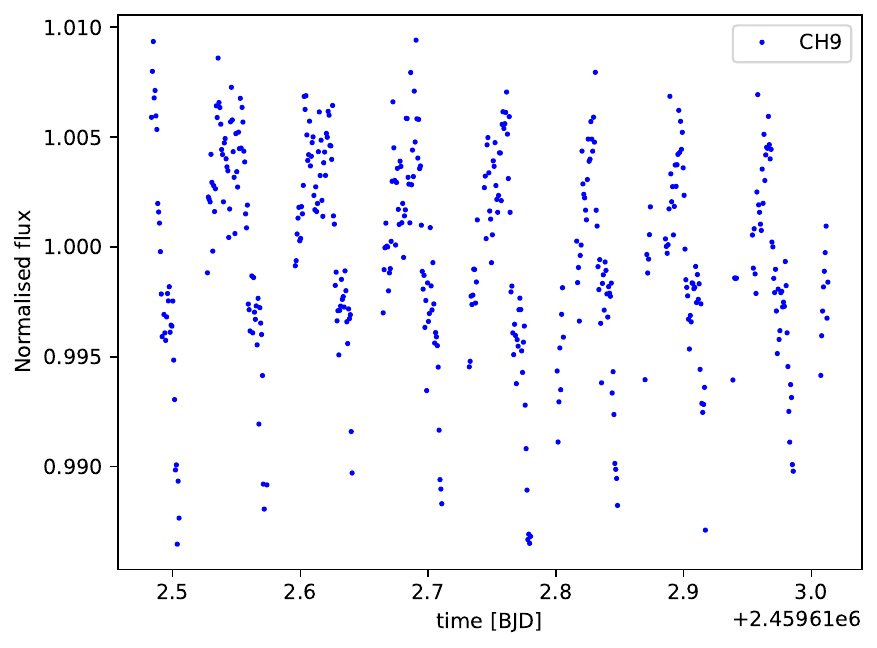}
\includegraphics[width=0.495\textwidth]{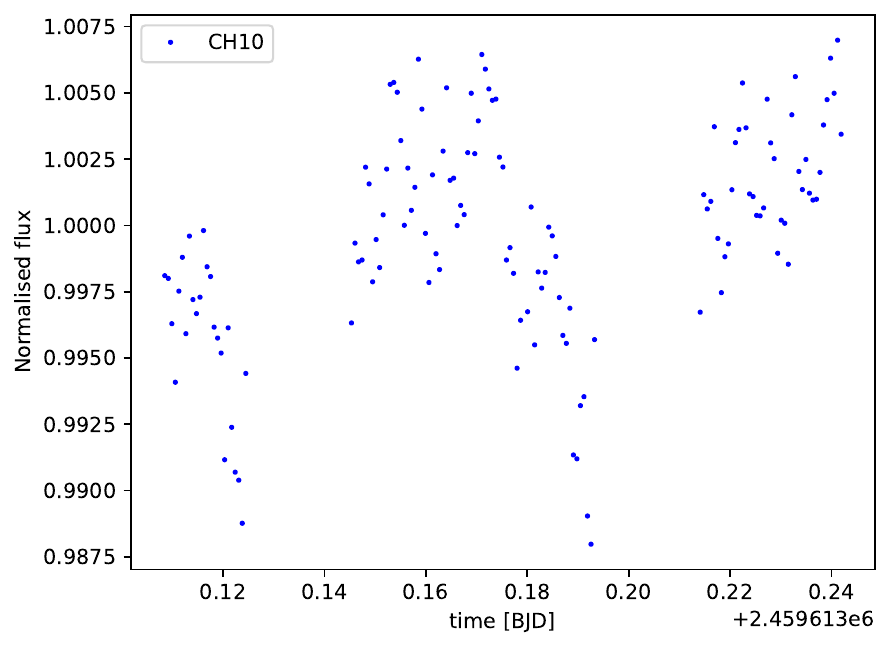} \\
\includegraphics[width=0.495\textwidth]{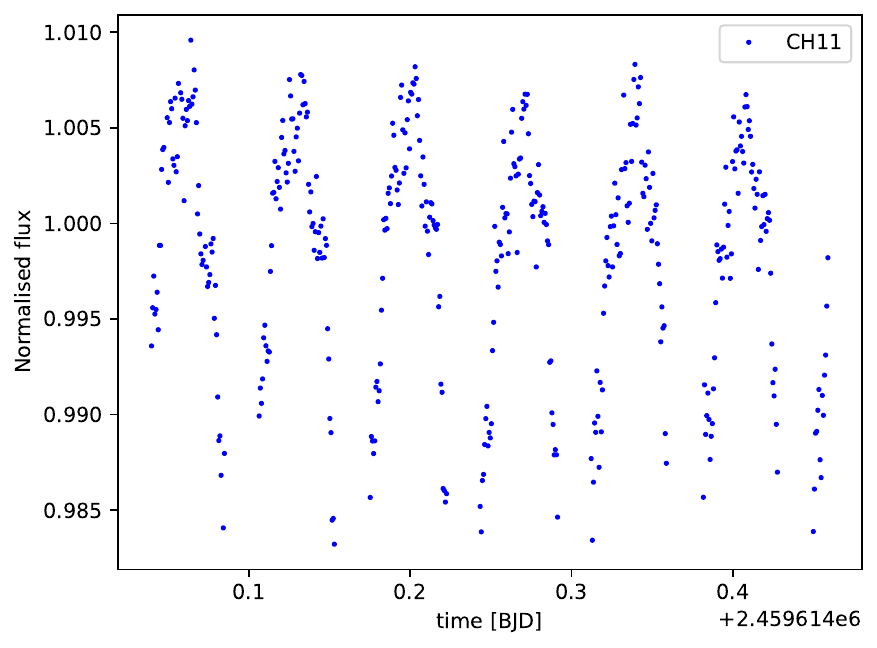}
\includegraphics[width=0.495\textwidth]{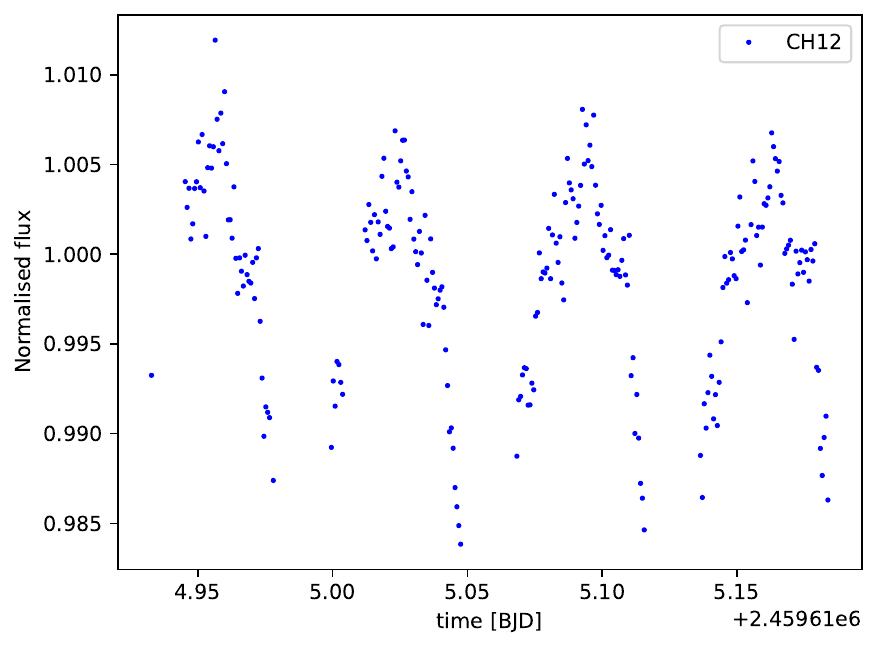}
\caption{Same as Fig.~\ref{fig:rawCheopsLC1-6}, but for CHEOPS LCs from CH~7 up to CH~12.}
\label{fig:rawCheopsLC7-12}
\end{figure*}

\clearpage

\begin{figure*}
\centering
\includegraphics[width=0.495\textwidth]{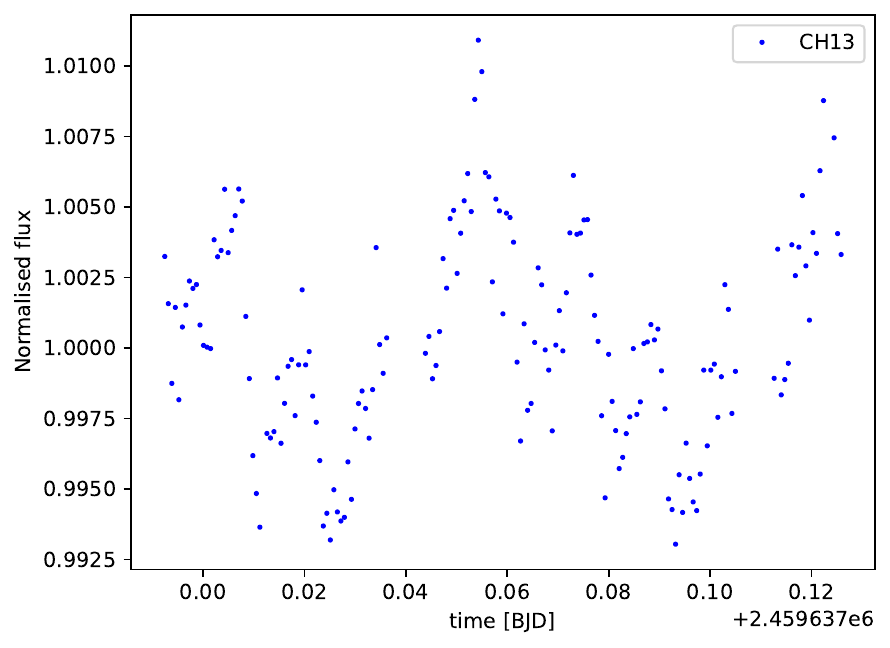}
\includegraphics[width=0.495\textwidth]{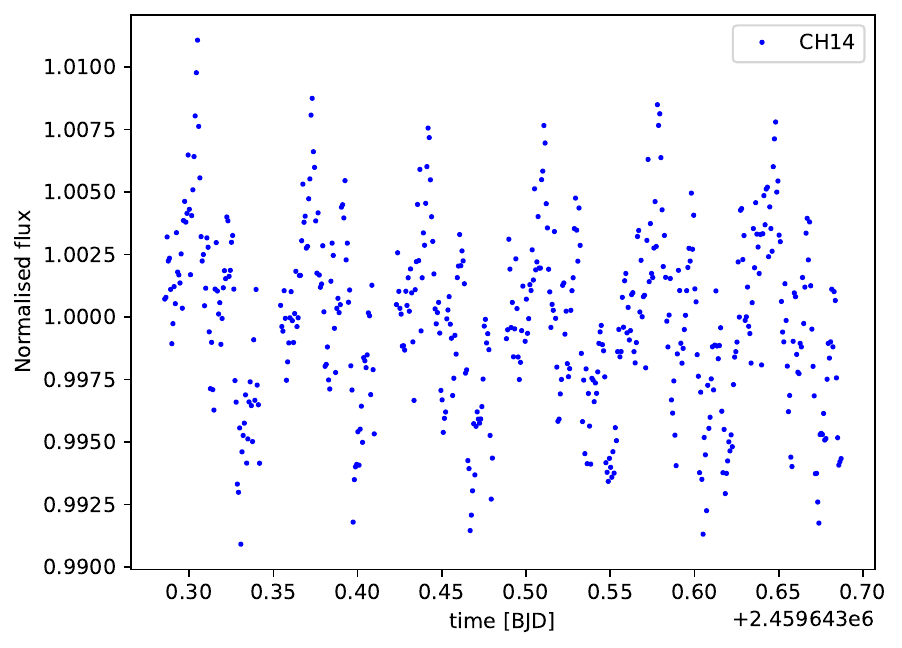} \\
\includegraphics[width=0.495\textwidth]{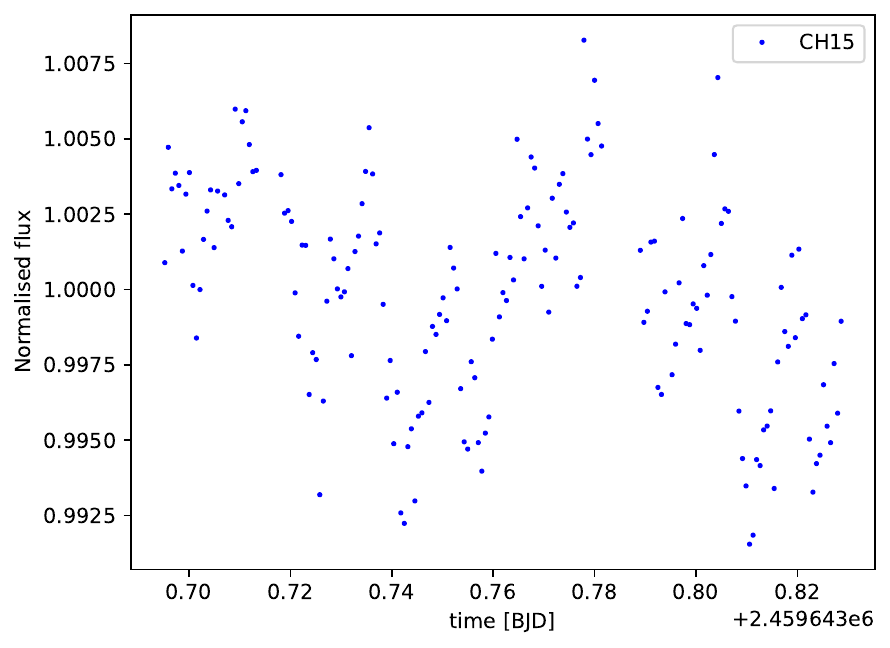}
\includegraphics[width=0.495\textwidth]{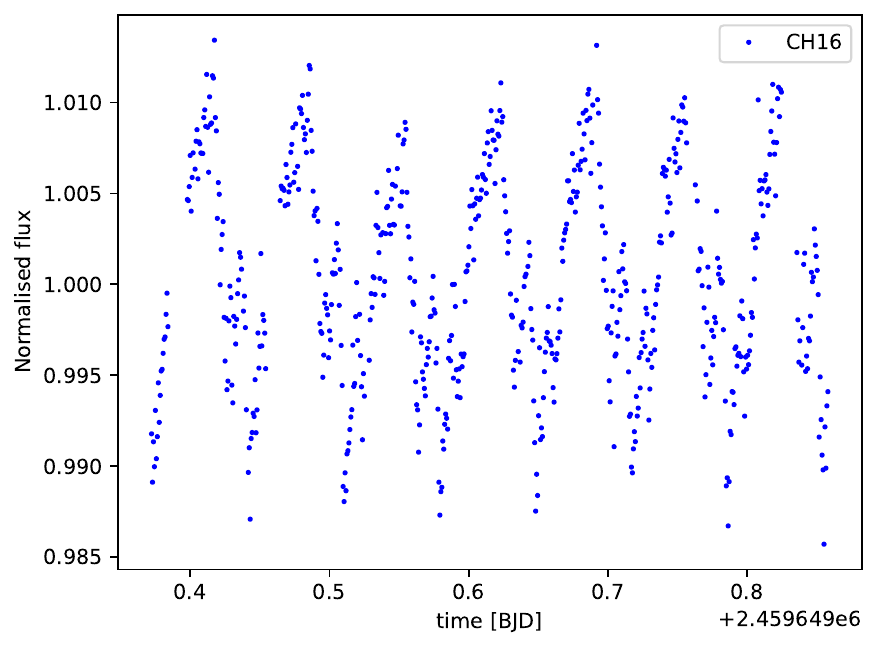} \\
\includegraphics[width=0.495\textwidth]{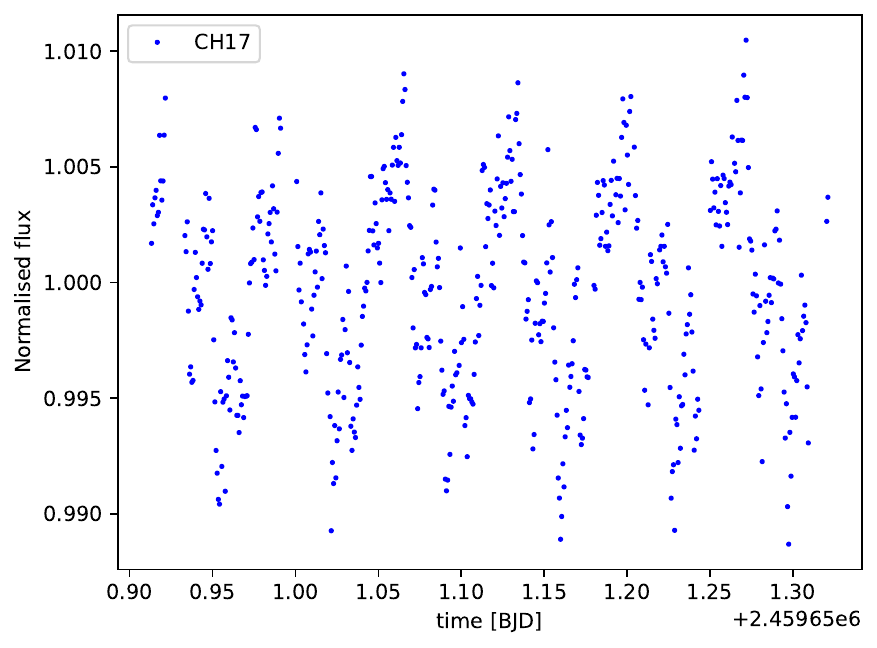}
\includegraphics[width=0.495\textwidth]{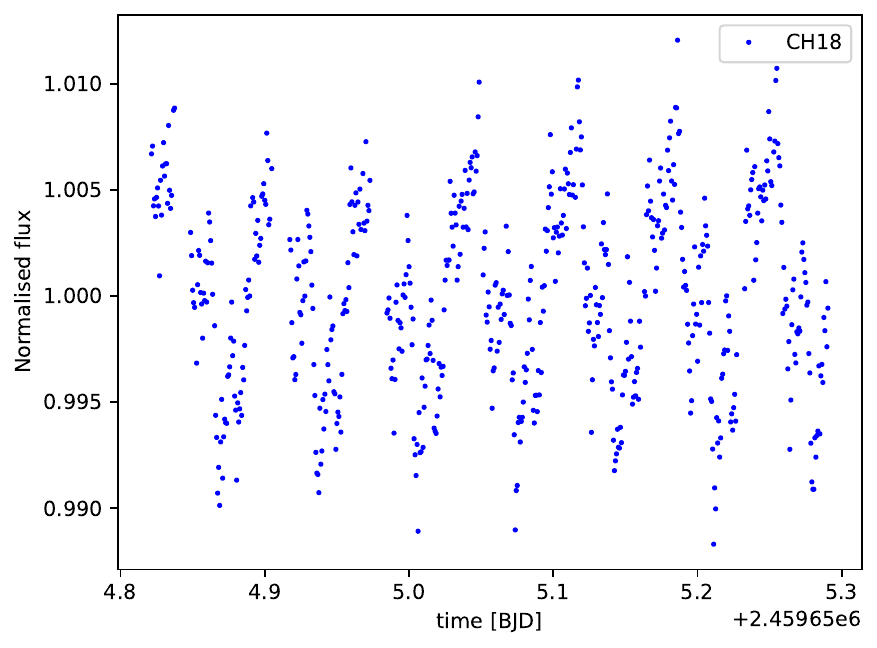}
\caption{Same as Fig.~\ref{fig:rawCheopsLC1-6}, but for CHEOPS LCs from CH~13 up to CH~18.}
\label{fig:rawCheopsLC13-18}
\end{figure*}

\clearpage

\begin{figure*}
\centering
\includegraphics[width=0.495\textwidth]{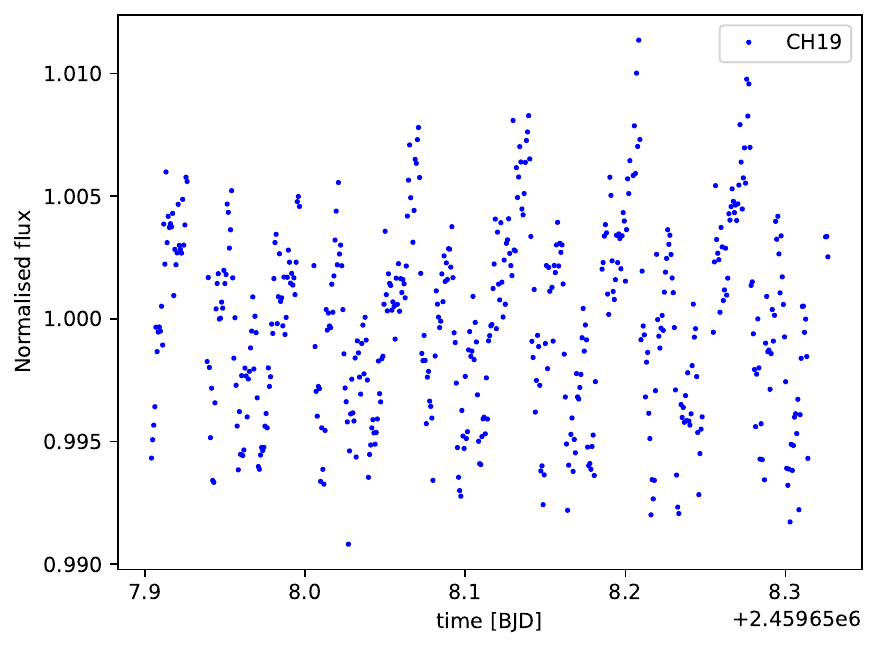}
\includegraphics[width=0.495\textwidth]{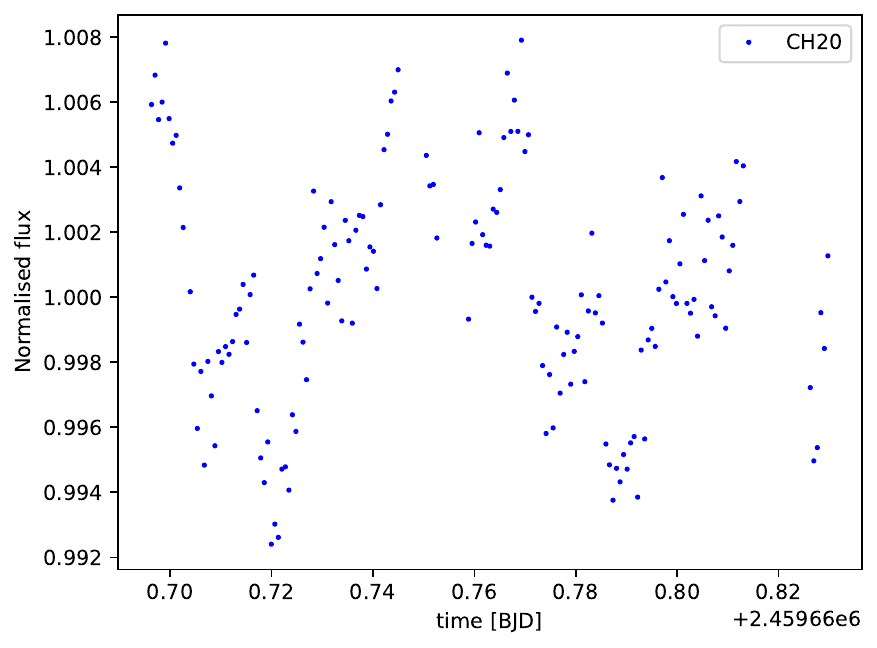} \\
\includegraphics[width=0.495\textwidth]{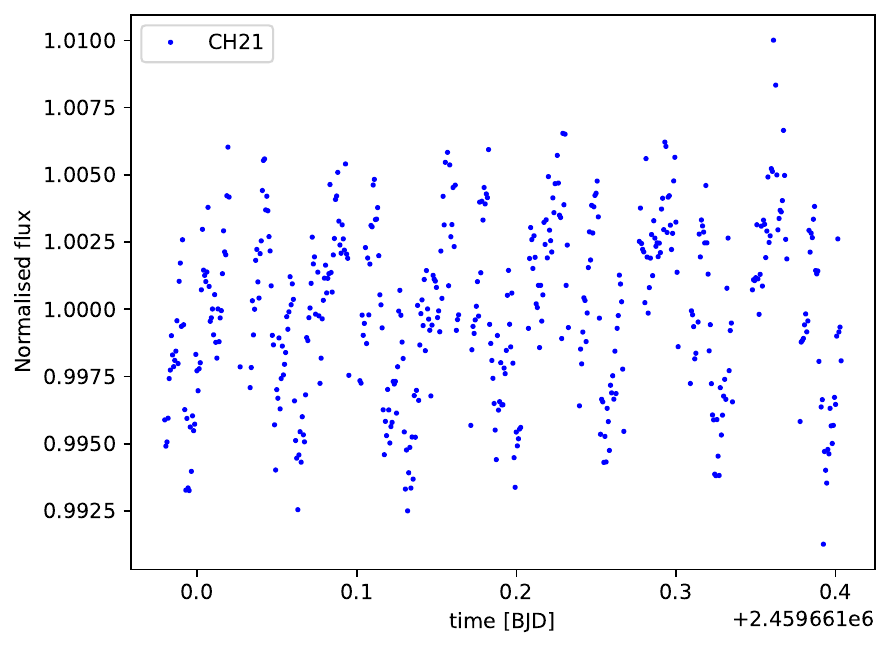}
\includegraphics[width=0.495\textwidth]{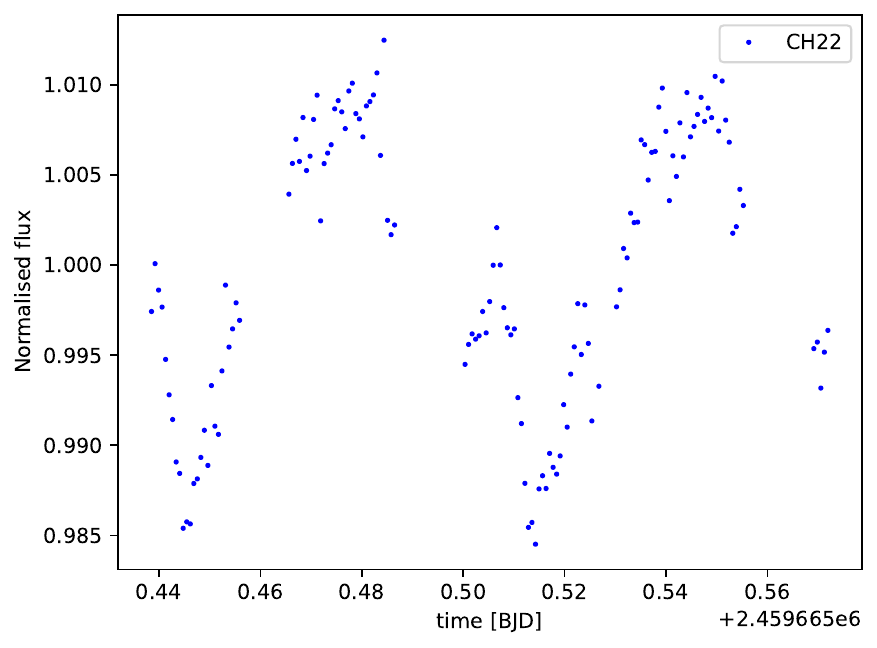} \\
\includegraphics[width=0.495\textwidth]{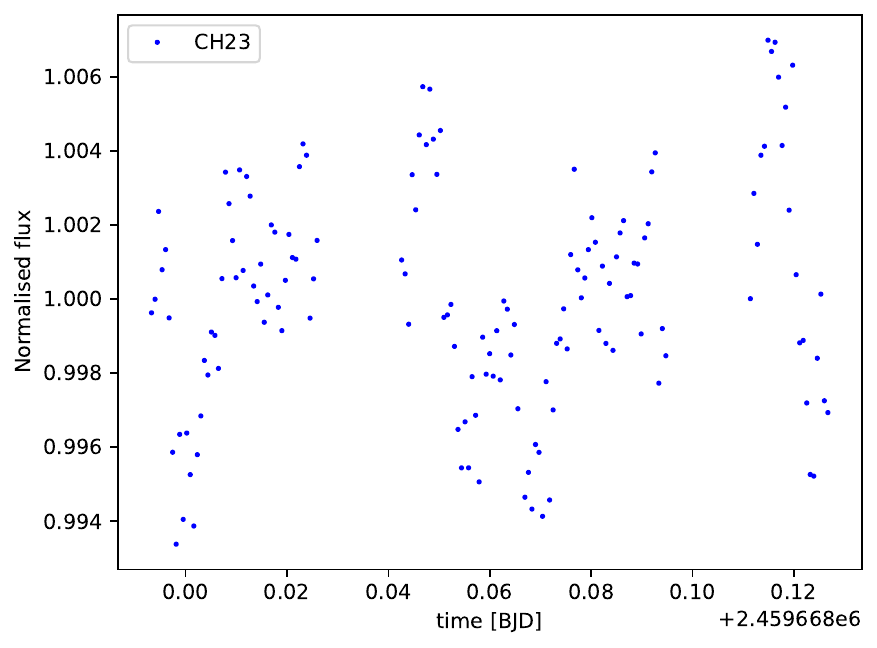}
\includegraphics[width=0.495\textwidth]{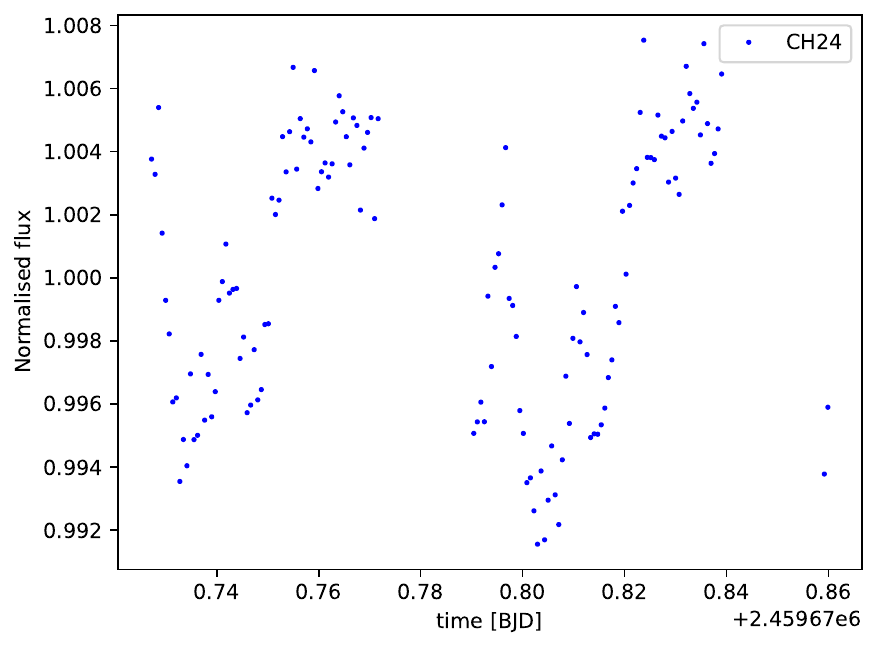}
\caption{Same as Fig.~\ref{fig:rawCheopsLC1-6}, but for CHEOPS LCs from CH~19 up to CH~24.}
\label{fig:rawCheopsLC19-24}
\end{figure*}

\clearpage

\begin{figure}
\centering
\includegraphics[width=\columnwidth]{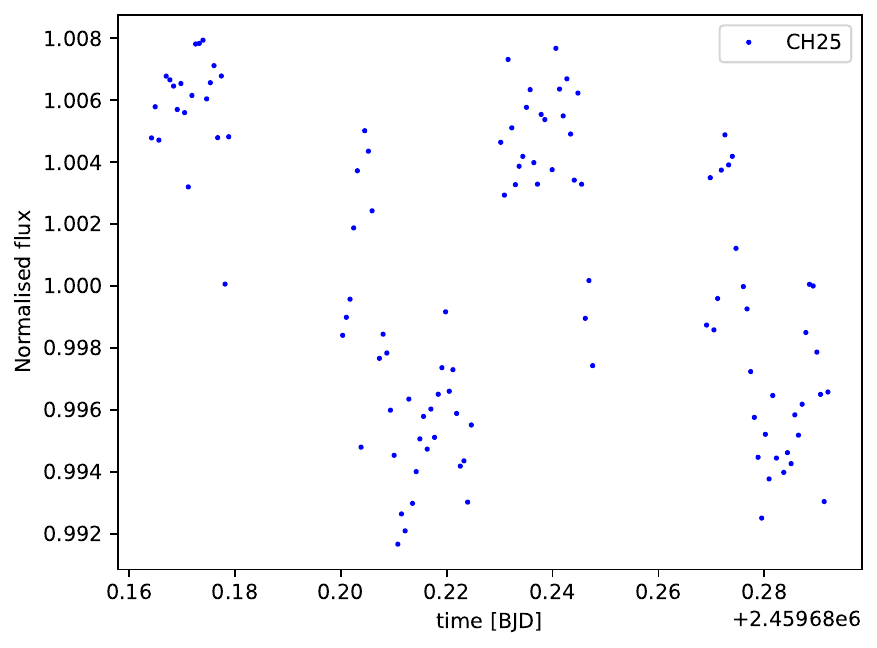}
\caption{Same as Fig.~\ref{fig:rawCheopsLC1-6}, but for CHEOPS LCs CH~25.}
\label{fig:rawCheopsLC25}
\end{figure}

\newpage
\vfill\null
\section{Ground-based facilities light curves}\label{sec:groundLC}

\begin{minipage}{\textwidth}
\centering
\includegraphics[height=0.238\textheight]{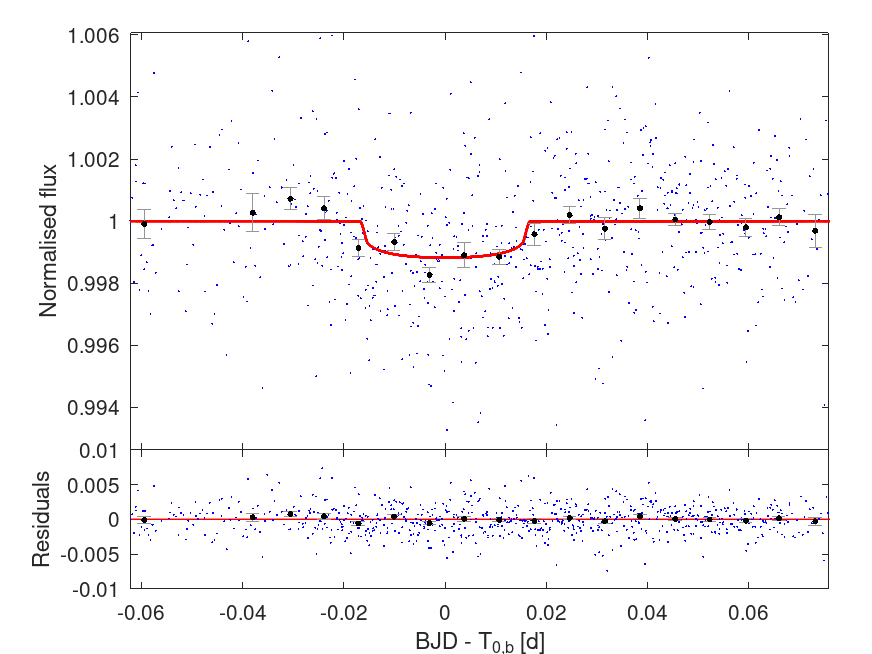}
\includegraphics[height=0.238\textheight]{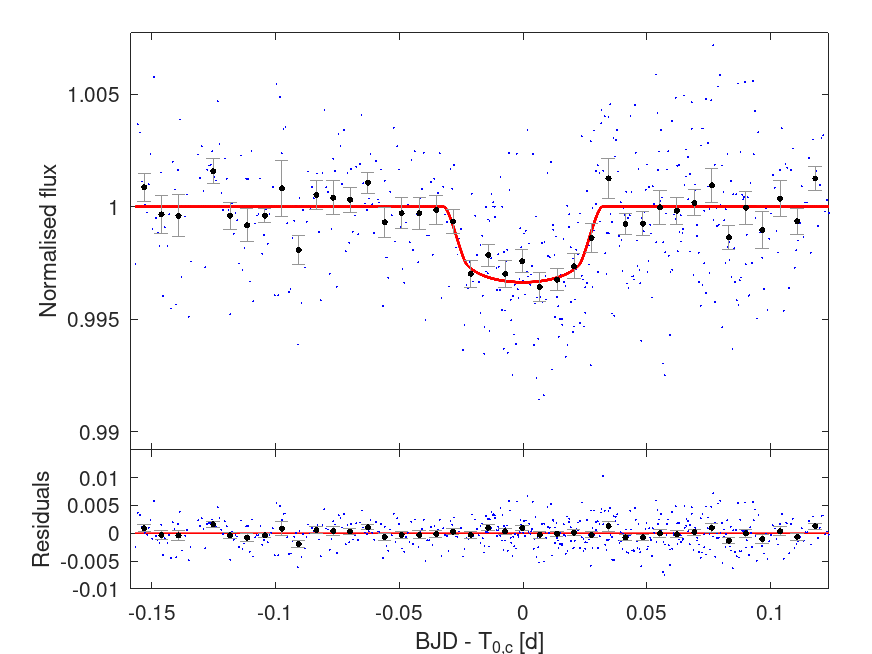} \\
\includegraphics[height=0.238\textheight]{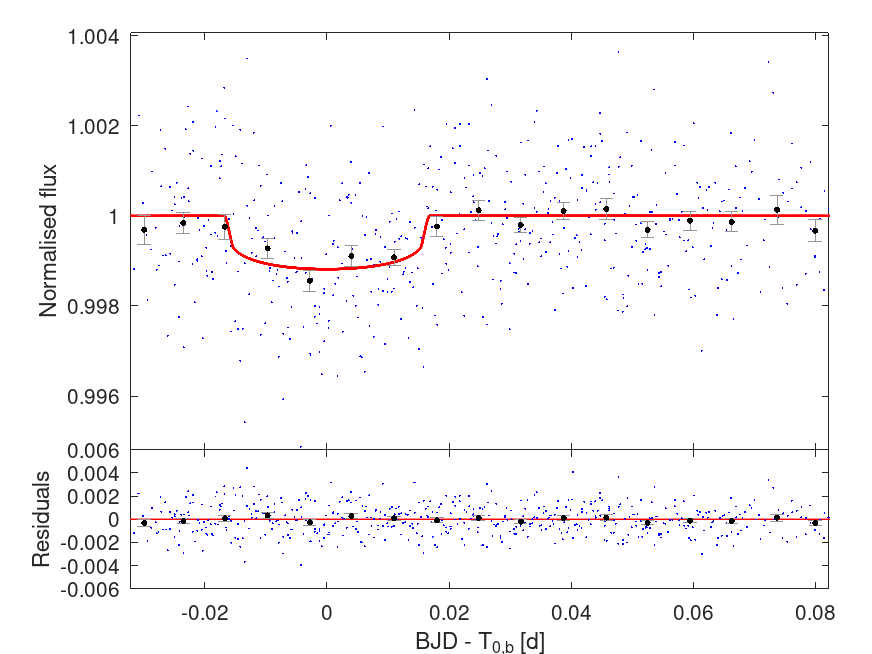}
\includegraphics[height=0.238\textheight]{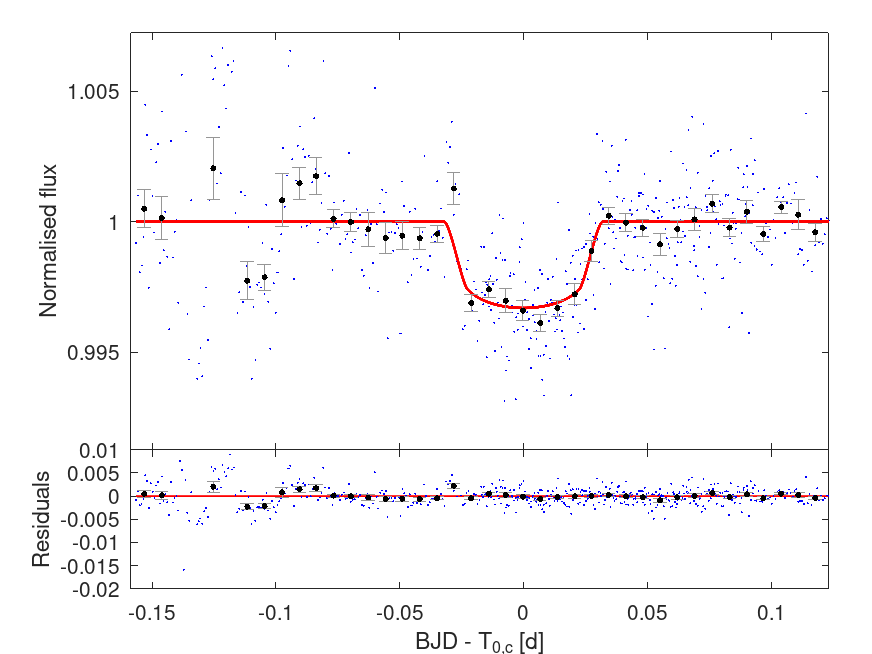} \\
\includegraphics[height=0.238\textheight]{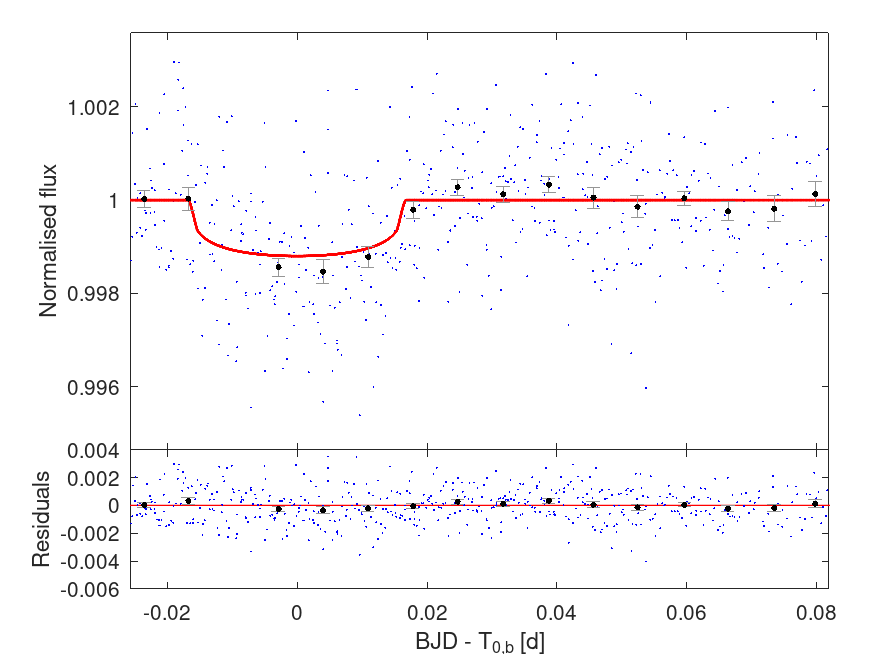}
\includegraphics[height=0.238\textheight]{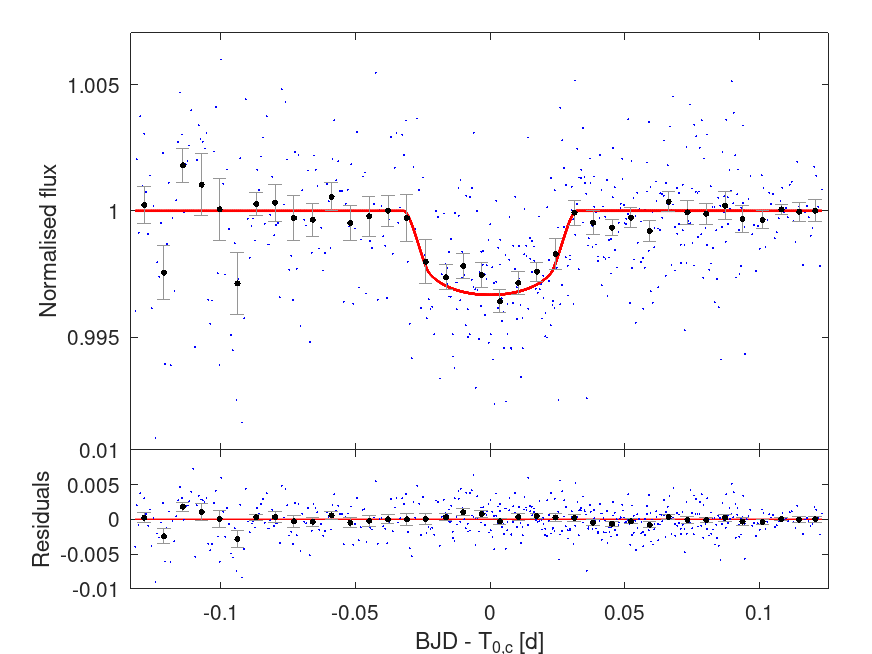} \\
\includegraphics[height=0.238\textheight]{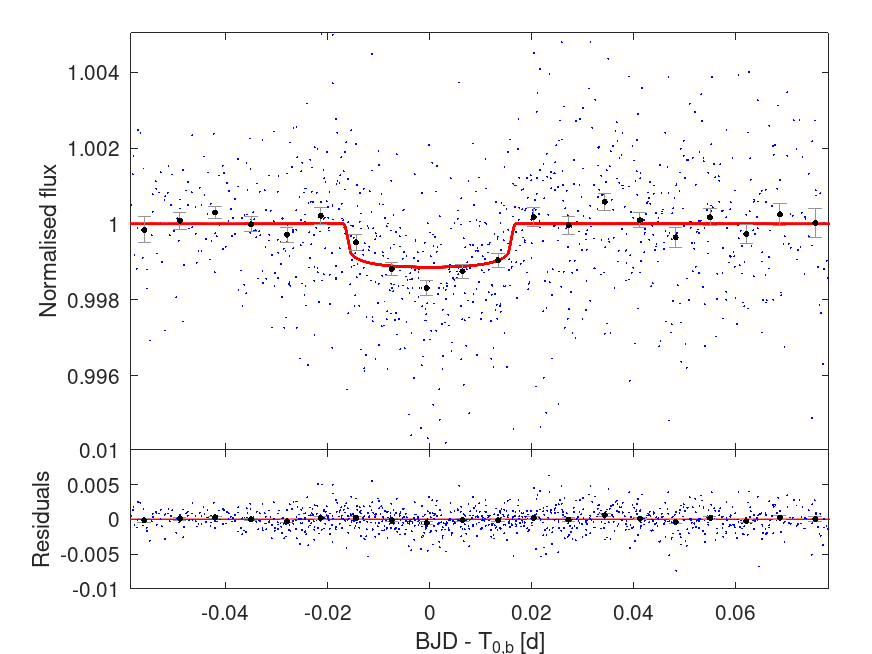}
\includegraphics[height=0.238\textheight]{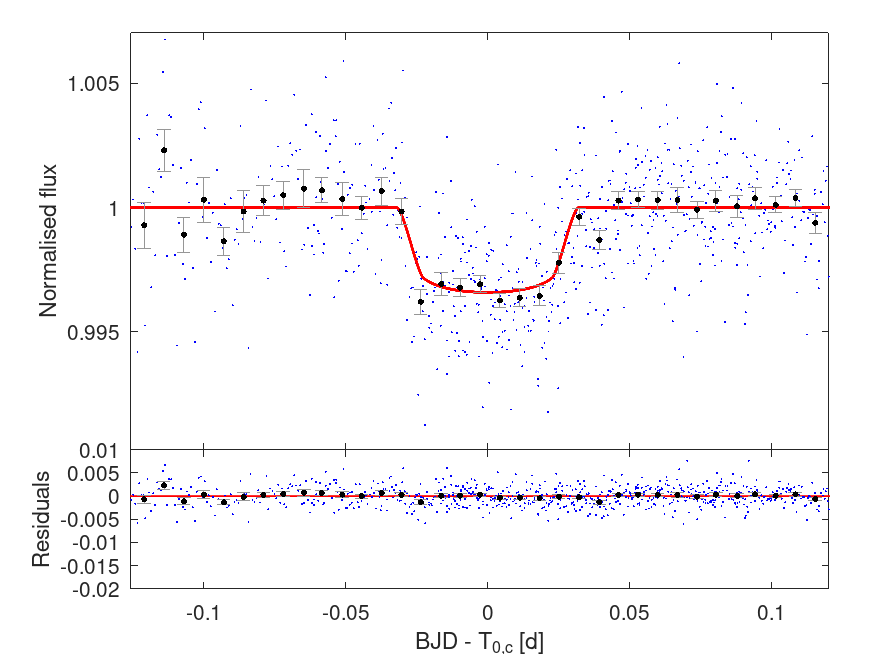}
\captionof{figure}{Phase-folded LCs of TOI-732\,b (first column) and TOI-732\,c (second column) observed by ground-based facilities in the following filters: g' (first row), r' (second row), i' (third row), and z' (fourth row).}
\label{fig:grizFilters}
\end{minipage}

\begin{figure*}
\centering
\includegraphics[width=0.49\textwidth]{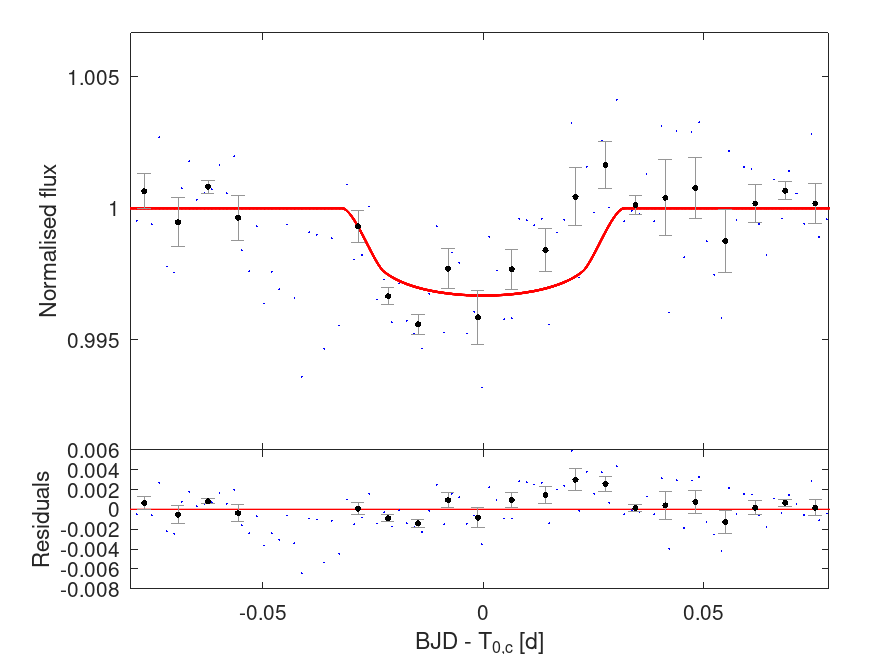}
\includegraphics[width=0.49\textwidth]{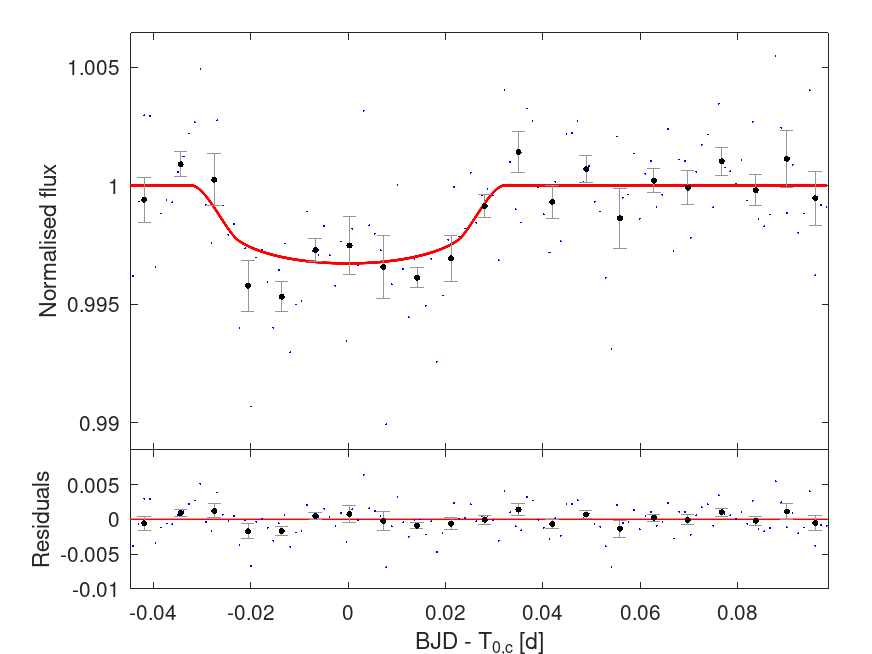} \\
\includegraphics[width=0.49\textwidth]{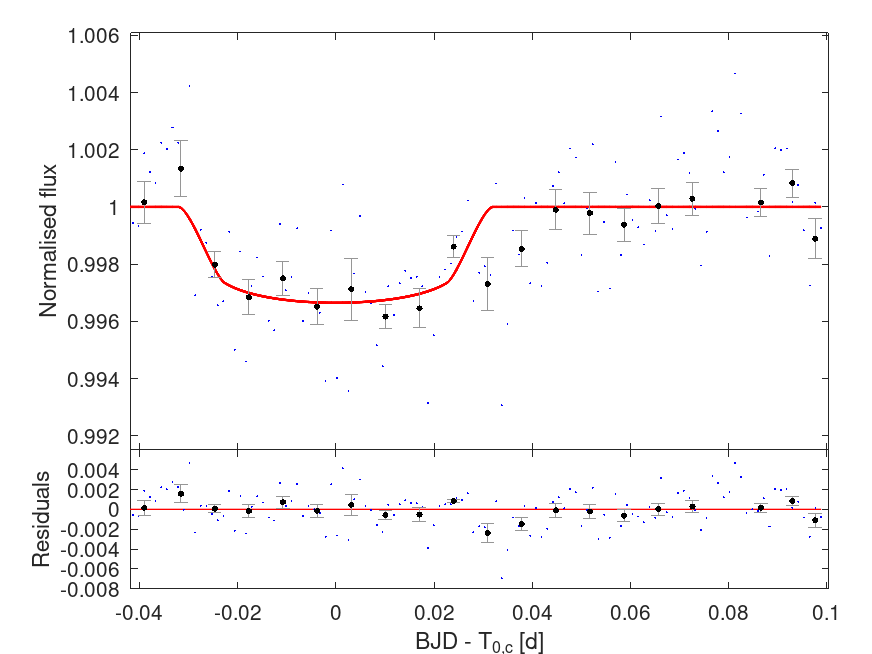}
\includegraphics[width=0.49\textwidth]{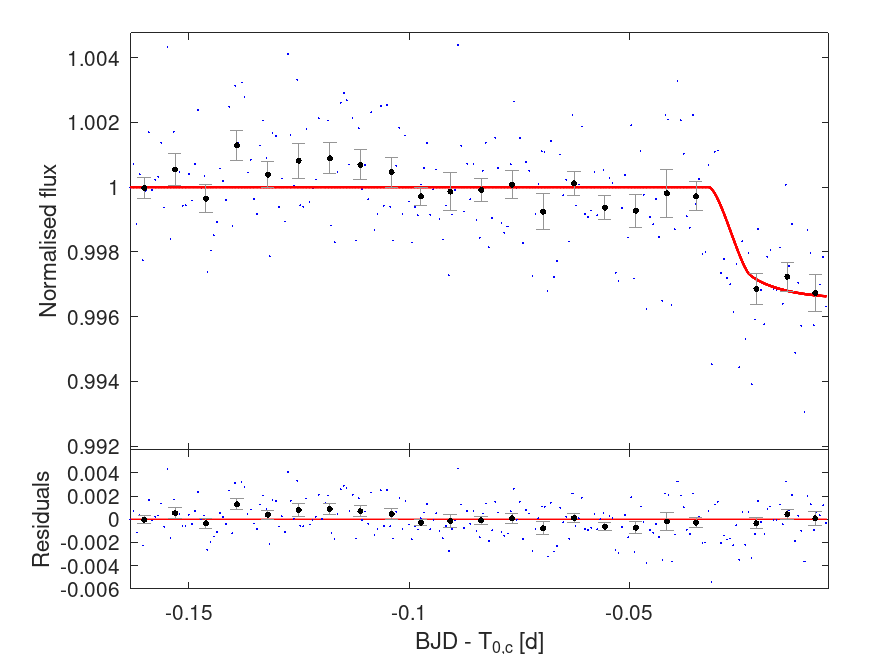} \\
\caption{LCs of TOI-732\,c taken by ground-based facilities. From \textit{Top} to \textit{Bottom} going row wise the observation filters are B, V, R, and RG715.}
\label{fig:BVRMeFiltersPlaC}
\end{figure*}

\begin{figure}
\centering
\includegraphics[width=\columnwidth]{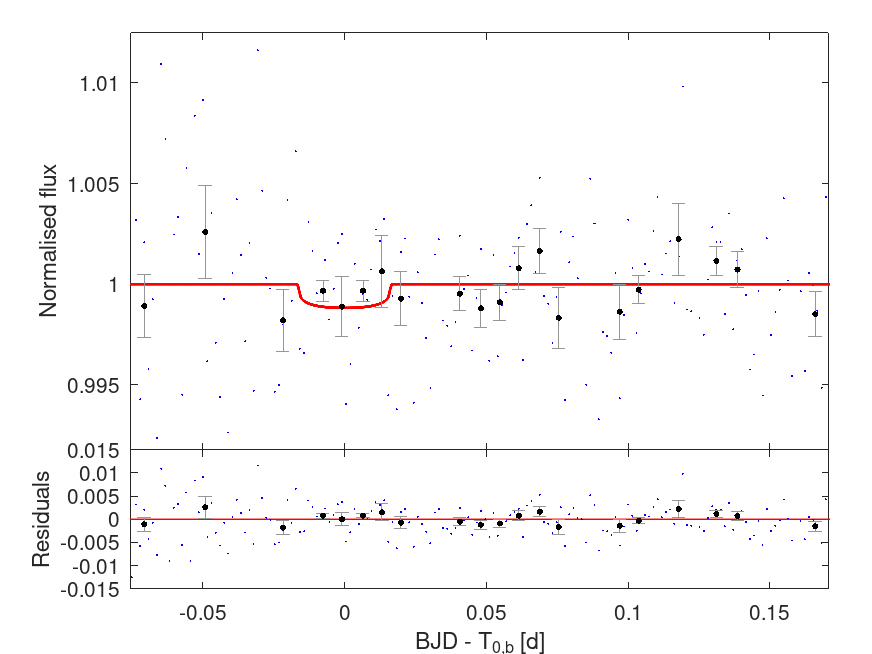}
\caption{LC of TOI-732\,b as observed by OAA in the I filter.}
\label{fig:IfilterPlaB}
\end{figure}

\clearpage

\section{Additional tables and figures}
\begin{table}[h!]
\renewcommand{\arraystretch}{1.2}
\caption{Quadratic limb darkening (LD) coefficients $(u_1,u_2)$ for each photometric filter.}
\label{tab:LD}
\centering
\begin{tabular}{l c c}
\hline\hline
LD & Prior & Posterior \\
\hline
\noalign{\smallskip}
CHEOPS $u_1$ & $\mathcal{N}(0.311,0.054)$ & $0.330_{-0.054}^{+0.053}$ \\
CHEOPS $u_2$ & $\mathcal{N}(0.383,0.041)$ & $0.387\pm0.042$ \\
TESS $u_1$   & $\mathcal{N}(0.208,0.042)$ & $0.217\pm0.043$ \\
TESS $u_2$   & $\mathcal{N}(0.415,0.030)$ & $0.418\pm0.032$ \\
g' $u_1$     & $\mathcal{N}(0.408,0.047)$ & $0.398_{-0.047}^{+0.048}$ \\
g' $u_2$     & $\mathcal{N}(0.386,0.029)$ & $0.382\pm0.031$ \\
r' $u_1$     & $\mathcal{N}(0.444,0.078)$ & $0.434_{-0.077}^{+0.078}$ \\
r' $u_2$     & $\mathcal{N}(0.326,0.061)$ & $0.324_{-0.066}^{+0.065}$ \\
i' $u_1$     & $\mathcal{N}(0.310,0.048)$ & $0.320_{-0.052}^{+0.051}$ \\
i' $u_2$     & $\mathcal{N}(0.346,0.043)$ & $0.345_{-0.045}^{+0.044}$ \\
z' $u_1$     & $\mathcal{N}(0.162,0.043)$ & $0.153_{-0.044}^{+0.043}$ \\
z' $u_2$     & $\mathcal{N}(0.439,0.028)$ & $0.437\pm0.030$ \\
B $u_1$      & $\mathcal{N}(0.399,0.034)$ & $0.400\pm0.034$ \\
B $u_2$      & $\mathcal{N}(0.401,0.018)$ & $0.401_{-0.020}^{+0.019}$ \\
V $u_1$      & $\mathcal{N}(0.400,0.063)$ & $0.393_{-0.064}^{+0.066}$ \\
V $u_2$      & $\mathcal{N}(0.386,0.041)$ & $0.385\pm0.045$ \\
R $u_1$      & $\mathcal{N}(0.413,0.070)$ & $0.412_{-0.075}^{+0.072}$ \\
R $u_2$      & $\mathcal{N}(0.325,0.055)$ & $0.322_{-0.060}^{+0.059}$ \\
I $u_1$      & $\mathcal{N}(0.270,0.052)$ & $0.271\pm0.054$ \\
I $u_2$      & $\mathcal{N}(0.379,0.042)$ & $0.377_{-0.044}^{+0.045}$ \\
RG715 $u_1$  & $\mathcal{N}(0.193,0.040)$ & $0.193\pm0.042$ \\
RG715 $u_2$  & $\mathcal{N}(0.427,0.028)$ & $0.426\pm0.030$ \\
\hline
\end{tabular}
\tablefoot{$\mathcal{N}(\mu,\sigma)$ denotes a Normal prior with mean $\mu$ and standard deviation $\sigma$.}
\end{table}

\begin{table*}
\caption{Polynomial detrending baselines applied to the space-based light curves within the MCMC scheme.}
\label{tab:LCdetrending}
\centering
\begin{minipage}{0.49\textwidth}
\begin{tabular}{l c c}
\hline\hline
Time series & Planet & Detrending model \\
\hline
\noalign{\smallskip}
  CH 1 & b c & \ttGProll{} + \ttt$^3$ + \ttxy$^2$ \\
  CH 2 & b & \ttGProll{} + \ttxy$^2$ \\
  CH 3 & b c & \ttGProll{} + \ttsme$^1$ + \ttxy$^1$ \\
  CH 4 & b & \ttGProll{} + \ttt$^1$ + \ttxy$^1$ \\
  CH 5 & b & \ttGProll{} + \ttxy$^2$ \\
  CH 6 & 0 & \ttGProll{} + \ttsky$^1$ \\
  CH 7 & b & \ttGProll{} + \ttxy$^2$ \\
  CH 8 & b & \ttGProll{} +\ttt$^1$ + \ttsky$^1$ + \ttxy$^2$ \\
  CH 9 & b c & \ttGProll{} + \ttt$^1$ + \ttxy$^2$ \\
  CH 10 & 0 & \ttGProll{} + \ttxy$^1$ \\
  CH 11 & b & \ttGProll{} + \ttt$^4$ + \ttxy$^2$ \\
  CH 12 & b & \ttGProll{} + \ttt$^1$ \\
  CH 13 & 0 & \ttGProll{} + \ttxy$^2$ \\
  CH 14 & b & \ttGProll{} +\ttt$^2$ + \ttxy$^2$ \\
  CH 15 & 0 & \ttGProll{} + \ttxy$^2$ \\
  CH 16 & b c & \ttGProll{} + \ttxy$^2$ \\
  CH 17 & b & \ttGProll{} + \ttxy$^2$ \\
  CH 18 & b & \ttGProll{} + \ttxy$^2$ \\
  CH 19 & b & \ttGProll{} \\
  CH 20 & 0 & \ttGProll{} + \ttt$^2$ \\
  CH 21 & b & \ttGProll{} \\
  CH 22 & 0 & \ttGProll{} \\
  CH 23 & b & \ttGProll{} \\
  CH 24 & 0 & \ttGProll{} + \ttt$^2$ \\
  CH 25 & 0 & \ttGProll{} + \ttt$^2$ \\
  TE 1 & b & \ttdx$^1$ \\
  TE 2 & b & \ttdx$^1$ \\
  TE 3 & b & \ttdx$^1$ \\
  TE 4 & b c & \ttdx$^1$ \\
  TE 5 & b & \ttdx$^1$ \\
  TE 6 & b & $c$ \\
  TE 7 & b & \ttdx$^1$ \\
  TE 8 & b & \ttdx$^1$ \\
  TE 9 & b & \ttdx$^1$ \\
  TE 10 & b & $c$ \\
  TE 11 & b & $c$ \\
  TE 12 & b & $c$ \\
  TE 13 & b & $c$ \\
  TE 14 & b & \ttdx$^1$ \\
  TE 15 & b & \ttt$^1$ \\
  TE 16 & b & $c$ \\
  TE 17 & b c & \ttdy$^1$ \\
  TE 18 & b & \ttdx$^1$ \\
  TE 19 & b & $c$ \\
  TE 20 & b & $c$ \\
  TE 21 & b & $c$ \\
  TE 22 & b & $c$ \\
  TE 23 & b & $c$ \\
  TE 24 & b & $c$ \\
  TE 25 & b & $c$ \\
  TE 26 & b & \ttdx$^1$ \\
  TE 27 & b & $c$ \\
  TE 28 & b & $c$ \\
\hline
\end{tabular}
\end{minipage}
\begin{minipage}{0.49\textwidth}
\begin{tabular}{l c c}
\hline\hline
Time series & Planet & Detrending model \\
\hline
\noalign{\smallskip}
  TE 29 & b & \ttdx$^1$ + \ttxy$^1$ \\
  TE 30 & b & \ttdx$^1$ + \ttxy$^1$ \\
  TE 31 & b c & \ttt$^1$ + \ttdx$^1$ + \ttxy$^1$ \\
  TE 32 & b & \ttt$^1$ + \ttdx$^1$ \\
  TE 33 & b & \ttt$^3$ + \ttdx$^1$ \\
  TE 34 & b & \ttt$^3$ + \ttdx$^1$ + \ttdy$^1$ \\
  TE 35 & b & \ttdx$^1$ \\
  TE 36 & b & \ttdx$^1$ \\
  TE 37 & b & \ttt$^1$ + \ttdx$^1$ + \ttxy$^1$ \\
  TE 38 & b & \ttt$^1$ + \ttdx$^2$ \\
  TE 39 & b & \ttdx$^1$ \\
  TE 40 & b & \ttdx$^1$ + \ttxy$^1$ \\
  TE 41 & b & \ttdx$^1$ \\
  TE 42 & b & \ttt$^2$ + \ttdx$^3$ + \ttxy$^1$ \\
  TE 43 & b & \ttdx$^1$ \\
  TE 44 & b & \ttdx$^1$ \\
  TE 45 & b & \ttt$^1$ + \ttdx$^1$ \\
  TE 46 & b & \ttt$^1$ + \ttdx$^1$ \\
  TE 47 & b & \ttdx$^1$ \\
  TE 48 & b & \ttdx$^1$ \\
  TE 49 & b & \ttt$^2$ + \ttdx$^1$ + \ttxy$^1$ \\
  TE 50 & b & \ttdx$^1$ \\
  TE 51 & b & \ttt$^2$ + \ttdx$^1$ \\
  TE 52 & b & \ttdx$^1$ \\
  TE 53 & b & \ttt$^1$ + \ttdx$^1$ \\
  TE 54 & b & \ttdx$^1$ \\
  TE 55 & b & \ttdx$^1$ \\
  TE 56 & b c & \ttt$^1$ + \ttdx$^1$ \\
  TE 57 & b & \ttdx$^1$ \\
  TE 58 & b & \ttt$^1$ + \ttdx$^1$ \\
  TE 59 & b & \ttdx$^1$ \\
  TE 60 & b & \ttt$^2$ + \ttdx$^1$ \\
  TE 61 & b & \ttt$^1$ + \ttdx$^1$ \\
  TE 62 & b & \ttt$^1$ + \ttdx$^1$ \\
  TE 63 & b & \ttt$^3$ + \ttdx$^1$ \\
  TE 64 & b & \ttdx$^1$ \\
  TE 65 & b & \ttt$^1$ + \ttdx$^1$ \\
  TE 66 & b & \ttdx$^1$ \\
  TE 67 & b & \ttt$^2$ + \ttdx$^1$ \\
  TE 68 & b & \ttt$^1$ + \ttdx$^1$ \\
  TE 69 & b & \ttdx$^1$ \\
  TE 70 & b c & \ttt$^1$ + \ttdx$^1$ \\
  TE 71 & b & \ttdx$^1$ \\
  TE 72 & b & \ttdx$^1$ \\
  TE 73 & b & \ttt$^1$ + \ttdx$^1$ + \ttxy$^1$ \\
  TE 74 & b & \ttdx$^1$ \\
  TE 75 & b & \ttt$^1$ + \ttdx$^1$ \\
  TE 76 & b & \ttdx$^1$ \\
  TE 77 & b & \ttt$^3$ + \ttdx$^1$ \\
  TE 78 & b & \ttdx$^1$ \\
  TE 79 & b & \ttt$^1$ + \ttdx$^1$ \\
  TE 80 & b & \ttt$^1$ + \ttdx$^1$ \\
  TE 81 & b & \ttt$^2$ + \ttdx$^1$ \\
\hline
\end{tabular}
\end{minipage}
\tablefoot{CHEOPS LCs further required a GP-based pre-detrending against the roll angle, here denoted with \ttGProll. The LC counter refers to the CHEOPS (CH) and TESS (TE) light curves, extracted as detailed in the text in chronological order of observations. In particular, TE LC from 1 to 28, from 29 to 52, and from 53 to 81 are extracted from Sector 9, 35, and 62, respectively. All the ground-based observations reduced as explained in the text only required a normalisation scalar ($c$). See text for further details.}
\end{table*}

\begin{table}
\renewcommand{\arraystretch}{1.2}
\caption{Radial velocity jitter for each RV time series as inferred from the MCMC global analysis, after applying the polynomial detrending as specified in the third column. $c$ denotes a scalar offset; see text for further details}
\label{tab:RVjitter}
\centering
\begin{tabular}{l c c}
\hline\hline
Instrument & RV jitter [m\,s$^{-1}$] & Detrending \\
\hline
\noalign{\smallskip}
HARPS & $1.593_{-0.026}^{+0.045}$ & \ttt$^4$ \\
IRD & $0.8391_{-0.0063}^{+0.0064}$ & $c$ \\
HARPS-N & $2.151_{-0.040}^{+0.043}$ & \ttt$^3$ \\
CARMENES & $2.033_{-0.057}^{+0.060}$ & \ttt$^3$ \\
iSHELL & $4.05\pm0.30$ & $c$ \\
MAROON-X blue Feb\,2021 & $1.12_{-0.23}^{+0.21}$ & $c$ \\
MAROON-X blue Apr\,2021 & $0.007_{-0.007}^{+0.038}$ & $c$ \\
MAROON-X blue May\,2021 & $0.11_{-0.11}^{+0.16}$ & $c$ \\
MAROON-X red Feb\,2021 & $0.07_{-0.07}^{+0.15}$ & $c$ \\
MAROON-X red Apr\,2021 & $0.06_{-0.06}^{+0.20}$ & $c$ \\
MAROON-X red May\,2021 & $0.616_{-0.039}^{+0.052}$ & \texttt{dlw}$^1$ \\
\hline
\end{tabular}
\end{table}

\begin{table*}
\renewcommand{\arraystretch}{1.2}
\caption{\textbf{Main planetary parameters of the \texttt{Msample}.}}
\label{tab:Msample}
\centering
\small
\begin{tabular}{l c c c c c c}
\hline\hline
Planet & $P$ [d] & $R_p$ [$R_{\oplus}$] & $M_p$ [$M_{\oplus}$] & $\rho_p$ [$\rho_{\oplus}$] & $R_p$-location & Reference \\
\hline
\noalign{\smallskip}
TOI-732 b & $0.76837931_{-0.00000042}^{+0.00000039}$ & $1.325_{-0.058}^{+0.057}$ & $2.46\pm0.19$ & $1.06_{-0.14}^{+0.18}$ & below & This work \\
TOI-732 c & $12.252284\pm0.000013$ & $2.39_{-0.11}^{+0.10}$ & $8.04_{-0.48}^{+0.50}$ & $0.59_{-0.08}^{+0.10}$ & above & This work \\
GJ 1132 b & $1.628931\pm0.000027$ & $1.13\pm0.056$ & $1.66\pm0.23$ & $1.15\pm0.23$ & below & \citet{bonfils2018} \\
GJ 1214 b & $1.58040433\pm0.00000013$ & $2.74_{-0.053}^{+0.050}$ & $8.17\pm0.43$ & $0.396\pm0.031$ & above & \citet{cloutier2021GJ1214} \\
GJ 1252 b & $0.51824160\pm0.00000069$ & $1.180\pm0.078$ & $1.32\pm0.28$ & $0.80\pm0.23$ & below & \citet{crossfield2022} \\
GJ 3090 b & $2.853136_{-0.000038}^{+0.000064}$ & $2.13\pm0.11$ & $3.34\pm0.72$ & $0.346\pm0.092$ & above & \citet{almenara2022} \\
GJ 3473 b & $1.1980035_{-0.0000019}^{+0.0000018}$ & $1.264_{-0.049}^{+0.050}$ & $1.86\pm0.30$ & $0.92\pm0.18$ & below & \citet{kemmer2020} \\
GJ 357 b & $3.93072_{-0.00006}^{+0.00008}$ & $1.217_{-0.083}^{+0.084}$ & $1.84\pm0.31$ & $1.02\pm0.27$ & below & \citet{luque2019} \\
GJ 367 b & $0.321962_{-0.000012}^{+0.000010}$ & $0.718\pm0.054$ & $0.546\pm0.078$ & $1.48\pm0.39$ & below & \citet{lam2021} \\
GJ 486 b & $1.467119_{-0.000030}^{+0.000031}$ & $1.305_{-0.067}^{+0.063}$ & $2.82_{-0.12}^{+0.11}$ & $1.27\pm0.20$ & below & \citet{trifonov2021} \\
HD 260655 b & $2.76953\pm0.00003$ & $1.24\pm0.023$ & $2.14\pm0.34$ & $1.12\pm0.19$ & below & \citet{luque2022HD260655} \\
HD 260655 c & $5.70588\pm0.00007$ & $1.533_{-0.046}^{+0.051}$ & $3.09\pm0.48$ & $0.86\pm0.16$ & below & \citet{luque2022HD260655} \\
K2-146 b & $2.6698\pm0.0001$ & $2.25\pm0.10$ & $5.6\pm0.7$ & $0.492\pm0.090$ & above & \citet{lam2020} \\
K2-18 b & $32.940045\pm0.000010$ & $2.61\pm0.087$ & $8.63\pm1.35$ & $0.485\pm0.090$ & above & \citet{benneke2019} \\
K2-25 b & $3.48456408_{-0.00000050}^{+0.00000060}$ & $3.44\pm0.12$ & $24.5_{-5.2}^{+5.7}$ & $0.60\pm0.15$ & above & \citet{stefansson2020} \\
K2-3 b & $10.05465350_{-0.00000091}^{+0.00000088}$ & $2.078_{-0.067}^{+0.076}$ & $5.11_{-0.64}^{+0.65}$ & $0.569\pm0.093$ & above & \citet{diamondLowe2022} \\
Kepler-138 c & $13.78150_{-0.00009}^{+0.00007}$ & $1.51\pm0.04$ & $2.3_{-0.5}^{+0.6}$ & $0.67\pm0.17$ & below & \citet{piaulet2023} \\
L 168-9 b & $1.40150\pm0.00018$ & $1.39\pm0.09$ & $4.6\pm0.56$ & $1.71\pm0.39$ & below & \citet{astudilloDefru2020} \\
L 98-59 c & $3.6904\pm0.0003$ & $1.35\pm0.07$ & $2.42_{-0.34}^{+0.35}$ & $0.98\pm0.21$ & below & \citet{cloutier2019} \\
L 98-59 d & $7.4507245_{-0.0000046}^{+0.0000081}$ & $1.521_{-0.10}^{+0.12}$ & $1.94\pm0.28$ & $0.55\pm0.14$ & below & \citet{demangeon2021} \\
LHS 1140 b & $24.73694_{-0.00040}^{+0.00041}$ & $1.635\pm0.046$ & $6.38_{-0.44}^{+0.46}$ & $1.46\pm0.16$ & below & \citet{lilloBox2020} \\
LHS 1140 c & $3.77792\pm0.00003$ & $1.169_{-0.038}^{+0.037}$ & $1.76_{-0.16}^{+0.17}$ & $1.10\pm0.15$ & below & \citet{lilloBox2020} \\
LHS 1478 b & $1.9495378_{-0.0000041}^{+0.0000040}$ & $1.242_{-0.049}^{+0.051}$ & $2.33\pm0.20$ & $1.22\pm0.18$ & below & \citet{soto2021} \\
LP 791-18 c & $4.9899093_{-0.0000072}^{+0.0000074}$ & $2.438\pm0.096$ & $7.1\pm0.7$ & $0.490\pm0.075$ & above & \citet{peterson2023} \\
LTT 1445 A b & $5.3587657_{-0.0000042}^{+0.0000043}$ & $1.305_{-0.061}^{+0.066}$ & $2.87_{-0.25}^{+0.26}$ & $1.29\pm0.22$ & below & \citet{winters2022} \\
LTT 1445 A c & $3.1239035_{-0.0000036}^{+0.0000034}$ & $1.147_{-0.054}^{+0.055}$ & $1.54_{-0.19}^{+0.20}$ & $1.02\pm0.19$ & below & \citet{winters2022} \\
TOI-1075 b & $0.6047328\pm0.0000032$ & $1.791_{-0.08}^{+0.12}$ & $9.95_{-1.3}^{+1.4}$ & $1.73\pm0.37$ & below & \citet{essack2023} \\
TOI-1201 b & $2.4919863_{-0.0000031}^{+0.0000030}$ & $2.415_{-0.090}^{+0.091}$ & $6.28_{-0.88}^{+0.84}$ & $0.446\pm0.079$ & above & \citet{kossakowski2021} \\
TOI-1231 b & $24.245586_{-0.000066}^{+0.000064}$ & $3.65_{-0.15}^{+0.16}$ & $15.4\pm3.3$ & $0.317\pm0.079$ & above & \citet{burt2021} \\
TOI-1235 b & $3.444717_{-0.000042}^{+0.000040}$ & $1.694_{-0.077}^{0.080}$ & $5.9_{-0.61}^{+0.62}$ & $1.21\pm0.21$ & below & \citet{bluhm2020} \\
TOI-1634 b & $0.9893436\pm0.0000020$ & $1.749\pm0.079$ & $10.14\pm0.95$ & $1.90\pm0.31$ & below & \citet{hirano2021} \\
TOI-1695 b & $3.1342791_{-0.0000063}^{+0.0000071}$ & $1.9_{-0.14}^{+0.16}$ & $6.36\pm1.0$ & $0.93\pm0.26$ & below & \citet{cherubim2023} \\
TOI-244 b & $7.397225_{-0.000026}^{+0.000023}$ & $1.52\pm0.12$ & $2.68\pm0.3$ & $0.76\pm0.20$ & below & \citet{castroGonzalez2023} \\
TOI-269 b & $3.6977104\pm0.0000037$ & $2.77\pm0.12$ & $8.8\pm1.4$ & $0.414\pm0.085$ & above & \citet{cointepas2021} \\
TOI-270 b & $3.3601538\pm0.0000048$ & $1.206\pm0.039$ & $1.58\pm0.26$ & $0.901\pm0.172$ & below & \citet{vanEylen2021} \\
TOI-270 c & $5.6605731\pm0.0000031$ & $2.355\pm0.064$ & $6.15\pm0.37$ & $0.471\pm0.048$ & above & \citet{vanEylen2021} \\
TOI-270 d & $11.379573\pm0.000013$ & $2.133\pm0.058$ & $4.78\pm0.43$ & $0.493\pm0.060$ & above & \citet{vanEylen2021} \\
TOI-776 b & $8.24661_{-0.00004}^{+0.00005}$ & $1.85\pm0.13$ & $4.0\pm0.9$ & $0.63\pm0.19$ & below & \citet{luque2021} \\
TRAPPIST-1 b & $1.510826\pm0.000006$ & $1.116_{-0.012}^{+0.014}$ & $1.374\pm0.069$ & $0.989\pm0.060$ & below & \citet{agol2021} \\
TRAPPIST-1 c & $2.421937\pm0.000018$ & $1.097_{-0.012}^{+0.014}$ & $1.308\pm0.056$ & $0.991\pm0.055$ & below & \citet{agol2021} \\
TRAPPIST-1 d & $4.049219\pm0.000026$ & $0.788_{-0.010}^{+0.011}$ & $0.388\pm0.012$ & $0.793\pm0.040$ & below & \citet{agol2021} \\
TRAPPIST-1 e & $6.101013\pm0.000035$ & $0.920_{-0.012}^{0.013}$ & $0.692\pm0.022$ & $0.889\pm0.046$ & below & \citet{agol2021} \\
TRAPPIST-1 f & $9.20754\pm0.000032$ & $1.045_{-0.012}^{+0.013}$ & $1.039\pm0.031$ & $0.910\pm0.042$ & below & \citet{agol2021} \\
TRAPPIST-1 g & $12.352446\pm0.000054$ & $1.129_{-0.013}^{+0.015}$ & $1.321\pm0.038$ & $0.920\pm0.043$ & below & \citet{agol2021} \\
TRAPPIST-1 h & $18.772866\pm0.000021$ & $0.755\pm0.014$ & $0.326\pm0.020$ & $0.757\pm0.063$ & below & \citet{agol2021} \\
\hline
\end{tabular}
\tablefoot{$R_p$-location refers to the location of the exoplanets with respect to the radius valley as derived by the SVM algorithm described in Sec.~\ref{sec:Rvalley}.}
\end{table*}

\begin{figure}
    \centering
    \includegraphics[width=\columnwidth]{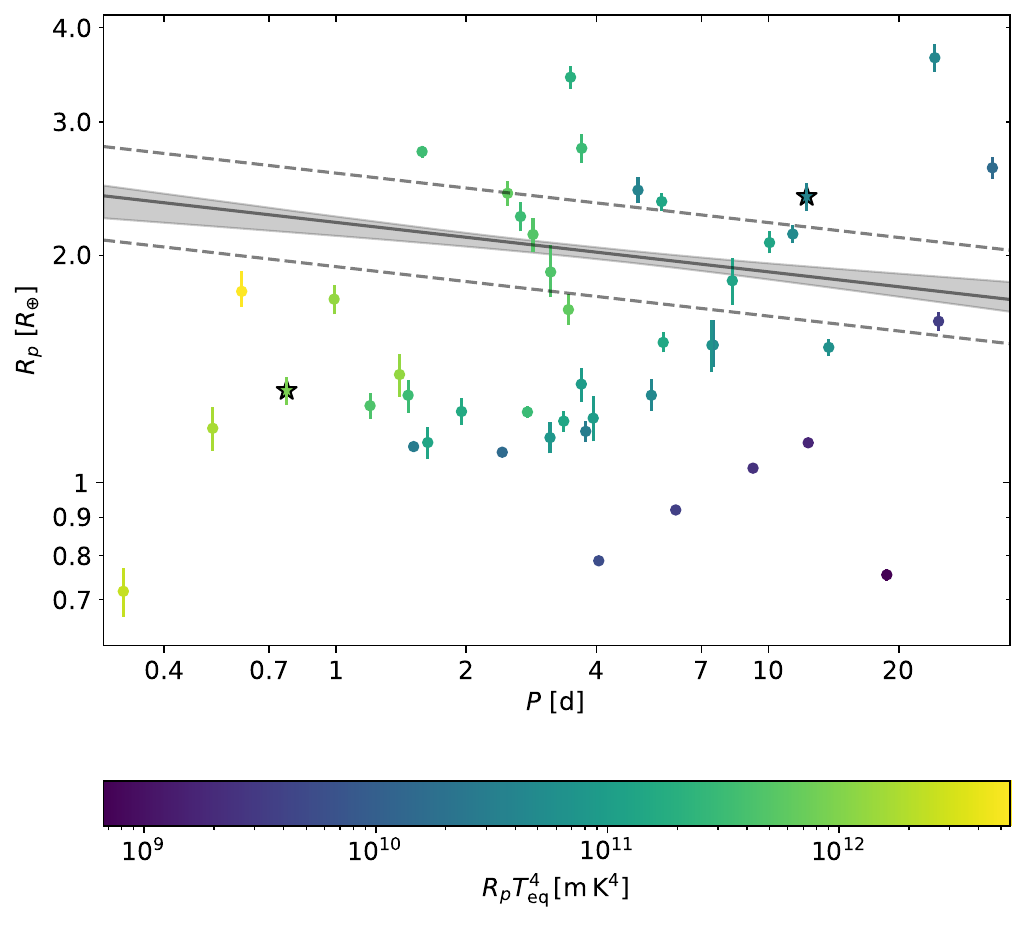}
    \caption{Same as Fig.~\ref{fig:RvalleyP}, but with the markers colour-coded against $R_p T_{\mathrm{eq}}^4$, which correlates with the core-powered mass loss strength \citep{gupta2019}.}
    \label{fig:RvalleyPcbarCorePow}
\end{figure}

\end{appendix}

\end{document}